\pdfoutput=1
\documentclass[usenatbib]{mn2e}
\usepackage{apjfonts}
\usepackage{amssymb}
\usepackage{amsmath}
\usepackage{ctable}
\usepackage{url}
\usepackage{fixltx2e} 
\newcommand{\Msun}{\mathrm{M}_{\odot}}
\newcommand{\Msunyr}{\Msun\ \mathrm{yr}^{-1}}

\newcommand{\Rvir}{R_{\rm vir}}
\newcommand\altaffilmark[1]{$^{#1}$}
\newcommand\altaffiltext[1]{$^{#1}$}
\newcommand{\etal}{et al.}

\title[Gusty, gaseous flows of FIRE]{Gusty, gaseous flows of FIRE: galactic winds in cosmological simulations with explicit stellar feedback\vspace{-0.5cm}}

\vspace{-0.2cm}
\author[Muratov \etal]{
\parbox[t]{\textwidth}{ 
Alexander L.~Muratov\thanks{E-mail:amuratov@ucsd.edu}\altaffilmark{1}, 
Du\v{s}an Kere\v{s}\altaffilmark{1},
Claude-Andr{\'e} Faucher-Gigu{\`e}re\altaffilmark{2},
Philip F.~Hopkins\altaffilmark{3},
Eliot Quataert\altaffilmark{4}, \&\
Norman Murray\altaffilmark{5,6}  
}
\vspace*{6pt} \\
\altaffiltext{1}{Department of Physics, Center for Astrophysics and Space Sciences, University of California at San Diego, 9500 Gilman Drive, La Jolla, CA 92093} \\
\altaffiltext{2}{Department of Physics and Astronomy and CIERA, Northwestern University, 2145 Sheridan Road, Evanston, IL 60208, USA} \\ 
\altaffiltext{3}{TAPIR, Mailcode 350-17, California Institute of Technology, Pasadena, CA 91125, USA} \\
\altaffiltext{4}{Department of Astronomy and Theoretical Astrophysics Center, University of California Berkeley, Berkeley, CA 94720} \\
\altaffiltext{5}{Canadian Institute for Theoretical Astrophysics, 
60 St.\ George Street, University of Toronto, ON M5S 3H8, Canada} \\
\altaffiltext{6}{Canada Research Chair in Astrophysics \vspace{-0.5cm}} \\
\vspace{-0.5cm}
}
\date{Submitted to MNRAS, December, 2014\vspace{-0.6cm}}
\begin{document}
\maketitle

\date{\today}

\begin{abstract}

We present an analysis of the galaxy-scale gaseous outflows from the FIRE (Feedback in Realistic Environments) simulations. This suite of hydrodynamic cosmological zoom simulations resolves formation of star-forming giant molecular clouds to $z=0$, and features an explicit stellar feedback model on small scales. Our simulations reveal that high redshift galaxies undergo bursts of star formation followed by powerful gusts of galactic outflows that eject much of the ISM and temporarily suppress star formation. At low redshift, however, sufficiently massive galaxies corresponding to L*-progenitors develop stable disks and switch into a continuous and quiescent mode of star formation that does not drive outflows far into the halo. Mass-loading factors for winds in L*-progenitors are $\eta \approx 10$ at high redshift, but decrease to $\eta \ll 1$ at low redshift. Although lower values of $\eta$ are expected as halos grow in mass over time, we show that the strong suppression of outflows with decreasing redshift cannot be explained by mass evolution alone. Circumgalactic outflow velocities are variable and broadly distributed, but typically range between one and three times the circular velocity of the halo. Much of the ejected material builds a reservoir of enriched gas within the circumgalactic medium, some of which could be later recycled to fuel further star formation. However, a fraction of the gas that leaves the virial radius through galactic winds is never regained, causing most halos with mass $M_h \le 10^{12} \Msun$ to be deficient in baryons compared to the cosmic mean by $z=0$.

\end{abstract}

\begin{keywords}
galaxies: formation --- galaxies: evolution --- stars: formation --- cosmology: theory\vspace{-0.5cm}
\end{keywords}

\vspace{-1.1cm}
\section{Introduction}

Young, hot stars release energy and momentum in the form of stellar winds, ionizing photons, radiation pressure, and powerful supernova explosions. Nearby gas absorbs this energy and momentum, is heated up and pushed away supersonically from the dense regions where stars formed, driving shocks that propagate through the interstellar medium (ISM). Working in unison, a population of young stars is capable of accelerating a substantial mass of gas beyond the escape velocity of the local gravitational potential, totally disrupting the local star-forming cloud, and potentially driving a large-scale galactic wind \citep{mathews_baker71, larson74, haehnelt95}. The effects of these energetic processes, both on the local state of the star-forming gaseous clouds, and on the dynamics and morphological structure of the galaxy as a whole, are collectively known as stellar feedback.

Winds attributed to stellar feedback have been directly observed in galaxies near and far using a variety of techniques in every window of the electromagnetic spectrum (see \citealt{veilleux_etal05} and references therein for a review). The approaches can generally be divided into either studying diffuse, heated gas following an outflow (e.g. \citealt{heckman_etal90}) or direct detection of kinematically distinct material from the rest frame of the host galaxy via Doppler shifted emission (e.g. \citealt{martin98}) and absorption lines (e.g. \citealt{heckman00, grimes_etal09, weiner_etal09, steidel10, martin_etal12, bouche_etal12, rubin_etal13}). Taken together, observations reveal that while galactic winds are nearly ubiquitous around starbursting galaxies, the relative strengths of the winds as measured by the amount of material they move can vary substantially from case to case. In addition the winds' kinematic properties (speeds range from tens to thousand of km/s) and phase compositions (cold molecular gas, warm gas detectable in the UV, and hot X-ray gas can often be seen simultaneously) are very diverse. 

The net result of these outflows is the transport of material from the dense star-forming regions to the so-called circumgalactic medium (CGM), which is loosely defined as the gas outside the galaxy but still within the gravitational potential of the dark matter halo that the galaxy inhabits. While it has been difficult to directly probe the total mass and phase composition of the CGM, detections of metal absorption everywhere from a few kpc from galactic centers to regions at their virial radius and beyond hints at a vast reservoir of gas that has been at least partially enriched with metals from galactic outflows \citep{chen_etal98, chen_etal01, gauthier09, tumlinson_etal11, kacprzak_churchill11, werk_etal14, lehner_etal14}. The high redshift CGM has also been extensively observed \citep{steidel10, rudie12} and provides important constraints for theoretical models which seek to determine the origin of gas in present-day galaxies \citep{keres05, faucher-giguere10, faucher-giguere11b, vandevoort_etal11b, shen_etal13, ford_etal14, faucher-giguere_etal14}. Filamentary infall from  the high-redshift intergalactic medium (IGM) must also traverse the CGM, where it can run into and interact with outflows, before it can accrete onto galaxies. At later times, the CGM of massive halos contains a hot atmosphere, which can interfere with further accretion while also gradually cooling and supplying the galaxy. Recognizing the CGM as an environment where the IGM interfaces with the galaxy through this complex series of interactions has been a crucial step in properly framing the essential unsolved questions of galaxy formation \citep{keres05}. 

Deciphering the nature of galactic winds and the dynamic between the ISM and CGM is particularly important in the context of regulating gas consumption during galaxy formation. Absent regulation through strong stellar feedback, the known universal abundance of baryons in the $\Lambda$CDM framework would suggest that there should be many more galaxies with high stellar mass than what is observed (e.g. \citealt{white91}). Even when adding together the mass of all stars to the mass of the ISM, observed galaxies generally do not appear to contain their share of available baryons in the cosmological context \citep{fukugita_etal98, mcgaugh_etal10}. Past efforts to account for the entire budget of "missing baryons" in the CGM and IGM have generally come up short, particularly if it is required that the gas would be at least as hot as the virial temperature of L* galaxies \citep{bregman07, anderson_bregman10, shull_etal12, anderson_etal13}. However, the latest results from the HST COS Halos Survey hint that when taking into account ionization corrections, massive amounts of gas may be hidden in warm and cool clouds which together could account for most, if not all of the missing baryons \citep{werk_etal14}, though the long-term stability of such a medium would likely require it to be in a complex dynamical state. In galaxies less massive than L*, winds generated by stellar feedback are thought to suppress star formation and eject gas from galaxies into the CGM and IGM \citep{larson74, white78, dekel86}. This hypothesis is used in all modern semi-analytical models of galaxy formation to match observed constraints on a wide range of galactic properties at various epochs \citep{benson03, bower06}.

Hydrodynamic cosmological simulations promise to provide a direct, versatile, and comprehensive way of studying galaxy formation, but in practice, properly implementing realistic stellar feedback physics that generates galactic winds has proven to be a significant challenge. Early efforts encountered the so-called "overcooling" problem, where thermal energy injection from supernovae dissipated quickly \citep{katz92a}, before it could generate a strong dynamical effect within the ISM (see \citet{kim_ostriker14} for a modern interpretation; see also \citet{martizzi_etal14}). Without strong dynamical effects from SNe that could generate galactic outflows, the star formation rates and stellar masses found in many cosmological galaxy simulations remain systematically too high (e.g. \citealt{keres09b}).

As resolving the relevant physical processes self-consistently has proven difficult, simulators have turned to physically-motivated subgrid recipes that allow the supernova energy from massive stars to couple effectively with the surrounding ISM and drive the galactic wind. In Smoothed Particle Hydrodynamics (SPH) simulations, the subgrid models which have been used can generally be divided into two classes. The first relies on injecting supernova energy as kinetic energy to selected SPH particles, and subsequently decoupling the particles from hydrodynamic interactions to ensure their escape from the galaxy (e.g. \citealt{springel03a, oppenheimer06, oppenheimer10, vogelsberger13}, see \citealt{dallavecchia08, schaye10}, for kinetic methods that do not decouple). The second approach uses the so-called "blastwave" model to account for the unresolved expansion phase of the supernova remnant following the explosion (e.g. \citealt{thacker00b, governato07, stinson06, guedes11, stinson13, shen_etal14, christensen_etal15}). In practice, this means that after the thermal energy is injected into surrounding particles, cooling is disabled for a fixed length of time (other groups e.g. \citealt{dallavecchia_schaye12, schaye_etal15} do not explicitly disable cooling but instead bottle up thermal energy until it is guaranteed to produce a blastwave). Simulations using either Adaptive Mesh Refinement \citep{agertz_etal11} or Moving Mesh \citep{vogelsberger_etal14} that have successfully run to $z=0$ have had to rely on similar subgrid models.

These subgrid approaches have, by construction, proven successful remedies for the overcooling problem, and have been used to produce galaxies with properties that are consistent with a variety of observational constraints of late-type galaxies at $z=0$. However, these simulations have limited predictive power for properties of the CGM. In large-volume cosmological simulations where star-forming regions are not well resolved (e.g. \citealt{ford_etal14, vogelsberger_etal14}), the sub-resolution model encapsulates a significant portion of the galaxy. Galactic winds from these coarse (\textasciitilde 1 kpc) regions are launched with unresolved phase structure, pre-determined uniform velocities, and pre-determined mass-loading efficiencies. Simulations that do resolve star-forming regions but employ the blastwave model (e.g. \citealt{stinson13}) may overheat the galactic wind material artificially as a consequence of temporarily disabled cooling \citep{agertz13, martizzi_etal14}. Gas that has joined the CGM through galactic winds generated by these prescriptions may arrive with unrealistic properties.

In order for simulations to play a role in improving our understanding of the formation and dynamics of the CGM, particularly given the complex, multi-phase picture emerging from the latest observations \citep{tumlinson_etal11, werk_etal14}, the level of detail and sophistication in stellar feedback models must improve. In this work, we analyze the outflowing (and infalling) gas seen in the galaxies and CGM of the Feedback In Realistic Environments (FIRE) simulations\footnote{Project website: http://fire.northwestern.edu}, first presented in \citet{hopkins_etal14}. Unlike the subgrid recipes which involve kinetically prescribed decoupled winds and cooling-suppressed blastwaves, the FIRE simulations solve the "overcooling" problem by explicitly modeling the radiation pressure, stellar winds and ionizing feedback from young stars as taken directly from the population synthesis code Starburst99 \citep{leitherer99}. These "early feedback" mechanisms act before SNe, heating and stirring the surrounding interstellar medium which is necessary to match conditions in star-forming regions such as Carina \citep{harper-clark_murray09} and 30 Dor \citep{lopez_etal11, pellegrini_etal11}. SNe are implemented by taking into account their energy and momentum input. When the cooling radius of SNe is resolved, SN energy injected is free to expand and generate momentum in the ISM before too much energy is radiated away. When this scale is poorly resolved, momentum accumulation from SN remnant evolution below the resolution scale is added to the surrounding gas. This model is physically realistic when it is applied on the scale of giant molecular clouds, meaning that a resolution of several to tens of parsecs is required. The physical feedback implementation in FIRE successfully regulates mass accumulation in galaxies and provides a physical explanation for the inefficiency of star formation in galactic disks \citep{kennicutt83, kennicutt98, genzel_etal10}. We stress that we allow hydrodynamical interactions and cooling of all gas at all times, unlike in typical sub-grid models. This is critical to make meaningful predictions for the phase structure of circumgalactic gas.
 
This feedback model has been developed and tested in idealized galaxy simulations (\citealt{hopkins11a, hopkins12a, hopkins12b, hopkins12c}; see also \citealt{agertz13}).  In \citet{hopkins_etal14} (H14) the model has been extended and appropriately modified to be applied in a cosmological framework. These simulations were able to reproduce the correct relationship between stellar mass and halo mass at a variety of epochs (see also \citealt{agertz_kravtsov14, agertz_kravtsov15, trujillo-gomez_etal15}) and provide a way to explain the under-abundance of baryons in galaxies, relative to the cosmic mean predicted by $\Lambda$CDM \citep{fukugita_etal98, mcgaugh_etal10}. Other successes of the simulations include reproducing HI covering fractions around Lyman-break galaxies \citep{faucher-giguere_etal14}, mass-metallicity relations for both gaseous and stellar components of galaxies \citep{ma_etal15}, and dark matter profiles that ease tension between $\Lambda$CDM and observed density structure of galaxies \citep{chan_etal15, onorbe_etal15}.

Using these simulations, we explore the degree to which this stellar feedback model drives winds through the galaxy and the halo, their kinematics and phases, and the ways that these outflows affect the long-term fate of the galaxy and the circumgalactic medium. We explore the relationship between gaseous infall, star formation, and outflow rates as measured in simulations. We will also attempt to bridge the gap between the methods typically employed by observers and simulators, and offer a framework for future comparisons between simulations employing various techniques to measure outflow rates.

The present work will focus on a measurement of the mass-loading factor $\eta$, which quantifies the ratio of the outflow rate to the star formation rate. We explore the nuances associated with measuring $\eta$ in simulations, and provide formulae for determining mass-loading factor and characteristic wind velocity as a function of halo properties and redshift. We also estimate the amount of gas each halo can deposit into its own CGM and how much of it leaves the virial radius and joins the IGM. The results of this analysis will be directly applicable to semi-analytic models and large-volume cosmological simulations.  Phase structure, kinematics, and metallicity evolution of winds will be explored in a companion paper. 
 
Section \ref{sec:sims} will give a brief overview of the FIRE simulations. In Section \ref{sec:measure}, we discuss our method for measuring inflow and outflow rates. We present analysis of inflow and outflow rates over time for a variety of halos in Section \ref{sec:timeevo}. We provide measurements of the mass-loading factor, $\eta$, as well as wind velocities in Section \ref{sec:eta}. We discuss the long-term impacts of galactic outflows on the evolution of galaxies and the CGM in Section \ref{sec:discussion} and summarize our conclusions in Section \ref{sec:conclusion}.

\section{Simulations \& Analysis}
\label{sec:sims}

\begin{footnotesize}
\ctable[
caption={{\normalsize Simulation Initial Conditions}\label{tbl:sims}},center,star
]{lccccccccl}{
\tnote[ ]{Parameters describing the initial conditions for our simulations (units are physical): \\
{\bf (1)} Name: Simulation designation. \\
{\bf (2)} $M_{h}(z=0)$: Mass of the $z=0$ ``main'' halo (most massive halo in the high-resolution region). \\
{\bf (3)} $M_{h}(z=2)$: Mass of the $z=2$ ``main'' halo. \\
{\bf (4)} $M_{*}(z=0)$: Stellar mass of the $z=0$ ``main'' halo. \\
{\bf (5)} $M_{*}(z=2)$: Stellar mass of the $z=2$ ``main'' halo. \\
{\bf (4)} $m_{b}$: Initial baryonic (gas and star) particle mass in the high-resolution region, in our highest-resolution simulations. \\ 
{\bf (5)} $\epsilon_{b}$: Minimum baryonic force softening (minimum SPH smoothing lengths are comparable or smaller).\\
{\bf (6)} $m_{dm}$: Dark matter particle mass in the high-resolution region, in our highest-resolution simulations. \\ 
{\bf (7)} $\epsilon_{dm}$: Minimum dark matter force softening (fixed in physical units at all redshifts). 
}
}{
\hline\hline
\multicolumn{1}{c}{Name} &
\multicolumn{1}{c}{$M_{h}(z=0)$} &
\multicolumn{1}{c}{$M_{h}(z=2)$} &
\multicolumn{1}{c}{$M_{*}(z=0)$} &
\multicolumn{1}{c}{$M_{*}(z=2)$} &
\multicolumn{1}{c}{$m_{b}$} & 
\multicolumn{1}{c}{$\epsilon_{b}$} & 
\multicolumn{1}{c}{$m_{dm}$} & 
\multicolumn{1}{c}{$\epsilon_{dm}$} & 
\multicolumn{1}{c}{Merger}  \\
\multicolumn{1}{c}{\ } &
\multicolumn{1}{c}{[$M_{\sun}$]} & 
\multicolumn{1}{c}{[$M_{\sun}$]} & 
\multicolumn{1}{c}{[$M_{\sun}$]} & 
\multicolumn{1}{c}{[$M_{\sun}$]} & 
\multicolumn{1}{c}{[$M_{\sun}$]} &
\multicolumn{1}{c}{[pc]} &
\multicolumn{1}{c}{[$M_{\sun}$]} &
\multicolumn{1}{c}{[pc]} &
\multicolumn{1}{c}{History} \\
\hline
{\bf m09} & 2.5e9 & 1.3e9 & 4.6e4 & 4.1e4 & 2.6e2 & 1.4 & 1.3e3 & 30 & normal\\
{\bf m10} & 7.8e9 & 3.8e9 & 2.3e6 & 1.7e6 & 2.6e2 & 3.0 & 1.3e3 & 30 & normal\\
{\bf m11} &  1.4e11 & 3.8e10 & 2.3e9 & 3.4e8 & 7.1e3 & 7.0 & 3.5e4 & 70 & quiescent  \\ 
{\bf m12v} &  6.3e11 & 2.0e11& 2.8e10 & 2.3e9 & 3.9e4 & 10 & 2.0e5 & 140 & violent\\ 
{\bf m12q} & 1.2e12 & 5.1e11 & 2.2e10 & 7.0e9 & 7.1e3 & 10 & 2.8e5 & 140 & late merger  \\ 
{\bf m12i} & 1.1e12 & 2.7e11 & 6.1e10 & 3.9e9 & 5.0e4 & 14 & 2.8e5 & 140 & normal \\ 
\hline \\
{\bf z2h350} & - & 7.9e11 & - & 9.0e9 & 5.9e4 & 9 & 2.9e5 & 143 & normal   \\
{\bf z2h400} & - & 7.9e11 & - & 7.0e9 & 5.9e4 & 9 & 2.9e5 & 143 & quiescent   \\
{\bf z2h450} & - & 8.7e11 & - & 1.3e10 & 5.9e4 & 9 & 2.9e5 & 143 & normal  \\
{\bf z2h506} & - & 1.2e12 & - & 1.8e10 & 5.9e4 & 9 & 2.9e5 & 143 & violent  \\
{\bf z2h550} & - & 1.9e11 & - & 4.4e9 &  5.9e4 & 9 & 2.9e5 & 143 & quiescent   \\
{\bf z2h600} & - & 6.7e11 & - & 1.7e10 & 5.9e4 & 9 & 2.9e5 & 143 & violent   \\
{\bf z2h650} & - & 4.0e11 & - & 6.6e9 & 5.9e4 & 9 & 2.9e5 & 143 & normal  \\
{\bf z2h830} & - & 5.4e11 & - & 1.4e10 & 5.9e4 & 9 & 2.9e5 & 143 & normal   \\
\hline\hline\\
}
\end{footnotesize}

The simulations described in this work were performed with the pressure-entropy formulation of SPH as described in \citet{hopkins13}, along with additional modifications detailed in H14. This formulation of SPH mitigates issues discovered in previous incarnations around contact discontinuities and due to artificial viscosity. These fixes are particularly crucial for detailed study of the CGM, as the numerical artifacts of some older SPH codes produced dense clumps in the halo, and underestimated cooling rates from the hot halo atmosphere (\citealt{keres12, hayward_etal14}; Kere\v{s} et al. in prep). 

The simulations employ the zoom-in technique to follow select Lagrangian regions around the halos of interest in considerably larger cosmological boxes. This technique limits our capacity to study the distant IGM, but captures the essential physics in circumgalactic gas flows, and gives us excellent resolution in the galaxy itself, including fully resolved structures in star forming regions. This enables us to use star forming threshold density of $10-100 cm^{-3}$ for all of the galaxies analyzed here and implement stellar feedback within giant molecular clouds. We analyze the same suite of simulations that was described in H14, many of which had initial conditions that were drawn from the AGORA suite \citep{kim_etal14}, which samples halos in decadal increments in mass from $10^9 \Msun$ to $10^{13} \Msun$ at $z=0$. Particular emphasis is placed on  halos that are in the L* range with mass \textasciitilde $10^{12} \Msun$ at z=0 - the AGORA sample provides two such initial conditions (\textbf{m12q} and \textbf{m12i}), which we supplement with an additional set of L*-progenitor initial conditions (\textbf{m12v}) studied extensively in prior work \citep{keres09c, faucher-giguere11a}. The isolated dwarf galaxies \textbf{m09} and \textbf{m10} are extensively described in \citet{onorbe_etal15}. To complement this sample, we also use the suite of simulations from \citet{faucher-giguere_etal14}, which only ran until $z=2$, but provide a sample of relatively high-mass halos (up to $10^{12} \Msun$ at $z=2$). We refer to this sample as the \textbf{z2h} sample. Although the zoom-in technique generally focuses on a single ``main'' halo at $z=0$, the multitude of progenitor halos at high redshift were also analyzed, and included in the results presented in Section \ref{sec:eta}. We exclude \textbf{m13} (described in H14) due to concern that the resolution is too low to properly capture galactic wind generation.  

The particle masses, softenings, and cosmological parameters employed in each run are give in Table \ref{tbl:sims}, which is in part reproduced from H14, and \citet{faucher-giguere_etal14}. We also include stellar and halo masses at $z=2$ and $z=0$ within $\Rvir$ for each halo. Note that a few values for halo mass differ slightly from the other papers due to different methods of halo finding. 

We use the public distribution of Amiga Halo Finder (AHF) \citep{knollmann_knebe09} to determine halo centers and measure central velocities using the adaptive mesh hierarchy method. We take the definition of virial overdensity from \citet{bryan98}, which evolves with redshift. We connect ``main'' halos to their progenitor halos using AHF's merger tree module. We mandate that the physical value of the virial radius must stay constant or increase monotonically with time, to ensure relatively smooth variation. We restrict most of our analysis to central galaxies, where this monotonic growth is expected, rather than satellite galaxies. For satellites, it is considerably more complicated to define quantities like the virial radius and the circular velocity, which we use to compare halos in Section \ref{sec:eta}. Furthermore, satellites are relatively poorly resolved in our simulations compared to zoomed-in isolated galaxies of the same mass. We have verified that satellites contribute relatively little star formation and CGM outflows compared to central galaxies.

Star formation rates are computed by adding up the mass of all stellar particles that have been formed since the previous snapshot, and then dividing by the time interval between the two snapshots. To account for mass loss between the exact formation time of each stellar particle and the snapshot time, we boost the measured star formation rate by 15\%, an approximation based on our typical snapshot interval (\textasciitilde50 Myr) and the expected mass loss due to stellar evolution from a \citet{kroupa01} IMF. In this work, we consider only star formation within 0.2$\Rvir$ when comparing to outflow rates from the halo center. The star formation rate outside 0.2$\Rvir$ is also tracked in case it is relevant for anomalous outflow episodes. We compute star formation rates on a fine temporal resolution scale equivalent to the snapshot spacing.

We quantify mass outflow rates from galaxies using the mass-loading factor, $\eta$: 

\begin{equation}
\eta = \frac {\dot{M}_{out}}{\dot{M}_{*}},
\end{equation}

\noindent where $\dot{M}_{out}$ is the gas outflow rate and $\dot{M}_{*}$ is the star formation rate. This quantity has been employed in both observational studies and in simulations to describe the strength of galactic winds.

In \citet{hopkins12b}, the mass-loading factors were measured in four isolated galaxy simulations, which featured a preliminary version of the feedback prescription employed in the FIRE simulations. 
In the present work, we aim for a more thorough investigation of outflows in the CGM in a cosmological framework. Using the large suite of simulations, our data provide enough halos to produce statistically meaningful fits for various properties of outflows for halos at various epochs throughout cosmic time, which can be applied directly in semi-analytic models and larger simulations which cannot resolve the sources of feedback directly. We focus on measuring a net flux at various points in the halo, including a measurement of how much gas leaves the virial radius.  

We also explore the nuances in measuring mass-loading factors using a variety of methods, and show how the derived values are sensitive to spatial and temporal factors. The next section will describe the methodologies we have adopted for computing outflow rates. Section \ref{sec:timeevo} will demonstrate how these outflow rates are correlated with star formation in a way that can be quantified with a mass-loading factor $\eta$. Section \ref{sec:eta} provides measurements of $\eta$, as well as wind velocity, for our large sample of halos. We also provide fits that can be used in semi-analytical models or large-scale simulations, as well as to guide observational studies.

\section{Measuring Outflow Rates} 
\label{sec:measure}

Using the Lagrangian property of SPH particles, it is possible to track parcels of gas directly as they drift through the simulated volume (e.g. \citealt{keres05, oppenheimer10, ford_etal13a}). Though the SPH particles employed in our simulations certainly can be analyzed with particle tracking, we elect to use kinematics to distinguish outflowing and infalling gas. We find this to be sufficient for determining instantaneous flow rates. Whether a particle is considered an outflow or an inflow is simply determined by the velocity in the radial direction relative to the halo center. In other words, an SPH particle is counted as an outflow if at the epoch of the snapshot, it has a velocity $\vec{v}$ relative to the halo center such that:

\begin{equation}
v_{rad} \equiv \vec{v} \cdot \frac{\vec{r}}{|r|} > v_{cut}, 
\label{eq:lala}
\end{equation} 

\noindent where $\vec{r}$ is the position vector from the halo center and $|r|$ is the distance from the halo center, and  $v_{cut}$ is the threshold velocity. Using specific values for $v_{cut}$ could allow a direct comparison with observations at a given resolution in velocity, or where a known velocity dispersion in gas unrelated to the galactic wind can contribute to the rates. We use the same threshold to identify infalling gas, only reversing the sign of the velocity vector. Infall rates are given with a negative sign. 

We have experimented with a variety of values for this threshold, but we will primarily present results that use a simple $v_{cut}=0$ threshold to ensure that we account for a complete census of outflowing and infalling gas. This simplifies comparison to and implementation of our formulas in semi-analytical and subgrid models. However, in the interest of a more meaningful observational comparison, and to quantify the amount of unbound gas, we also provide measurements using $v_{cut} = \sigma_{1D}$, the 1-D velocity dispersion of the halo. More information is found in Appendices \ref{sec:appendix:outflowrate}. 

Once outflowing particles are identified, we calculate outflow rates in $\Msunyr$ by computing the instantaneous mass flux through a thin spherical shell, in a manner similar to  \citet{faucher-giguere11b}. For simplicity, we divide the halo into finite spherical shells of thickness $dL$. The outflow rate is then computed with the following formula:

\begin{equation}
\frac{\partial M}{\partial t} = \sum \vec{v}  \cdot \frac{\vec{r}}{|r|}  m_{SPH} / dL. 
\label{eq:flux_otflow}
\end{equation}

This value is summed up for all particles within the shell. When discussing flux measurements with this approach, we denote shells by giving their median halocentric distance. For example, the shell that encapsulates 0.2-0.3$\Rvir$ is called the 0.25$\Rvir$ shell. For the analysis presented below, we use $dL = {\rm 0.1} \Rvir$. We treat the innermost shell, $0 < r < {\rm 0.1} \Rvir$,  as the shell that contains the dense galactic ISM, while all outer shells are considered CGM halo gas. This analysis relies on the assumption that the outflows are generated in this inner region of the galaxy by star formation that is also confined to the center. Alternate methods that explicitly treat geometry and filling factors of outflows and inflows will be explored in future work.

Our ability to do kinematic analysis relies on the time-spacing between snapshots being sufficiently short to capture each individual star-forming and outflow event. Our typical interval  between snapshots is between 10-40 Myr at $z>1$ and 20-70 Myr at $z<1$, which is sufficient for this analysis. In Appendix~\ref{sec:appendix:outflowrate} we demonstrate that the outflow rates derived through this \textit{Instantaneous Mass Flux} method are comparable to those computed using a comparable particle tracking method that explicitly accounts for material crossing the shell, and determines the flux simply by accounting for all of this material and dividing by the time interval (\textit{Interface Crossing} method). 

\begin{figure*}
\centering
\begin{minipage}{0.33\textwidth}
\centering
\vspace{0.35cm}
\hspace{-1.0cm}
\includegraphics[width=1.10\textwidth]{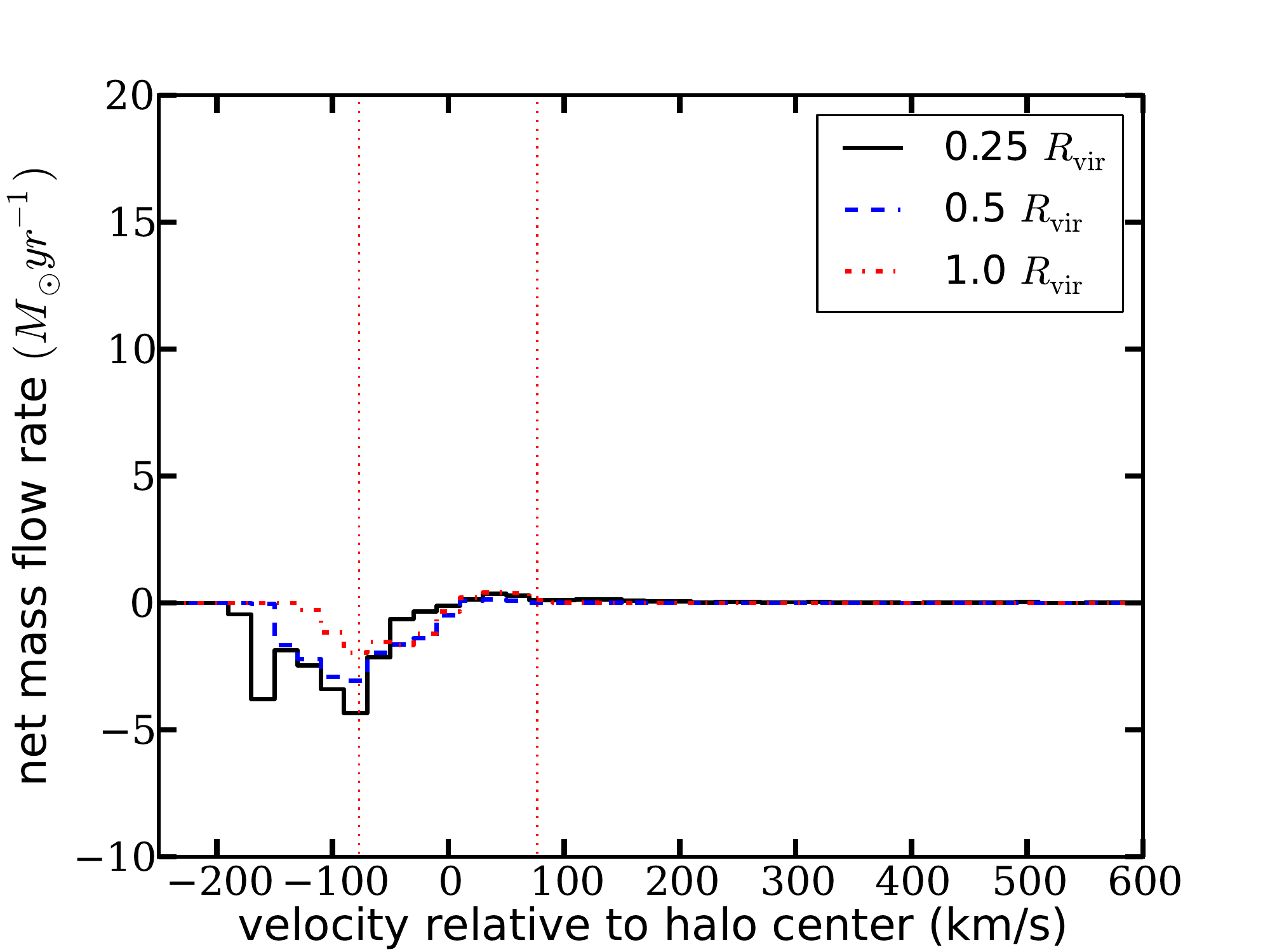}\\
\hspace{-1.0cm}
\vspace{0.25cm}
\includegraphics[width=1.10\textwidth]{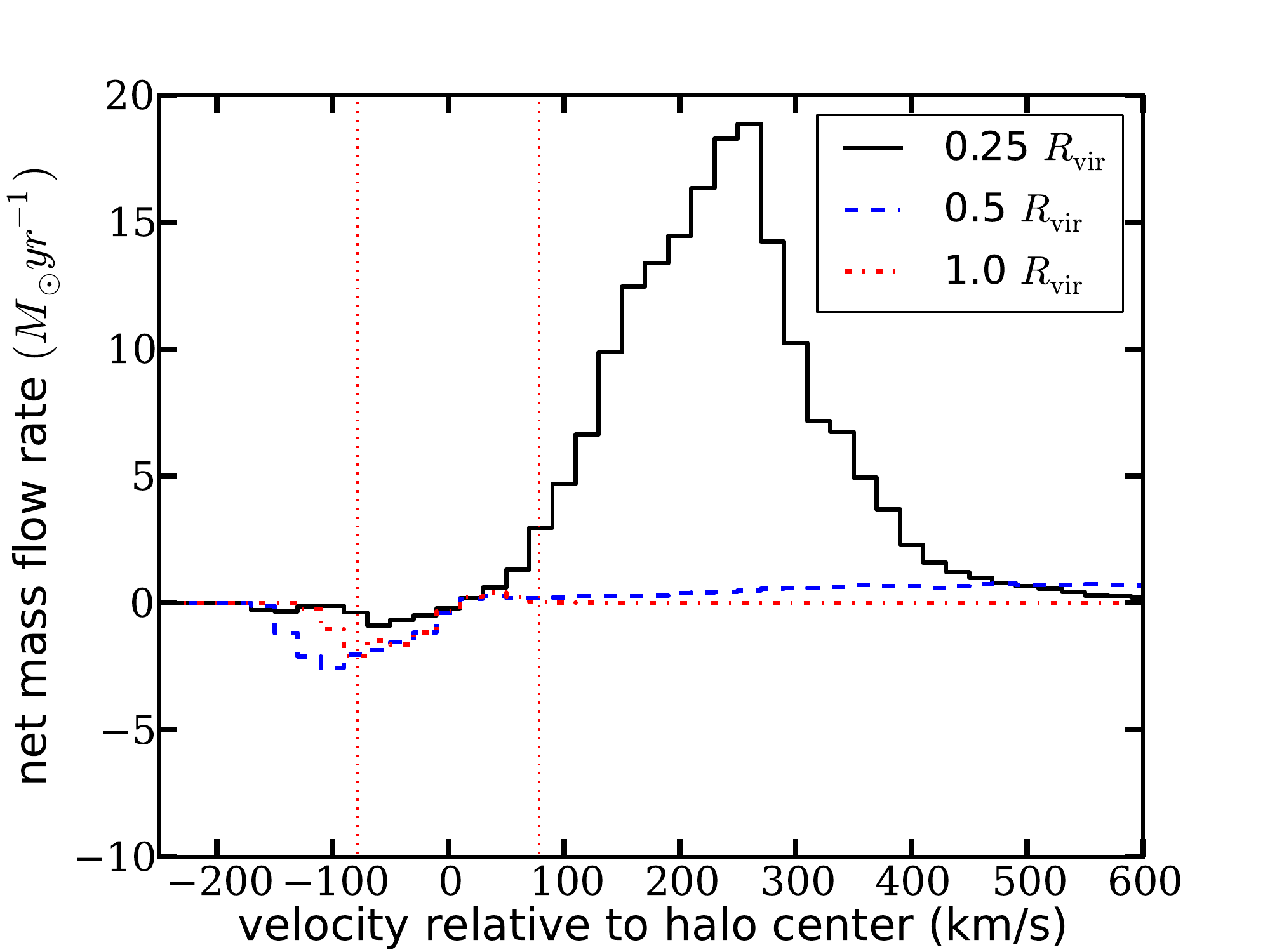}\\
\hspace{-1.0cm}
\vspace{0.2cm}
\includegraphics[width=1.10\textwidth]{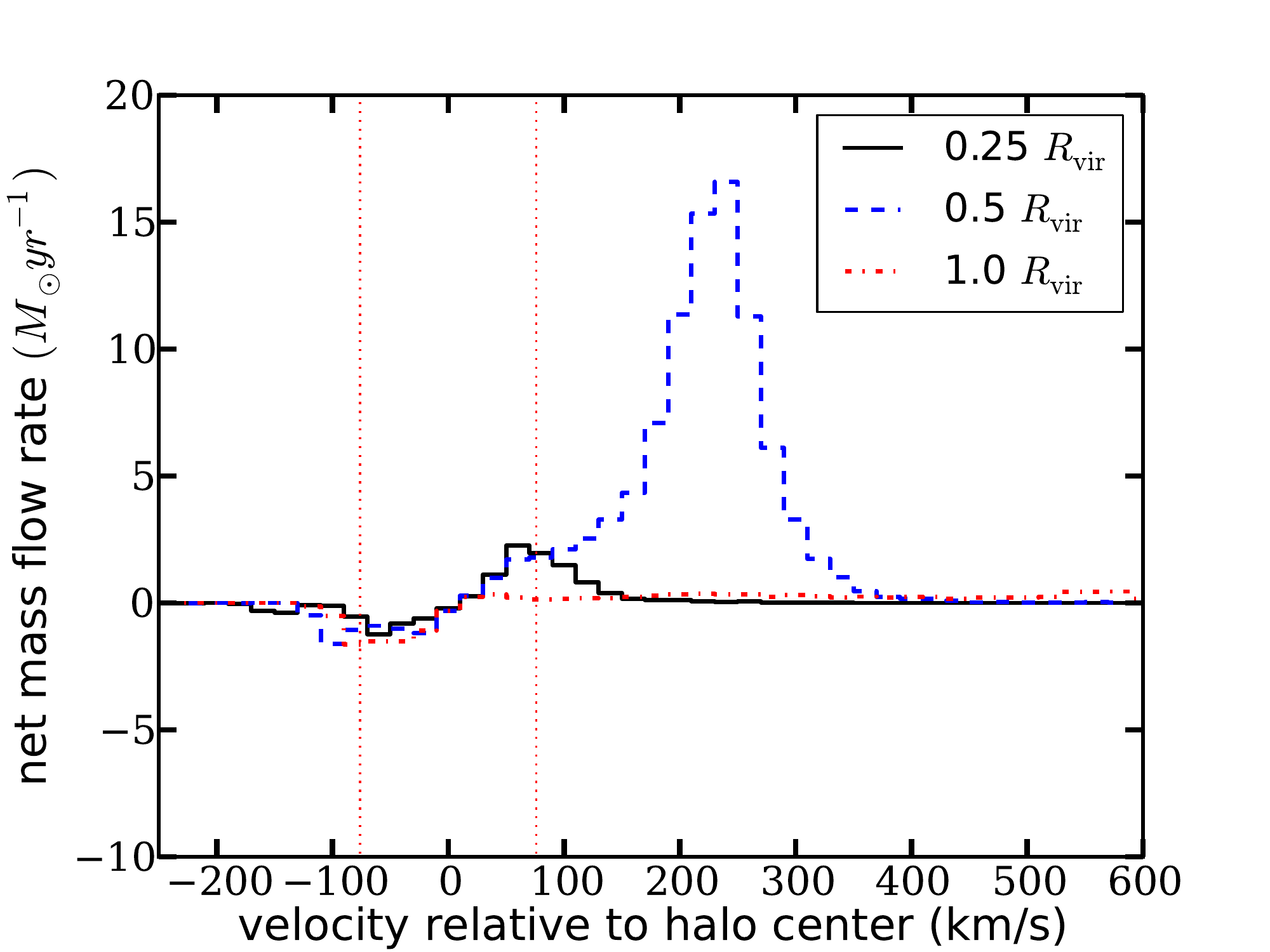}\\
\hspace{-1.0cm}
\vspace{0.2cm}
\includegraphics[width=1.10\textwidth]{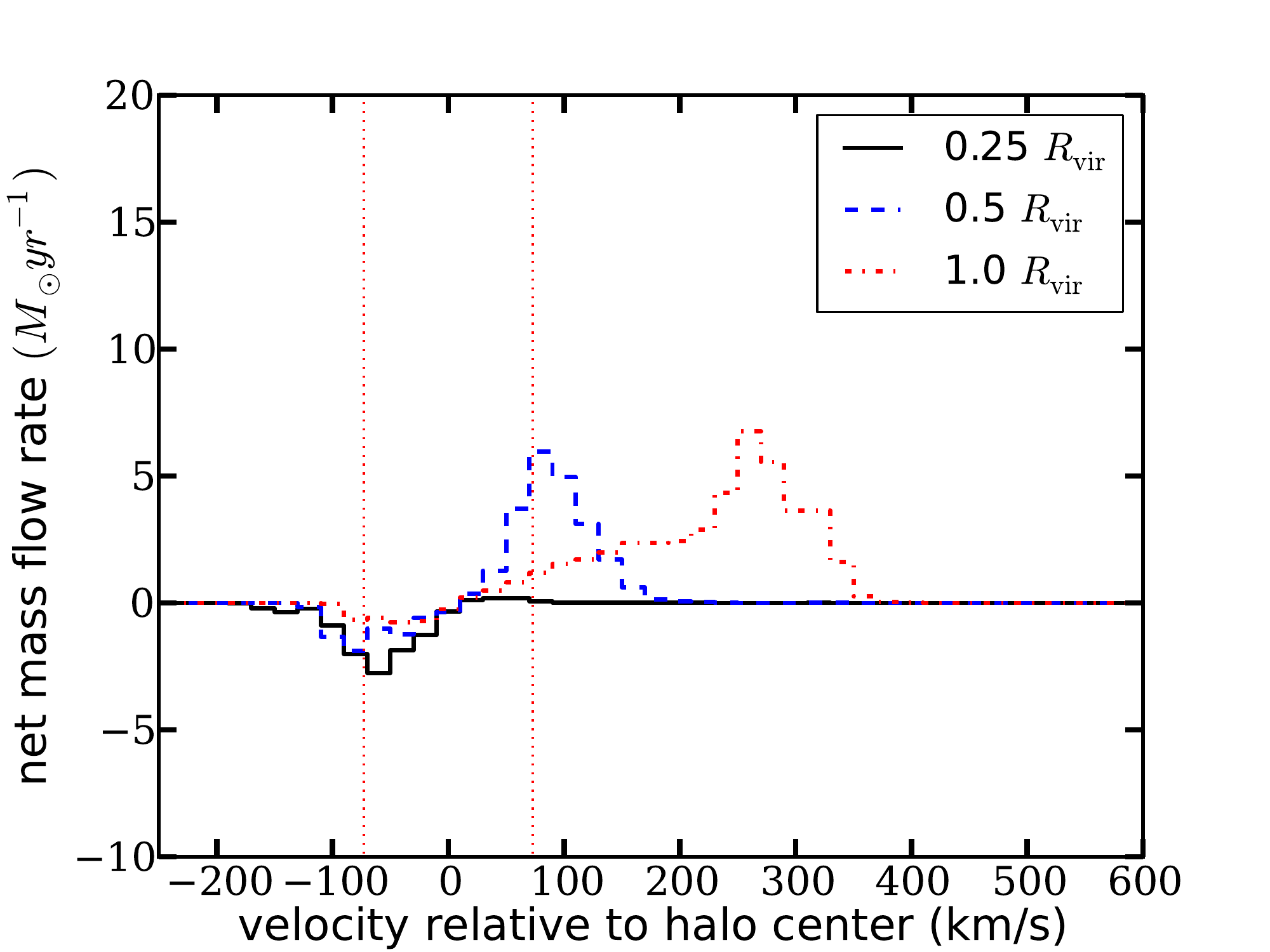}
\end{minipage}
\begin{minipage}{0.35\textwidth}
\includegraphics[width=1.10\textwidth]{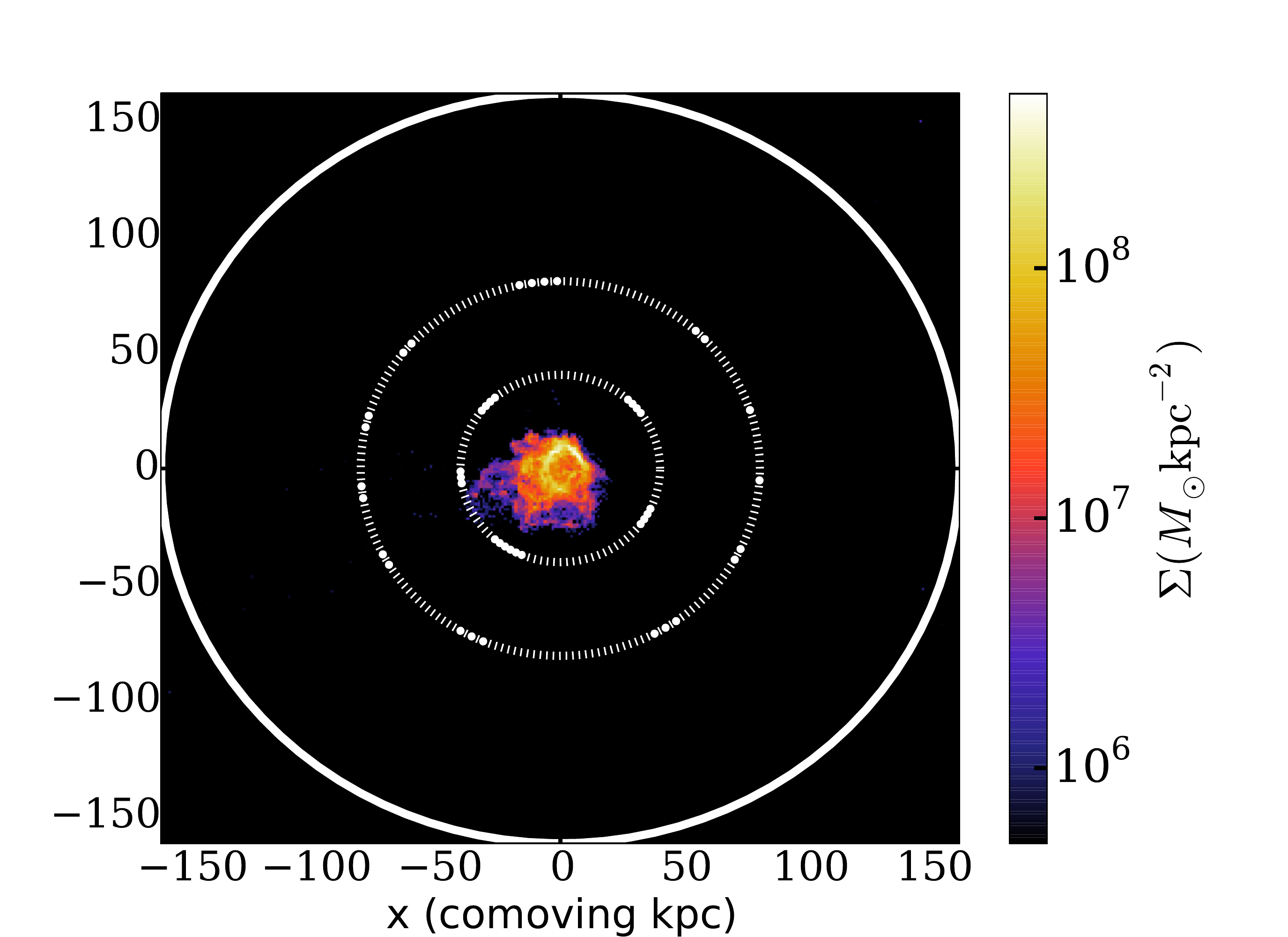}\\
\includegraphics[width=1.10\textwidth]{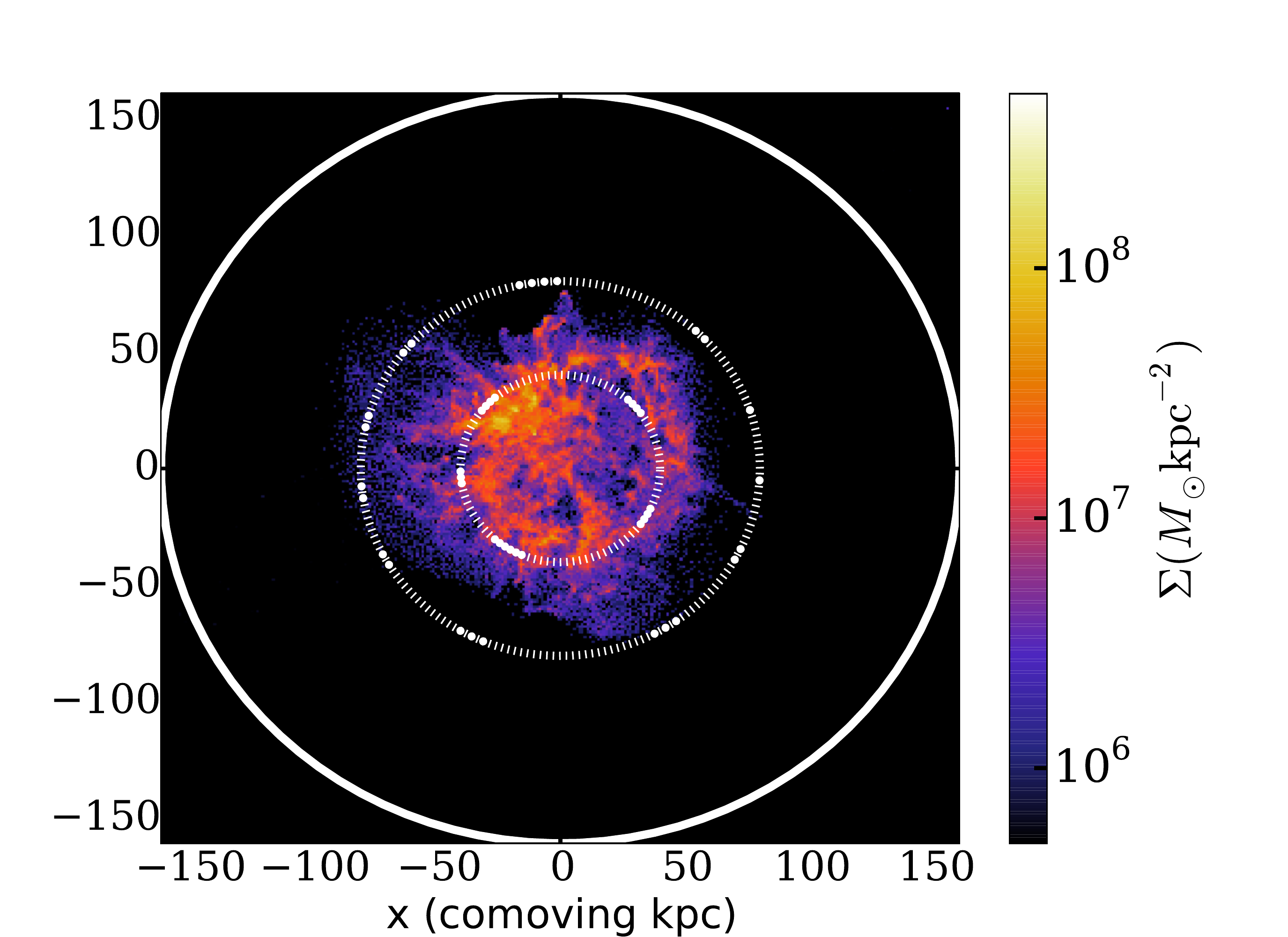}\\
\includegraphics[width=1.10\textwidth]{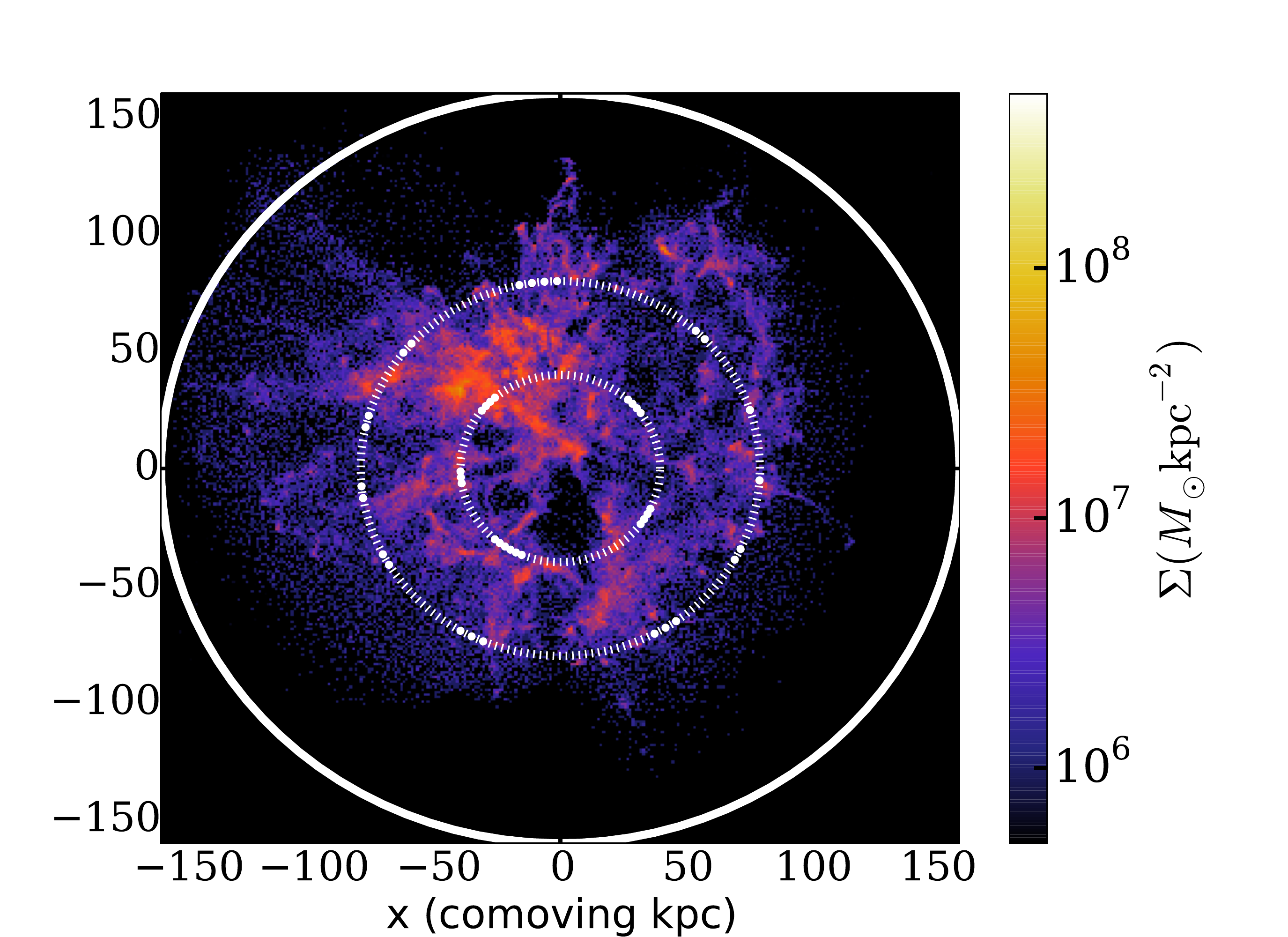}\\
\includegraphics[width=1.10\textwidth]{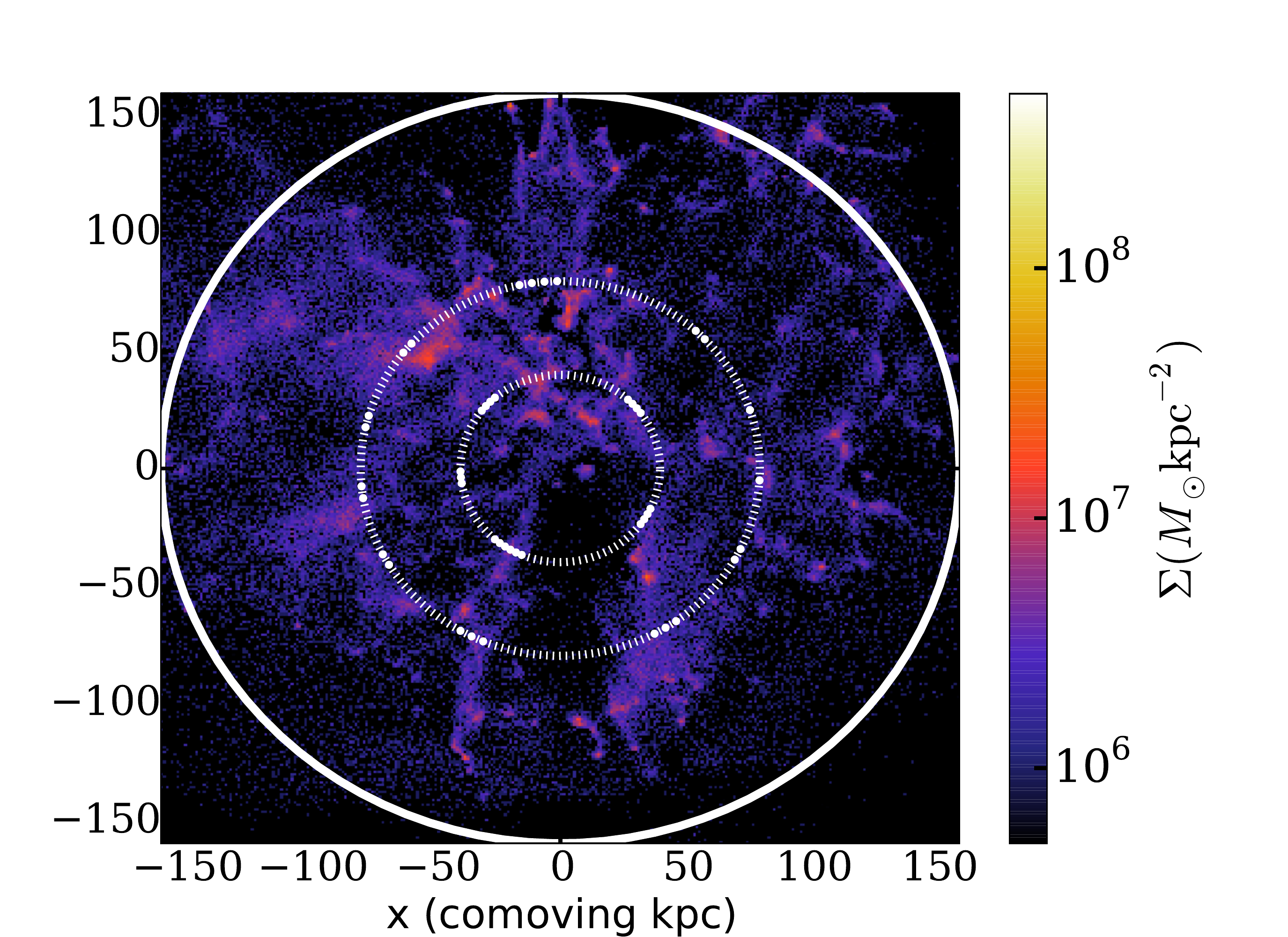}
\end{minipage}
\hspace{-1.2cm}
\begin{minipage}{0.35\textwidth}
\centering
\includegraphics[width=1.10\textwidth]{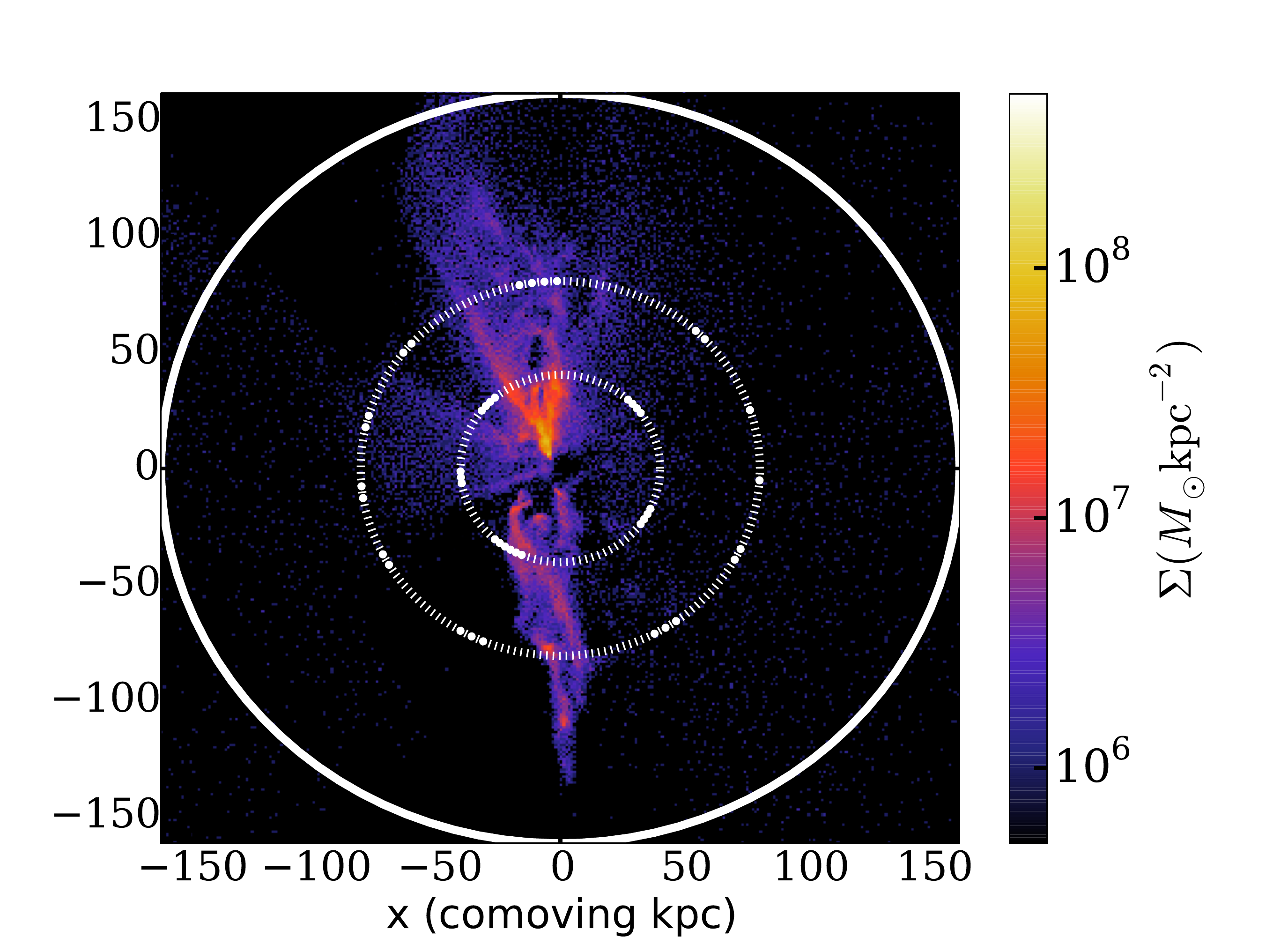}\\
\includegraphics[width=1.10\textwidth]{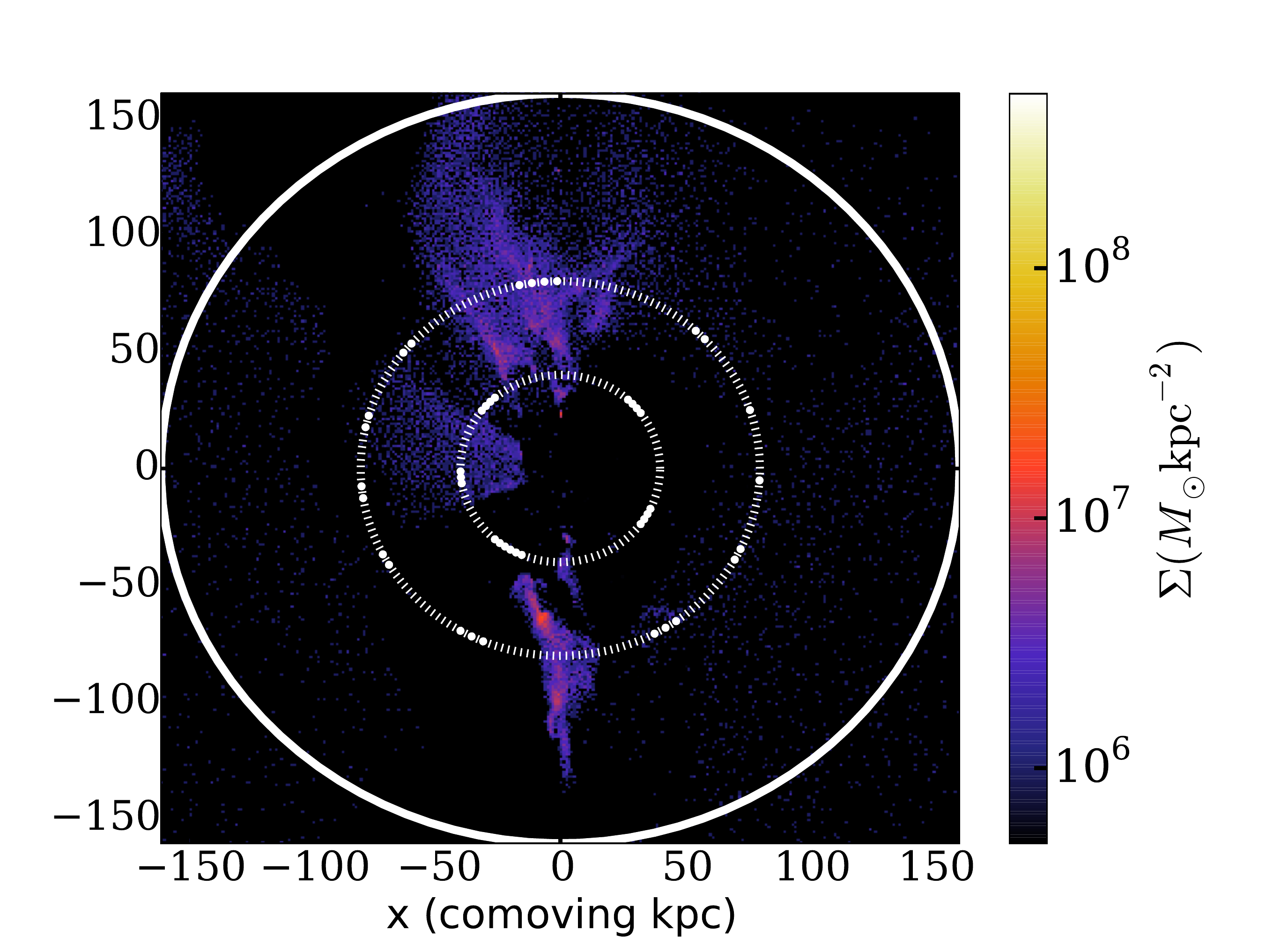}\\
\includegraphics[width=1.10\textwidth]{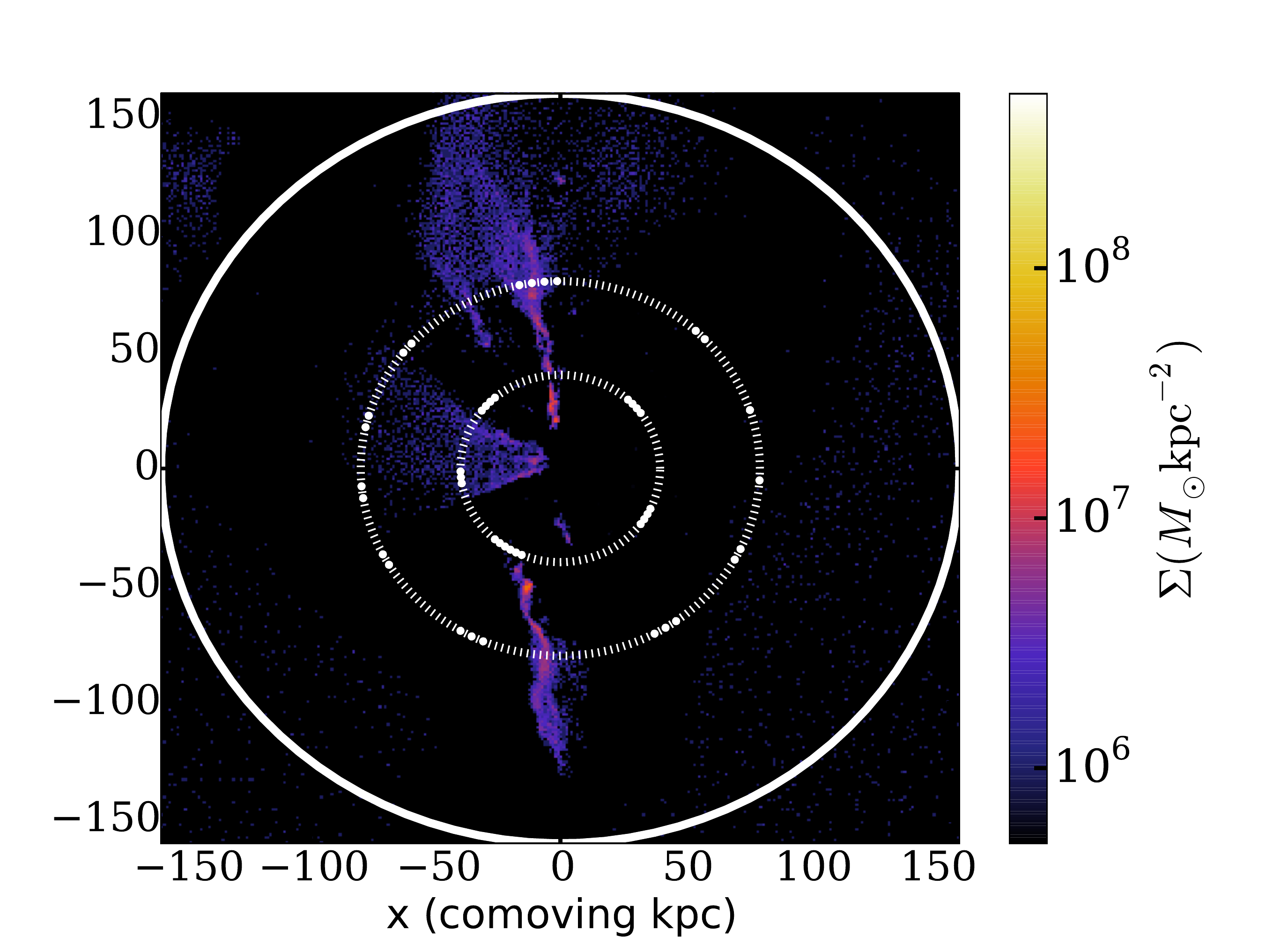}\\
\includegraphics[width=1.10\textwidth]{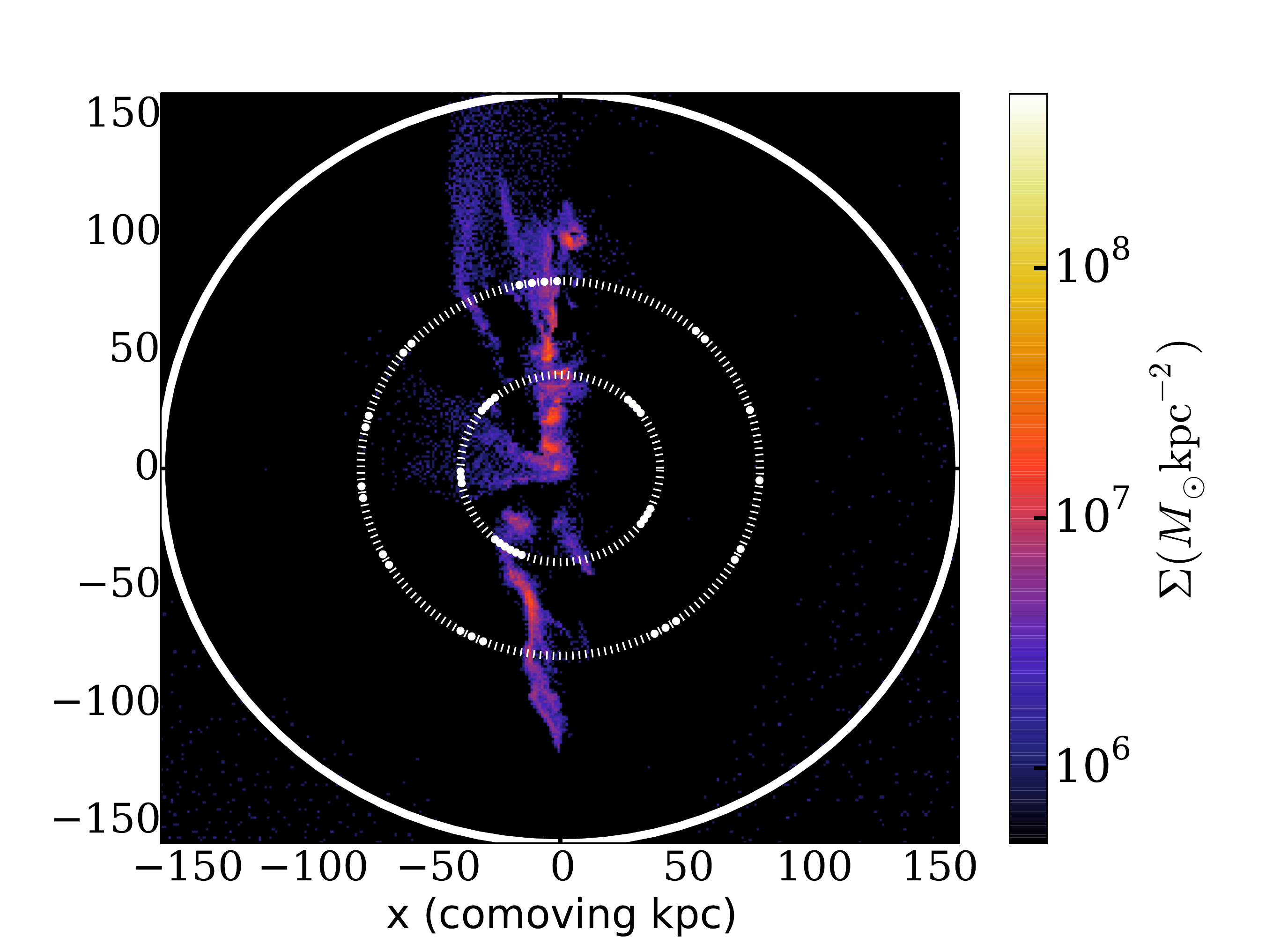}
\end{minipage}
\caption{Left: histogram showing the distribution of halo gas radial velocity, weighted by the flow rate before (first row) and during (second, third, and fourth rows) a major galactic wind flows through the CGM at $z\approx3$ in an L*-progenitor halo (\textbf{m12v}). Vertical dotted lines represent $v_{rad} = \pm \sigma_{1D}$ for the halo. Histograms are shown in three thin shells of width $dL = 0.1 \Rvir$, centered at $0.25 \Rvir$ (black, solid), $0.5 \Rvir$ (blue, dashed), and $1.0 \Rvir$ (red, dash-dot). Each row represents a snapshot at a particular epoch, with a spacing of \textasciitilde50 Myr between epochs. Center: corresponding to the same epoch as each radial velocity histogram, the mass surface density per 2-D spatial bin computed on a 300 $\times$ 300 grid between $-\Rvir < x,y < \Rvir$. Only particles with $|v_{rad}| > \sigma_{1D}$ are shown. Right: mass surface density of infalling gas computed with the same methodology as the outflowing particles in the central panel. White dotted circles denote $0.25 \Rvir$ and $0.5 \Rvir$. The thick white circle denotes $\Rvir$.}
\label{fig:traceoutflows1}

\end{figure*}

\section{Tracing Outflows Over Time}

\label{sec:timeevo}
Our method for measuring outflow rates requires that outflowing material has a detectable kinematic signature in the snapshots of our simulation. This method would not work if the interval between snapshots were long enough such that gas particles launched as winds during the interval would slow down and randomize their velocity vectors prior to the next output. To demonstrate that we avoid this pitfall, we must show that particles marked as outflows are both sufficiently kinematically distinct from "static" gas at the same galactocentric distance, and that "outflowing material" as marked by the detection method really does move outward, while inflowing material also has the correct kinematic and spatial structure. 

In Figure \ref{fig:traceoutflows1}, we show gas as binned by radial velocity ($v_{rad}$) and weighted by the "flow rate" from Equation \ref{eq:flux_otflow} at four epochs. Alongside each of these flow rate distribution diagrams, we show the surface density of outflowing and infalling gas, in a 2-D projection of the halo. The halo shown is one of the two main progenitors of \textbf{m12v} immediately following a major burst of star formation (peaking at \textasciitilde 12$\Msunyr$) at $z\approx3$. The first and second rows of Figure \ref{fig:traceoutflows1} show the halo at an epoch in the early stage of a major outflow episode, when outflowing material is still confined to the ISM (first row),  and approximately 50 Myr later (second row), when material has reached the CGM. In the early stage of the burst, the only mass flow seen in the CGM is a small gas inflow with $v_{rad} \approx \sigma_{1D}$. The filament that is feeding the galaxy with fresh material remains intact at this stage. This changes drastically in the second panel, as a vast swath of outflowing material has reached the inner CGM. This can be seen in the spatial 2D projections, as the $0.25 \Rvir$ shell now contains several clumps of high-column density outflowing material, while the filament has been disrupted in these inner regions. The $v_{rad}$ histogram shows that a large amount of material now occupies a broad distribution centered at $v_{rad} \approx 250 \mathrm{km/s}$, significantly offset from $\sigma_{1D}$. The fastest material has already streamed past that shell, as can be faintly seen in the $v_{rad}$ flow-rate diagrams in the $0.5 \Rvir$ shell. 

\begin{figure}
\vspace{-0.2cm}
\includegraphics[width=\columnwidth]{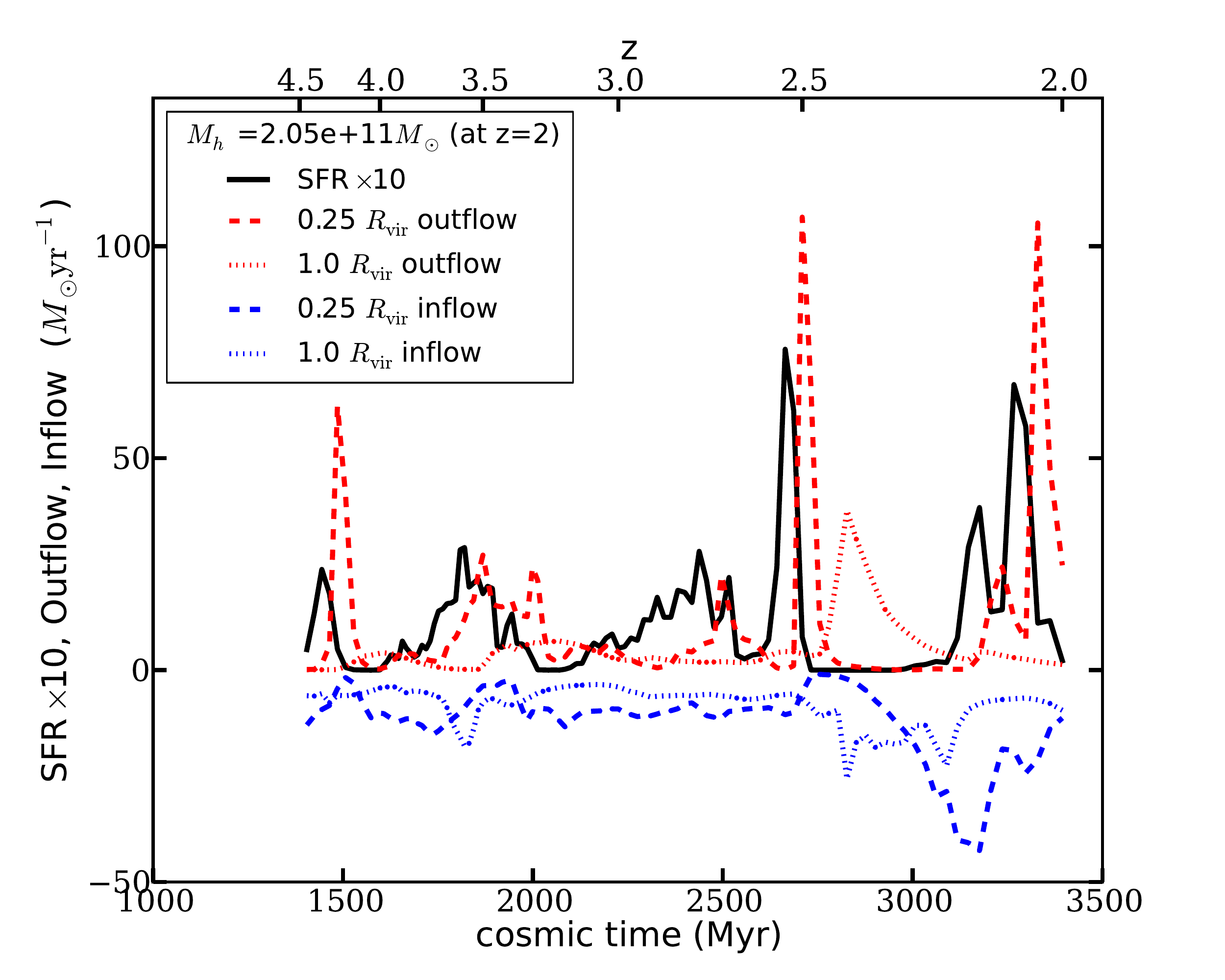}
\vspace{-0.0cm}
\caption{ Star formation, outflow, and inflow vs. cosmic time in one of the L* progenitor halos, \textbf{m12v}, for $4.5 > z > 2.0$. Star formation is shown in black, outflows rates in red at $0.25 \Rvir$ (dashed) and $1.0 \Rvir$ (dotted). Infall rates are shown for the same two halo-centric radii in blue, and are represented as negative flux. The star formation rate has been rescaled by a factor of 10 to be visible on the same linear scale. Outflow rates in the CGM are variable, and peak after bursts of star formation, though with a temporal offset. Inflow rates are much more steady, but can be disrupted temporarily by outflows. }
\vspace{0.3cm}
\label{fig:outandinflows_m12v}
\end{figure}

In the third and fourth rows of Figure \ref{fig:traceoutflows1}, we see the continuation of the outflow episode at \textasciitilde 100 Myr and \textasciitilde 150 Myr after the outflow began. In the $v_{rad}$ flow rate distribution, gas appears to be streaming into the $0.5 \Rvir$ shell in the third row, and  the $1.0 \Rvir$ shell in the bottom row. The velocity flow rates have the same characteristic broad distribution as was seen in the $0.25 \Rvir$ shell earlier, though with lower outflow rates, suggesting that a large fraction of the material does not escape the virial radius of the galaxy and either remains in the CGM or recycles. Infall rates over the interval of time shown remain low, but there is indication that a net flow of material is starting to trickle back into the $0.25 \Rvir$ shell, while the filamentary inflow geometry is restored at the last epoch considered. The characteristic velocity distribution of inflow never reaches the high speed of outflows.

To best understand the nature and prevalence of outflows in the CGM, and to connect them with galactic evolution, measurements over a time interval much longer than a single outflow episode is required. For each snapshot of each halo in our simulation suite, we calculate the star formation rate, outflow rate and inflow rate at various radii and plot them as a function of cosmic time. Inflow and outflow rates are measured using the \textit{Instantaneous Mass Flux} method (Equation \ref{eq:flux_otflow}) with $v_{cut} = 0 {\rm km/s}$ unless otherwise stated. The first of such time evolution plots is shown in Figure \ref{fig:outandinflows_m12v}. This Figure follows the progenitor of one of the L* galaxies, \textbf{m12v}, during the early stages of formation, $4.5 > z > 2.0$. The extremely bursty nature of star formation and "gusty," as opposed to steady, nature of galactic outflow rates are the most obvious results, and we shape the remainder of analysis in this paper based on this behavior. 
\begin{figure*}
\centering
\begin{minipage}{0.48\textwidth}
\centering
\includegraphics[width=\textwidth]{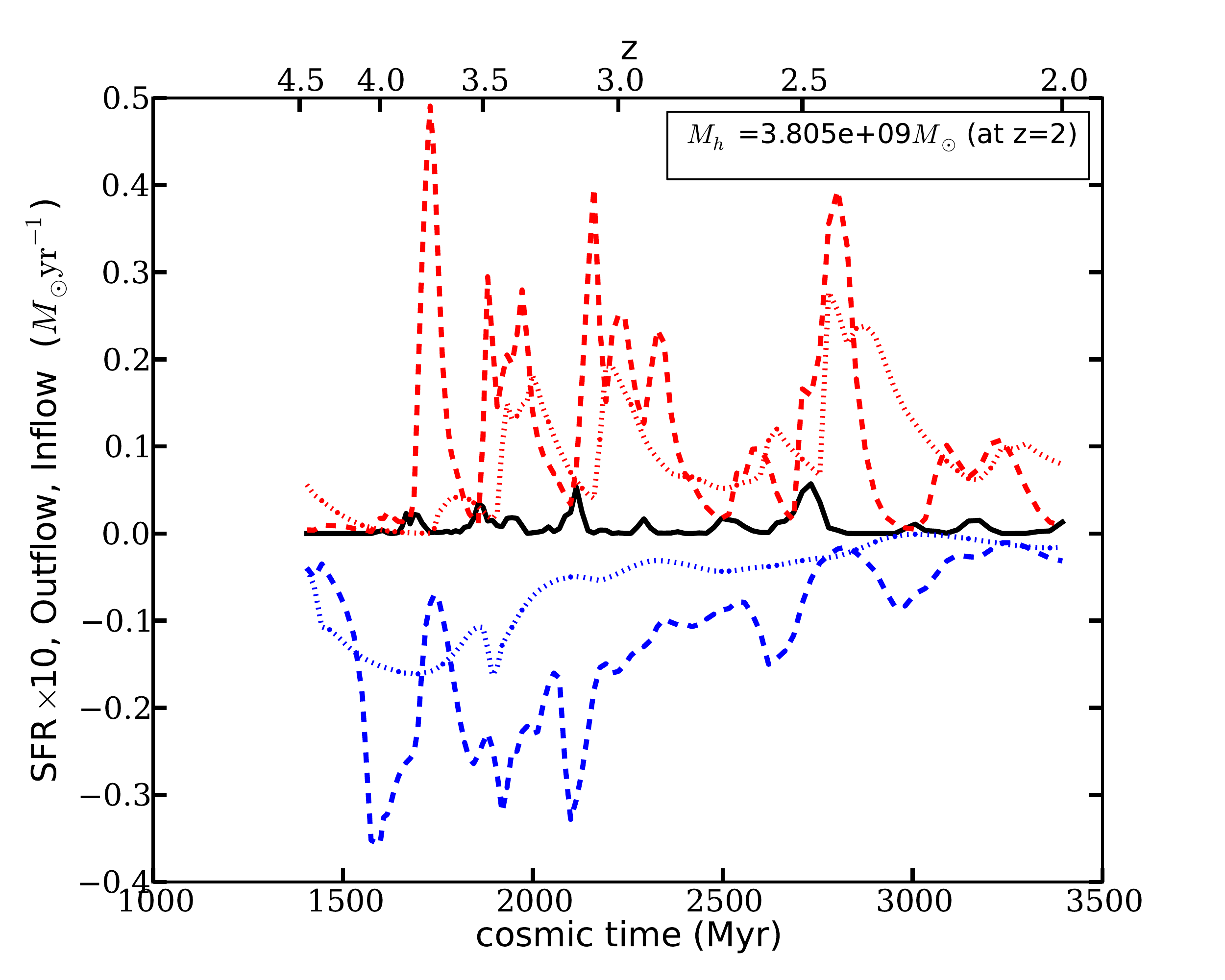}\\
\includegraphics[width=\textwidth]{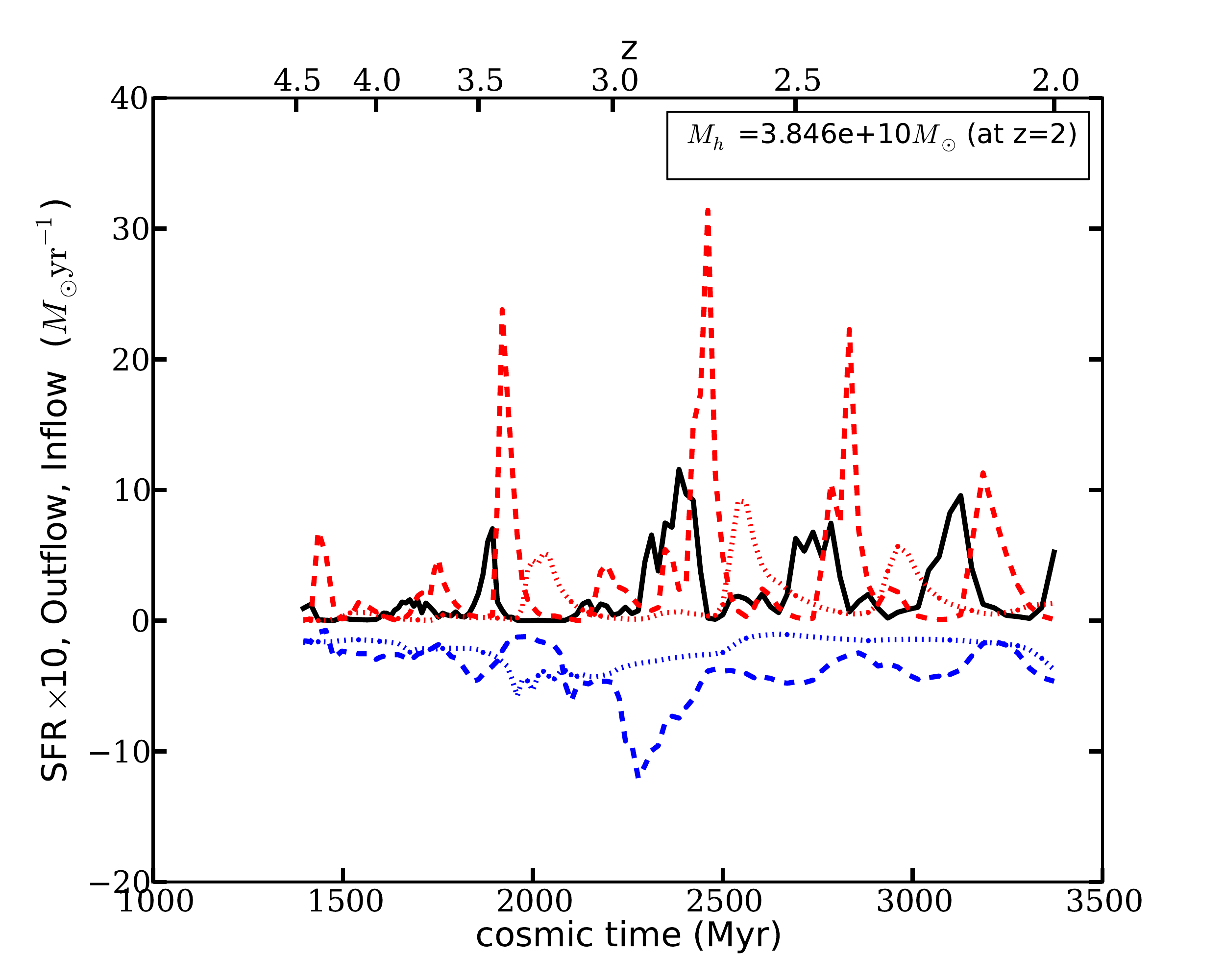}
\end{minipage}
\begin{minipage}{0.48\textwidth}
\includegraphics[width=\textwidth]{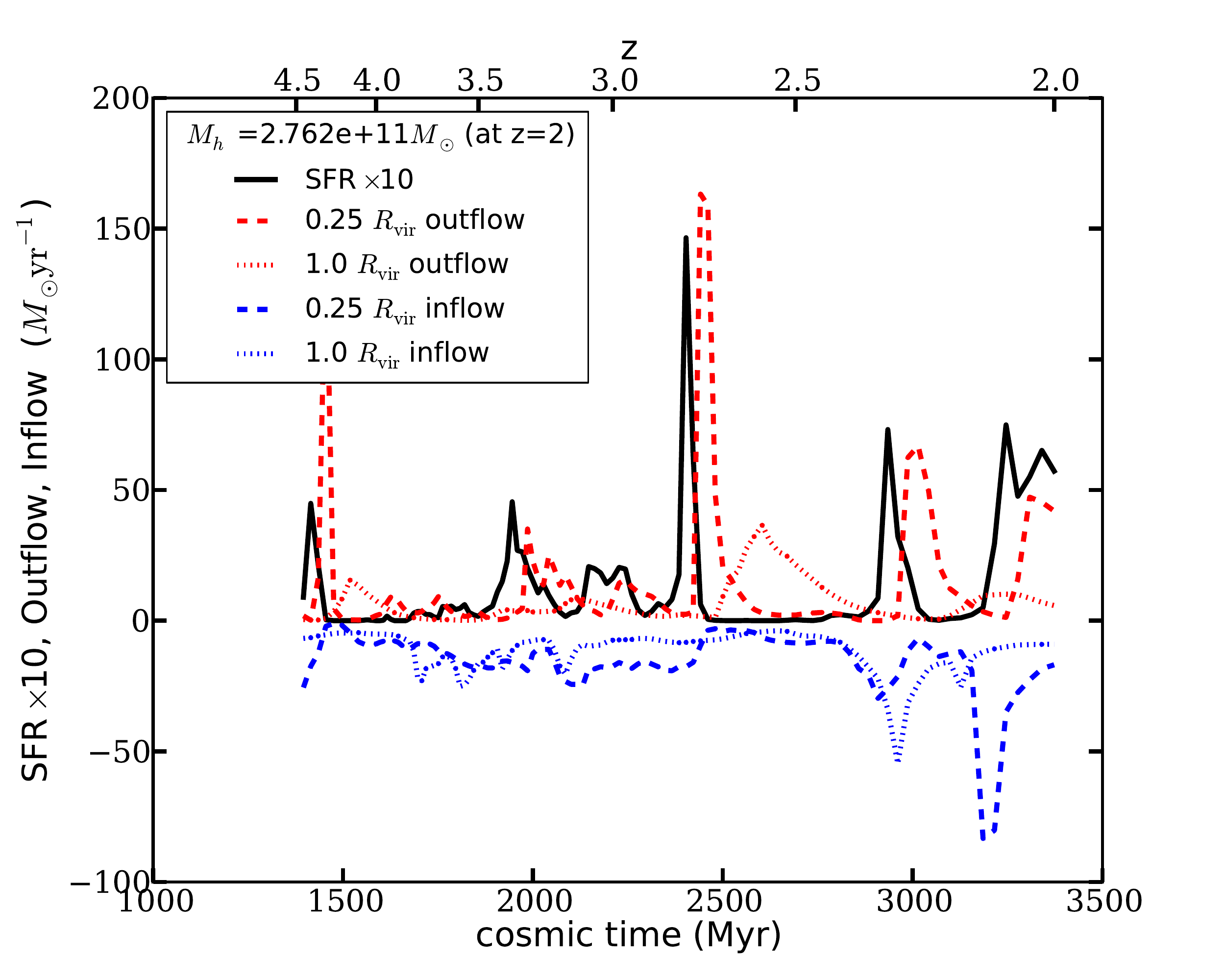}
\includegraphics[width=\textwidth]{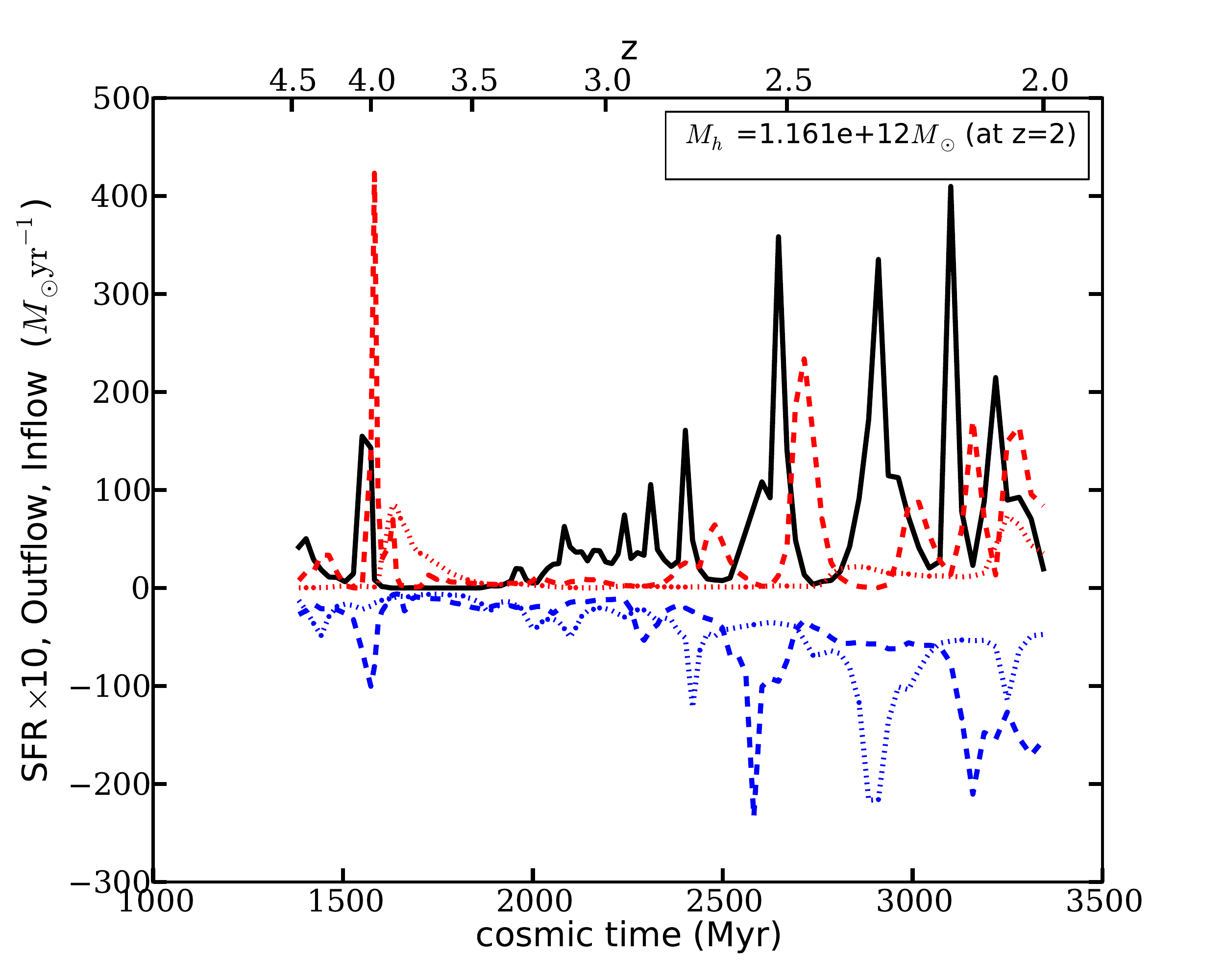}\\
\end{minipage}
\caption{Same as Figure \ref{fig:outandinflows_m12v}, but for four other halos  (upper-left: \textbf{m10}, lower-left: \textbf{m11}, upper-right: \textbf{m12i}, lower-right: \textbf{z2h506}). In all cases, the star formation rate is rescaled by a factor of 10 to appear visible on the linear scale. While the scales needed to show SFR and outflow rates for each figure differ by orders of magnitude, the pattern of bursty star formation and gusty outflows appears is generic. In the most massive halos, star formation does not shut off after $z=3$.}
\label{fig:wiggle-hiz}
\end{figure*}
We enumerate several interesting features of the plot:\\\\
1. The bulk of star formation occurs in bursts. The galaxy spends a significant fraction of time with near-zero star formation rates.\\
2. Bursts of star formation are always followed by outflows at $0.25 \Rvir$, and the outflows rates are typically higher than the star formation rates. \\
3. Bursts of star formation are preceded by inflow at $0.25 \Rvir$. \\
4. Outflows through $0.25 \Rvir$ are temporally offset from star formation. They would be observed as coincident within the timespan (100 Myr) of some star formation diagnostics (e.g. UV excess), but not within the timespan of others (e.g. $H\alpha$). We find outflow rates in the inner CGM at high redshift to be most correlated to star formation at a time $d\tau \approx 60 \mathrm{Myr}$ after each burst, although the scatter about this timescale is large (see Appendix \ref{sec:appendix:instant_app}). \\
5. Outflows through $1.0 \Rvir$ are even more temporally offset from star formation, sometimes by a longer period (\textasciitilde 250 Myr) than common star formation diagnostics \citep{kennicutt_evans12}. \\
6. Infall in the inner halo (0.25$\Rvir$) is temporarily suppressed when outflow rates are high. This suggests that inflows can geometrically overlap with outflows and interact with them (see also \citealt{oppenheimer10, vandevoort_etal11b, faucher-giguere11a, faucher-giguere11b, faucher-giguere_etal14}).\\
7. Outflows at $0.25 \Rvir$ do not always reach $1.0 \Rvir$. Even when the outflows do reach $1.0 \Rvir$, the outflow rates are typically not as high, and instead persist in the shell for a longer period of time, suggesting a slower flow rate in these outer regions.\\ 
8. Outflow, infall, and star formation rates are all highly variable. Throughout the $4.5 > z > 2.0$ interval shown, this galaxy never switches into a steady mode of star formation. We will later show that more massive galaxies may have continuous star formation, particularly at lower redshift. \\

Taken together, these features suggest a highly time-variable physical state of halo gas. Even the inner region of the halo (the galaxy) is prone to instability. Outflows are typically launched from within the galaxy, pushing out and dispersing the ISM. We find that during the $4.0 > z > 2.0$ interval represented in Figure \ref{fig:outandinflows_m12v}, the amount of dense gas within $\Rvir$ with $n > 1 {\rm cm}^{-3}$ only exceeds $10^8 \Msun$ for short periods of time, corresponding to the snapshots when the star formation rate is high, and the snapshots immediately prior. In fact, the galaxy can be completely devoid of this dense ISM for several hundred Myrs following a major outflow episode. 

The inflow rates at $0.25 \Rvir$ do not vary as drastically as outflow rates. The inflow rate at $1.0 \Rvir$ is even more steady, only changing significantly as the system approaches a merger event at $z\approx2$. The integrated mass flux going through $0.25 \Rvir$ as inflow slightly exceeds the integrated mass flux going out. This trend is even more pronounced when considering the integrated flux at $1.0 \Rvir$, which means that the overall amount of baryons contained within $\Rvir$ is increasing. The implications of the outflows and infall for the overall baryon content of galaxies will be explored in Section \ref{sec:discussion_galev}. 

We stress that this galaxy has a halo mass of approximately $2 \times 10^{11} \Msun$ and a stellar mass of $2.4 \times 10^9 \Msun$ at z=2, which is somewhat less than the mean Lyman-break galaxy halo mass (\textasciitilde$10^{12} \Msun$) and therefore would fall below the selection threshold of many surveys \citep{steidel03}. 
Therefore, the time variability seen in this object is not necessarily representative of current LBG samples. 
H14 show the star formation history of the same galaxy smoothed over 100 Myr timescale, which more closely corresponds to commonly used observational star formation rate indicators. They find that the measured star formation rates are broadly consistent with the star-forming main sequence \citep{noeske07}.

\subsection{Time evolution of outflows}

\label{sec:wiggle}

To follow the evolution of SFR, outflows, and inflows, we divide the redshift range being considered in our analysis, $4.0 > z > 0.0$, into three increments. The first, "high-z" interval is $4.0 > z > 2.0$, where our dataset is the most complete due to the inclusion of the \textbf{z2h} sample and large number of progenitors and satellite progenitors in the zoom-in region. The next division is somewhat arbitrary, but is justified based on the general nature of outflows and star formation in the L*-progenitor halos, which we will explore in subsequent sections. We use $2.0 > z > 0.5$ as the intermediate redshift, or "med-z" interval and $0.5 > z > 0$ as "low-z" interval.  We choose not to consider epochs at $z > 4.0$ to maintain a high standard for resolution elements per galaxy, and to confine analysis to typically observable redshifts.

During the high-z interval, all halos in our sample undergo bursty star formation and gusty outflows, interspersed with periods of quiescence as seen in Figure \ref{fig:outandinflows_m12v}. We plot several of the halos spanning a wide range of masses in Figure \ref{fig:wiggle-hiz}. Though the scales of activity are different, it is clear that high time variability is ubiquitous. This is generally consistent with observations of highly active low-mass star-forming galaxies at high redshift (e.g. \citealt{maseda_etal14}). 

The relative efficiency of galactic wind generation depends on the mass of the halo. At one end of the spectrum, dwarf galaxies like \textbf{m10} (upper left panel of Figure \ref{fig:wiggle-hiz}) and \textbf{m09} (not shown) need only form a small number of stars to expel the entire ISM and temporarily shut off star formation altogether. At the other end of the spectrum, we consider the most massive galaxy in the \textbf{z2h} sample, \textbf{z2h506} (lower right of Figure \ref{fig:wiggle-hiz}, which has a mass of $10^{12} \Msun$ at $z=2$). In this halo, star formation, inflow, and outflow seem to co-exist in a continuous, albeit highly variable state. The rest of the halos generally behave in a manner between these two extremes, and that behavior can largely be predicted by halo mass. This is expected, as halo mass is a good proxy for the depth of the gravitational potential well at the center of the galaxy, and for the rate of gas accretion from the IGM, both of which significantly impede galactic outflow launching and propagation. It is interesting that in the most massive halos in our sample at $z=2$ at around $10^{12} \Msun$, shutdown of star formation after bursts is not complete. While the majority of stars form in bursts, some level of continuous star formation rate is also being established.

\begin{figure}
\vspace{-0.2cm}
\includegraphics[width=\columnwidth]{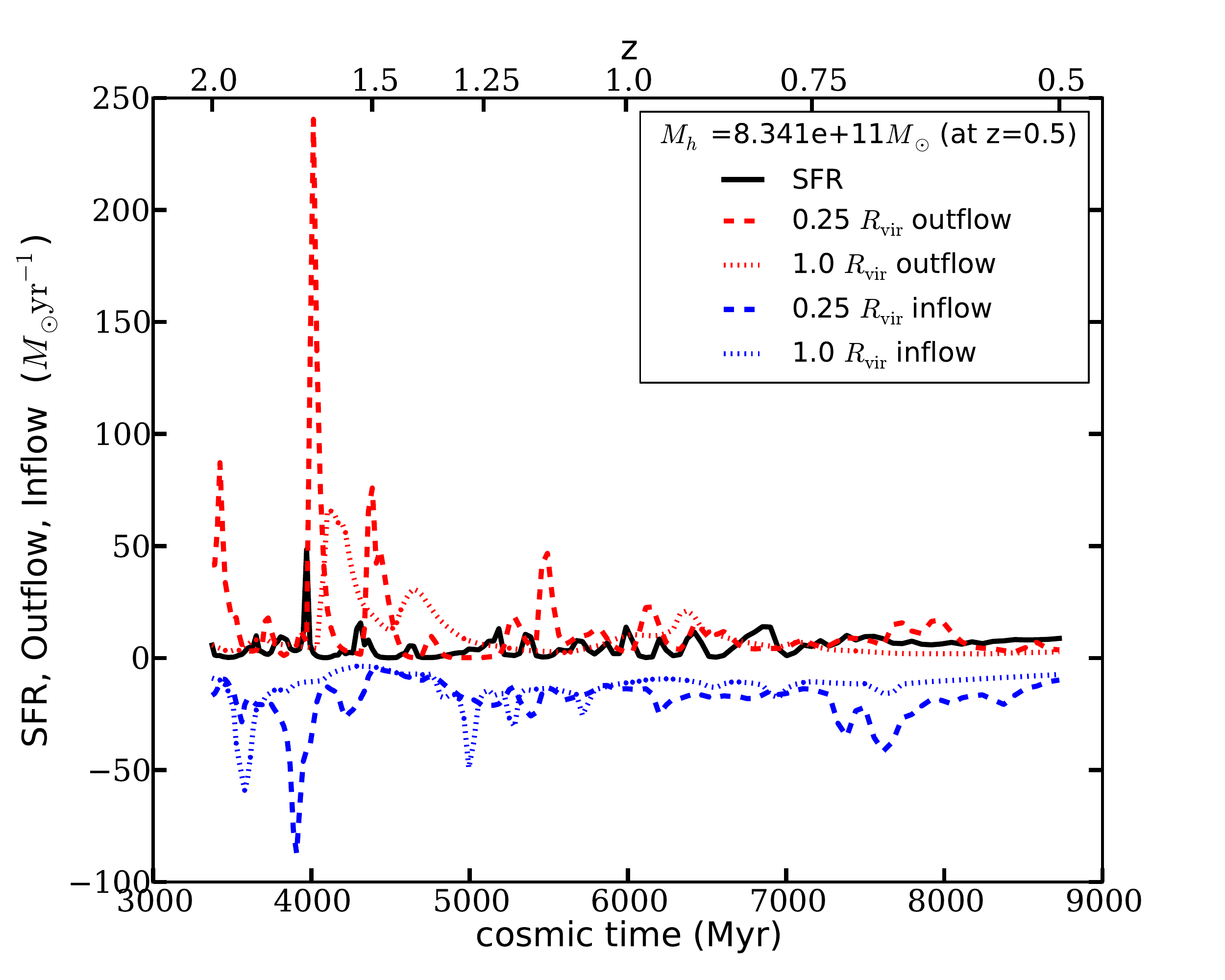}
\vspace{-0.0cm}
\includegraphics[width=\columnwidth]{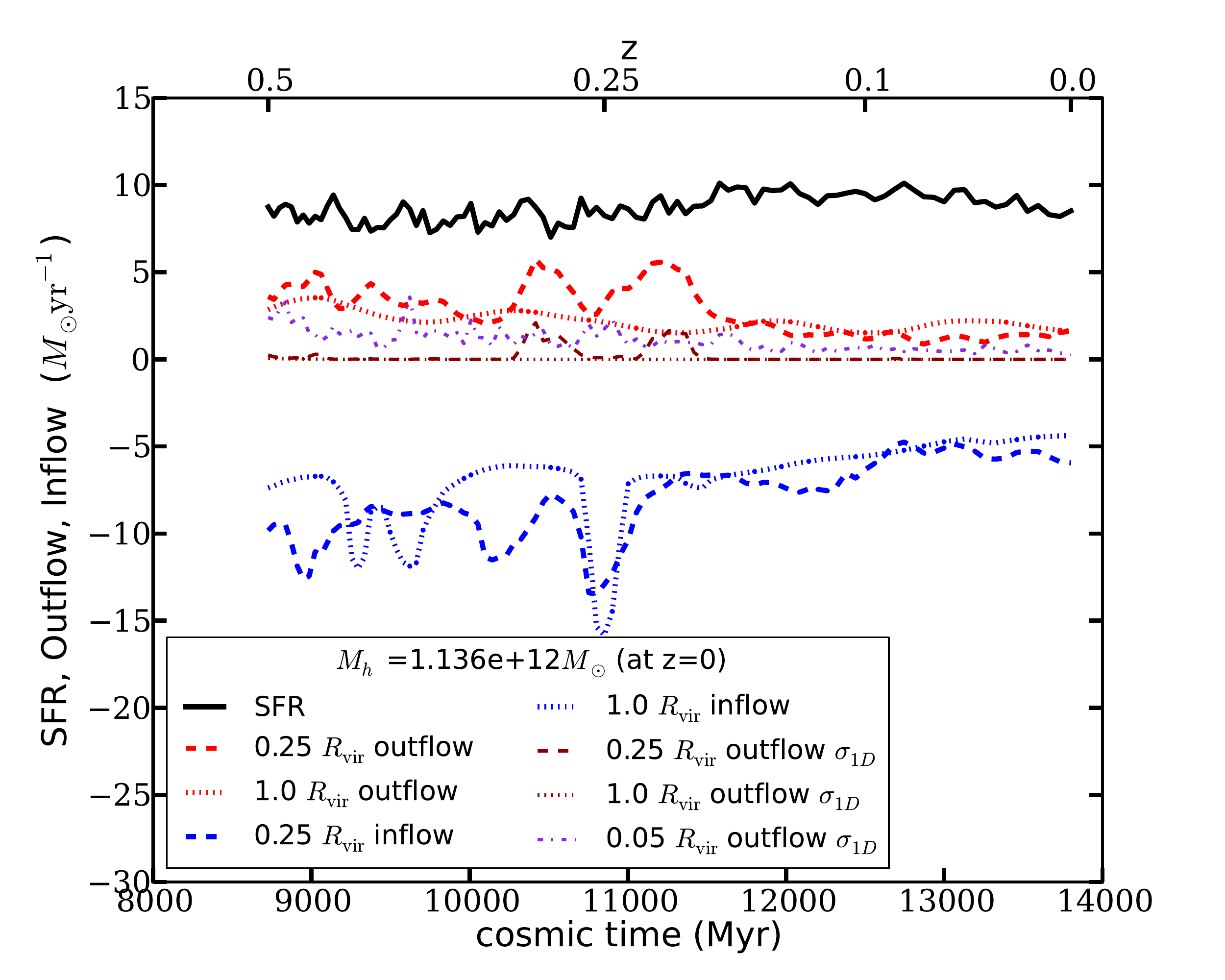}
\caption{ Same format as Figure \ref{fig:outandinflows_m12v}, but showing the $2.0 > z > 0.5$ and $0.5 > z > 0$ regime for the same halo (\textbf{m12i}). In the low-z regime, we also show outflow rates measured with $v_{cut} = \sigma_{1D}$. Compare this figure to the upper-right panel of Figure \ref{fig:wiggle-hiz} to see the full evolution of \textbf{m12i} with redshift. Outflows in the CGM become increasingly uncommon at low redshift, while star formation switches into a continuous mode. Note that unlike previous figures, star formation rate is not rescaled by a factor of 10. In lower mass galaxies such as \textbf{m10} and \textbf{m11}, the bursty nature of star formation and gusty outflows persist at low redshift.}
\vspace{0.3cm}
\label{fig:wiggle-loz}
\end{figure}

In the med-z regime, which is shown in the top panel of Figure \ref{fig:wiggle-loz} the L*-progenitor \textbf{m12i} continues the pattern of bursty star formation and gusty outflows seen in the high-z regime, especially at $z>1$. However, the outflow rate gradually decreases relative to the star formation rate suggesting a decreasing $\eta$. Furthermore star formation enters a continuous, non-bursty mode starting at $z\approx0.75$. This halo behaves significantly differently than the previously mentioned massive high-redshift \textbf{z2h506} halo, though the mass of \textbf{m12i} at $z=0.5$ is comparable to the mass of \textbf{z2h506} at $z=2$. While \textbf{z2h506} drove powerful, gusty outflows and simultaneously maintained ceaseless star formation at high redshift, \textbf{m12i} has entered into a continuous star formation mode with weak galactic winds.

Other L*-progenitors, \textbf{m12q, m12v}, and the massive dwarf \textbf{m11} also have reduced outflow rates relative to their star formation rates, but continue the bursty pattern of star formation and powerful outflows through the CGM during the interval. One interesting feature of this regime which will be elaborated on in Section \ref{sec:discussion_galev}, is that despite a lower efficiency in wind generation, all of these halos appear to lose more material through outflows than they gain through inflows at $\Rvir$. This is due to lower inflow rates from the intergalactic medium, combined with a few isolated incidents of intense star formation which generates the largest increases of stellar mass in the history of the halos, followed by the most intense galactic superwinds ever seen in the halos. The dwarf galaxy \textbf{m10} also continues to form stars and generate outflows and bursts, though they are relatively diminished compared to higher-redshift activity. Unlike the other dwarf galaxies, \textbf{m09} does not form a significant mass of stars near the galactic center after $z=2$.

In the low-z regime of \textbf{m12i}, which is shown in the bottom panel of Figure \ref{fig:wiggle-loz}, a qualitative shift is seen in the nature of star formation and outflows. Stars form at a moderate, steady rate, while outflow rates in the CGM are low and no longer gusty. Though they are not shown, the behaviors of \textbf{m12q} and \textbf{m12v} are both similar to that of \textbf{m12i} at this epoch, but with lower star formation rates. The inflow rates to all halos gradually taper off during this epoch, which coincides to the thinning nature of cosmic filaments and longer cooling times of the halo gas, which slows down gas infall to the inner CGM and the galaxy \citep{keres05, dekel06, faucher-giguere11b, vandevoort_schaye12, nelson13}. We will explore the nature of the remaining infall in future work. 

We consider one explanation for the drop-off in gusty outflows at $0.25 \Rvir$ below $z<0.5$ for all three L*-progenitor halos: that gas is still being ejected from the disk, but no longer able to traverse the CGM. In massive ($10^{12} \Msun$) halos, $0.25 \Rvir$ represents a great physical distance from the central galactic core, particularly at low redshift (\textasciitilde 50 kpc). To test this scenario we measure the flux at $0.05 \Rvir$  and find that while the outflow rates in the inner regions are greater, there is no indication that the outflow at either $0.05 \Rvir$ or $0.25 \Rvir$ represents a kinematically distinct component of gas. This is further evidenced by the drastically reduced values of outflow rates when they were measured with $v_{cut} = \sigma_{1D}$, where $\sigma_{1D} \approx 100 {\rm km/s}$ for this halo at $z=0$. The outflow rates using this threshold for both $0.05 \Rvir$ or $0.25 \Rvir$ are shown in Figure \ref{fig:wiggle-loz}. The strongly diminished flux measured with $v_{cut} = \sigma_{1D}$ indicates that the apparent ``outflows'' at late times can be explained by small random velocities of the halo gas as well as the motion of the material associated with and stripped from the satellite galaxies moving through the halo. Although the flux is somewhat higher at $0.05 \Rvir$, this can be entirely explained by the late-time accumulation and random motions of the ISM. In both cases apparent outflows are not connected to the star formation events. We have explicitly checked that both of these scenarios contribute to the measured outflow rates and will further discuss this point in Section \ref{sec:etaloz}. 

The drop-off in CGM outflows is likely caused by the deepening gravitational potential well in the center of the halo, the sparseness of gas-rich mergers (that trigger bursts of star formation), less geometrically concentrated star formation, and the overall more continuous, less bursty mode of star formation. We briefly explore the changing behavior of star formation by considering the presence and quantity of dense gas. In the high-redshift regime, the mass of ISM with number density $n > 1 {\rm cm}^{-3}$ fluctuates rapidly - ISM mass builds up, precipitates a burst of star formation, and becomes largely depleted after each burst. Specifically, the ratio of the mass of gas with number density $n > 1 {\rm cm}^{-3}$ to galaxy stellar mass reaches 0.2-1 at the epochs prior to and during each burst, but can fall below 0.01 afterwards, and remains low during quiescent periods. At low redshift, as \textbf{m12i} has significantly increased in mass and central density \citep{chan_etal15}, dense gas can persist. The ratio of dense ISM to stellar mass remains 0.1-0.2 at all times for $z<0.5$. Star formation is clearly no longer regulated by the same gas dispersal and ejection process that is so effective at high redshift. 



Instead, \textbf{m12i} appears to regulate star formation via a stable gaseous disk at low redshift, which can now persist for significant timespans, as merging activity has largely ceased for L* halos. The emergence of such stable gaseous disks at late times is consistent with observational studies \citep{kassin_etal12}. Feedback from star formation contributes to the stability of this disk by stirring turbulence \citep{thompson_etal05, ostriker_shetty11, faucher-giguere_etal13, hopkins_etal13b}. Gas affected by stellar feedback at low redshift can be launched as a local fountain, but generally does not propagate into the CGM. Measuring the prevalence of these local fountains and quantifying their effects on the galaxy requires sophisticated methodology and is left for future studies. In this stable disk, star formation no longer happens in a dense clump near the center, but is more extensively distributed. Extended gaseous disks are also seen in \textbf{m12v} and \textbf{m12q}, although they are considerably less massive, and the bulk of star formation is primarily confined to nuclear regions.

\begin{footnotesize}
\ctable[
caption={{\normalsize Mass-loading factors}\label{tbl:bigtable}},center,star
]{lcccccl}{
\tnote[ ] {Method used to calculate $\eta$ at $4 > z > 2$: \\
{\bf (1)} Flux T: Outflows measured by \textit{Instantaneous Mass Flux} method  for all gas with $v_{rad}>0$. $\eta$ is computed by the integrating total stars formed and total outflow rate over the whole interval. These values of $\eta$ were used in Section \ref{sec:etafit}. See Section \ref{sec:measure} and Appendix \ref{sec:appendix:flux}. \\
{\bf (2)} Cross T: Same as \textbf{(1)} but using \textit{Interface Crossing Method}  to calculate outflow rates over the interval. See Appendix \ref{sec:appendix:cross}. \\
{\bf (3} Flux T $\sigma_{1D}$: outflows rates over the interval are measured using the \textit{Instantaneous Mass Flux} method, but only particles with $v_{rad} > \sigma_{1D}$ are counted as outflows. See Appendix \ref{sec:appendix:sigma}. \\ 
{\bf (4)} Flux I: Using \textit{Instantaneous Mass Flux} with $v_{cut}=0$ to measure outflow rates over the high-z interval. $\eta$ is computed by finding correlation of star formation rate at each snapshot to outflow rate at a time $d\tau$ later (the \textit{instant} correlation method, see Appendix \ref{sec:appendix:instant_app}).\\
{\bf (5)} Flux E: Using \textit{Instantaneous Mass Flux} with $v_{cut}=0$ to measure outflow rates over the high-z interval. $\eta$ is computed by integrating total stars formed and mass expelled over individual episodes, and then deriving the correlation parameters (the \textit{episodic} correlation method, see Appendix \ref{sec:appendix:episodes_app}).
}
}{
\hline\hline
\multicolumn{1}{l}{Name } &
\multicolumn{1}{l}{Flux T} &
\multicolumn{1}{l}{Cross T}  &
\multicolumn{1}{l}{Flux T $\sigma_{1D}$}  &
\multicolumn{1}{l}{Flux I}  &
\multicolumn{1}{l}{Flux E} \\
\hline
m09 & 560 & 790  & 240  & 180 & 350  \\ 
m10 &  120 & 140 & 70 & 170 & 210   \\ 
m11 &  14 & 21 & 10  & 17 &  14    \\ 
m12v &  9.3 & 9.3 & 6.8  &  9.1 &  11   \\ 
m12i &  8.0 & 10 & 5.9  &  11 &   9.2  \\ 
m12q & 8.3 & 8.5 & 6.7  &  6.6 &   6.8   \\ 
z2h350 &  8.6 & 8.8 & 6.7  &  7.9 & 7.6   \\ 
z2h400 & 6.6 & 5.7 & 5.1  & 9.1 & 6.2  \\ 
z2h450 &  6.4 & 5.9 & 5.1 & 7.4 &   6.6 \\ 
z2h506 &  6.5 & 6.9 & 4.5 & 6.4 &  4.9   \\ 
z2h550 &  7.6 & 7.7 & 5.5 & 11  &   7.4\\
z2h600 &  6.0 & 6.3 & 4.6 & 7.2  &   6.7  \\
z2h650 &  6.8 & 8.8 & 5.1 & 7.1  &   7.9 \\
z2h830 &  4.8 & 4.2 & 3.6 & 5.3  &   3.6 \\
\hline\hline\\
}
\end{footnotesize}

We note that the dwarf galaxies in \textbf{m11} and \textbf{m10} maintain a bursty star formation history and outflows at low redshift, though the bursts acquire a new characteristic feature. Instead of a single burst of star formation followed by a corresponding outflow in the CGM, it now takes a series of small bursts interspersed with periods of quiescence to drive a sufficiently powerful outflow. This could indicate that gas is ejected from the ISM, but with insufficient velocity to kick it out to large radii, until finally there have been enough episodes that a threshold for velocity or temperature is reached, and the outflowing gas escapes to the CGM. In both cases, the magnitude of star formation and outflow activity is considerably weaker than at earlier epochs. 

\section{Measurements of mass loading in galactic winds}
\label{sec:eta}

We now provide measurements of the mass-loading factor, $\eta$, for all halos in our simulations. We note that there are at least two distinct approaches which can be used to measure $\eta$, each of which is appropriate for different purposes. The first approach is to make an instantaneous measurement of $\eta$ - a direct measurement of how much material is ejected for a particular star formation episode. This approach is useful for observational comparisons to data where $\eta$ is inferred from the current state of outflowing gas. On the other hand, this measurement is subject to stochastic effects, as each outflow episode is influenced by highly variable circumstances of the galaxy's dynamical state. As such, there is significant systematic scatter associated with these measurement. Nonetheless, we have made measurements of $\eta$ using several variations of this approach, and provide detailed methodology and results in Appendix \ref{sec:appendix:instant_app} and Table \ref{tbl:bigtable}. 

The second approach, which we will focus on here, is to measure an average cumulative $\eta$ over a time interval for each individual halo. This measurement provides a useful diagnostic that can be integrated to recover the gas mass ejected into the CGM and out of the halo for some portion of its evolution. The main caveat to this approach is that in order to get a statistically significant representative sample of star formation and outflow events for each halo, a relatively broad redshift interval must be used. 
\begin{figure*}
\centering
\begin{minipage}{0.44\textwidth}
\centering
\vspace{-0.4cm}
\includegraphics[width=\textwidth]{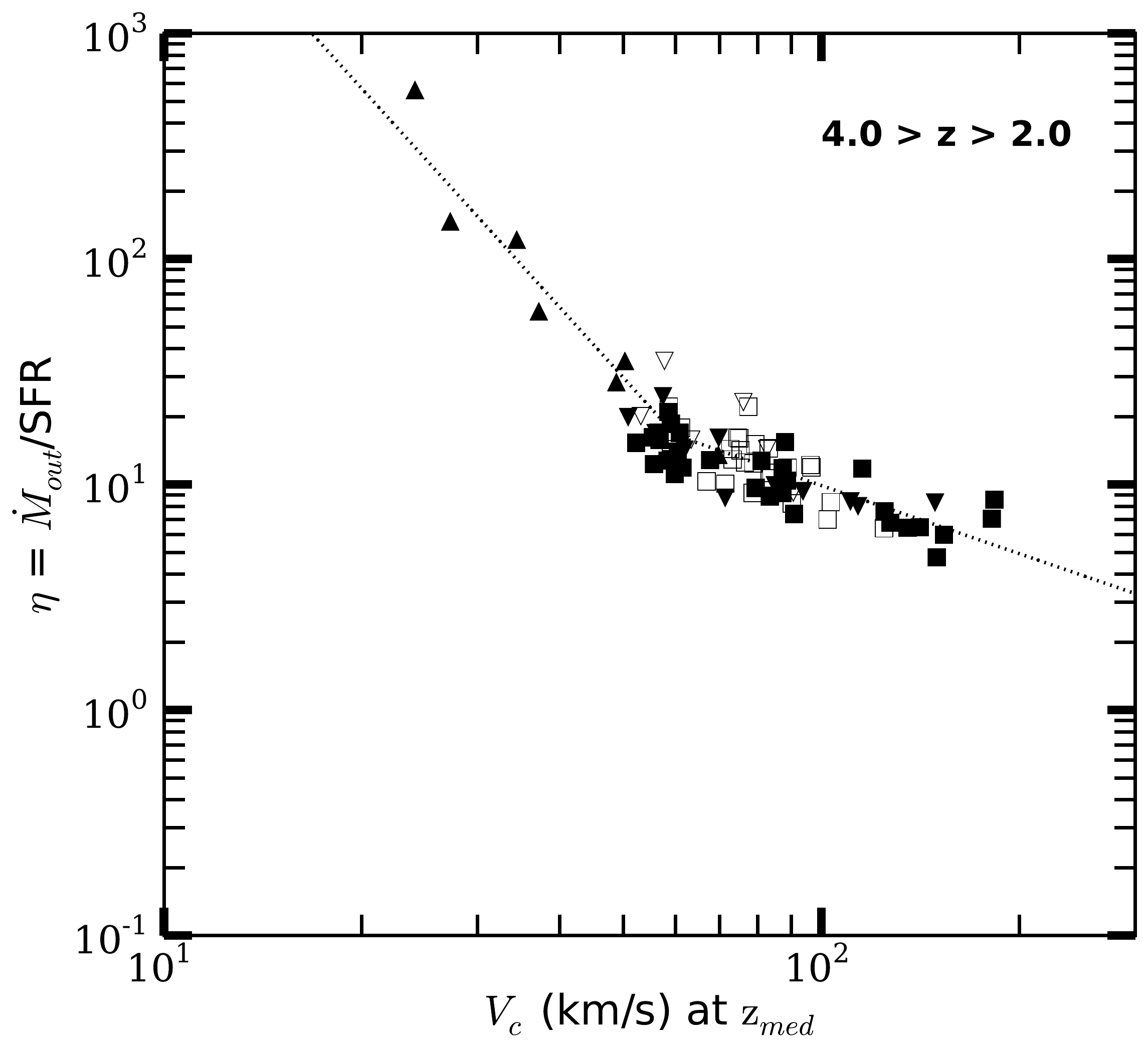}\\ \vspace{-0.12cm}
\includegraphics[width=\textwidth]{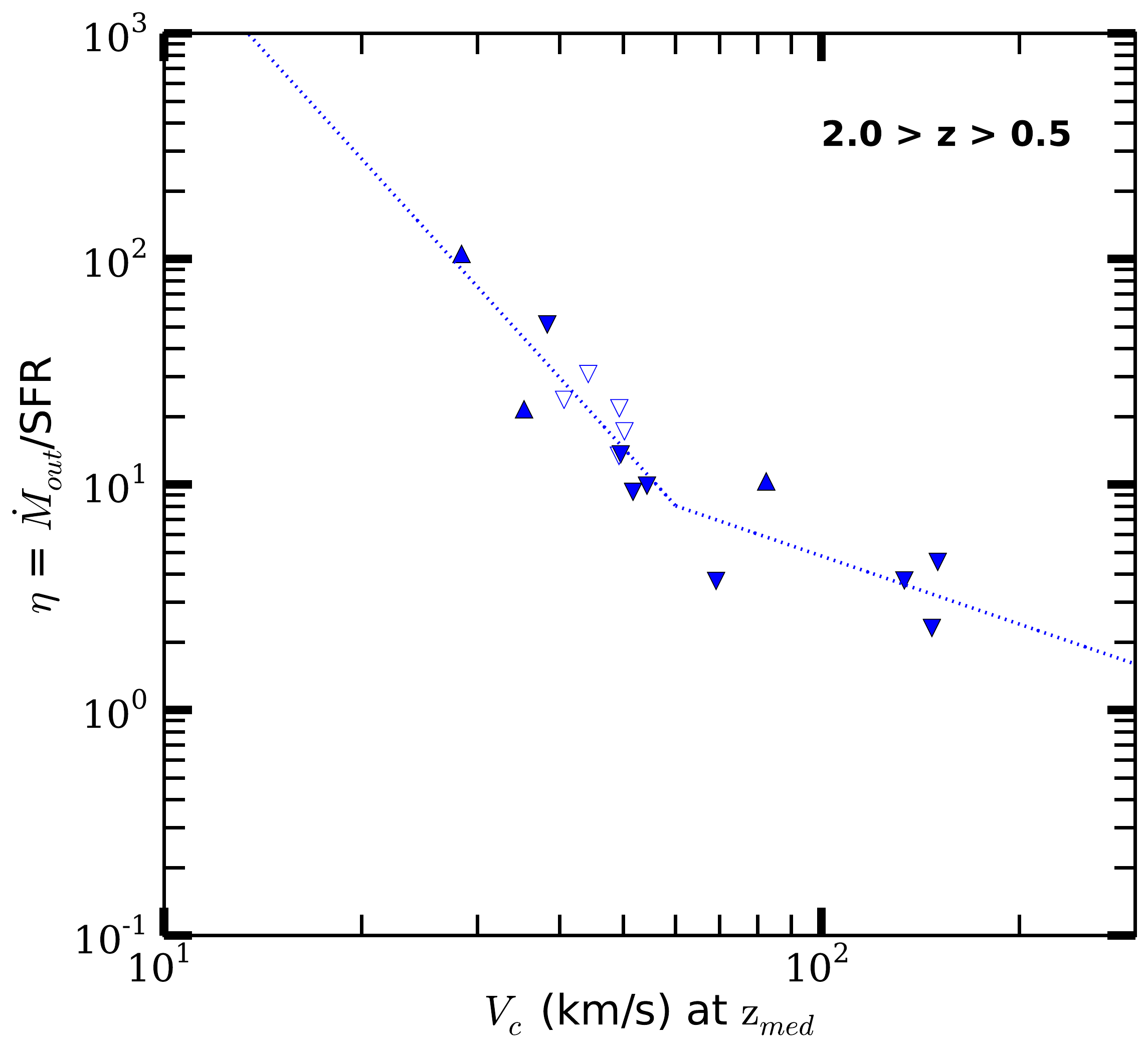}\\ \vspace{-0.12cm}
\includegraphics[width=\textwidth]{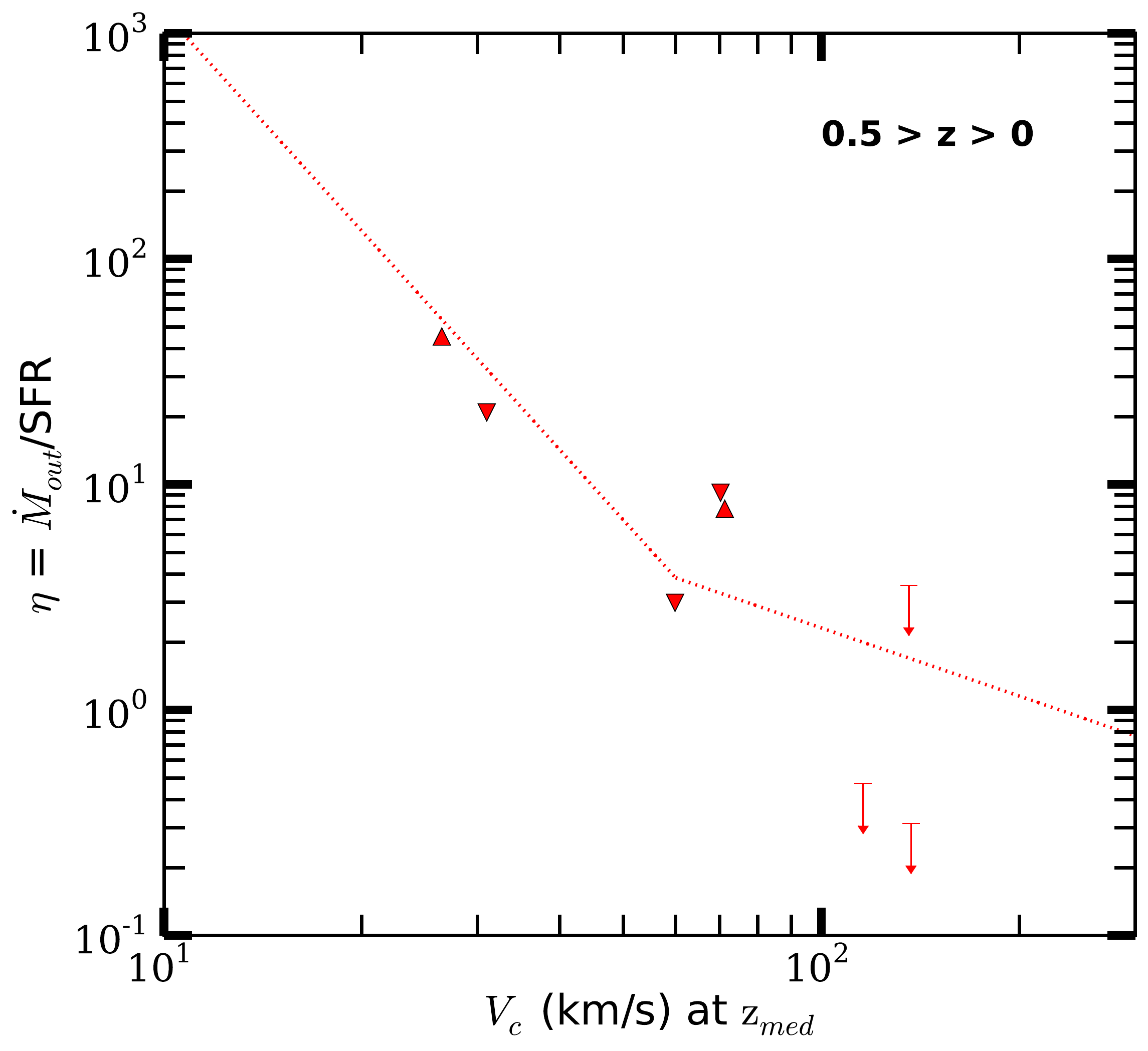}
\end{minipage}
\begin{minipage}{0.44\textwidth}
\vspace{-0.35cm}
\includegraphics[width=\textwidth]{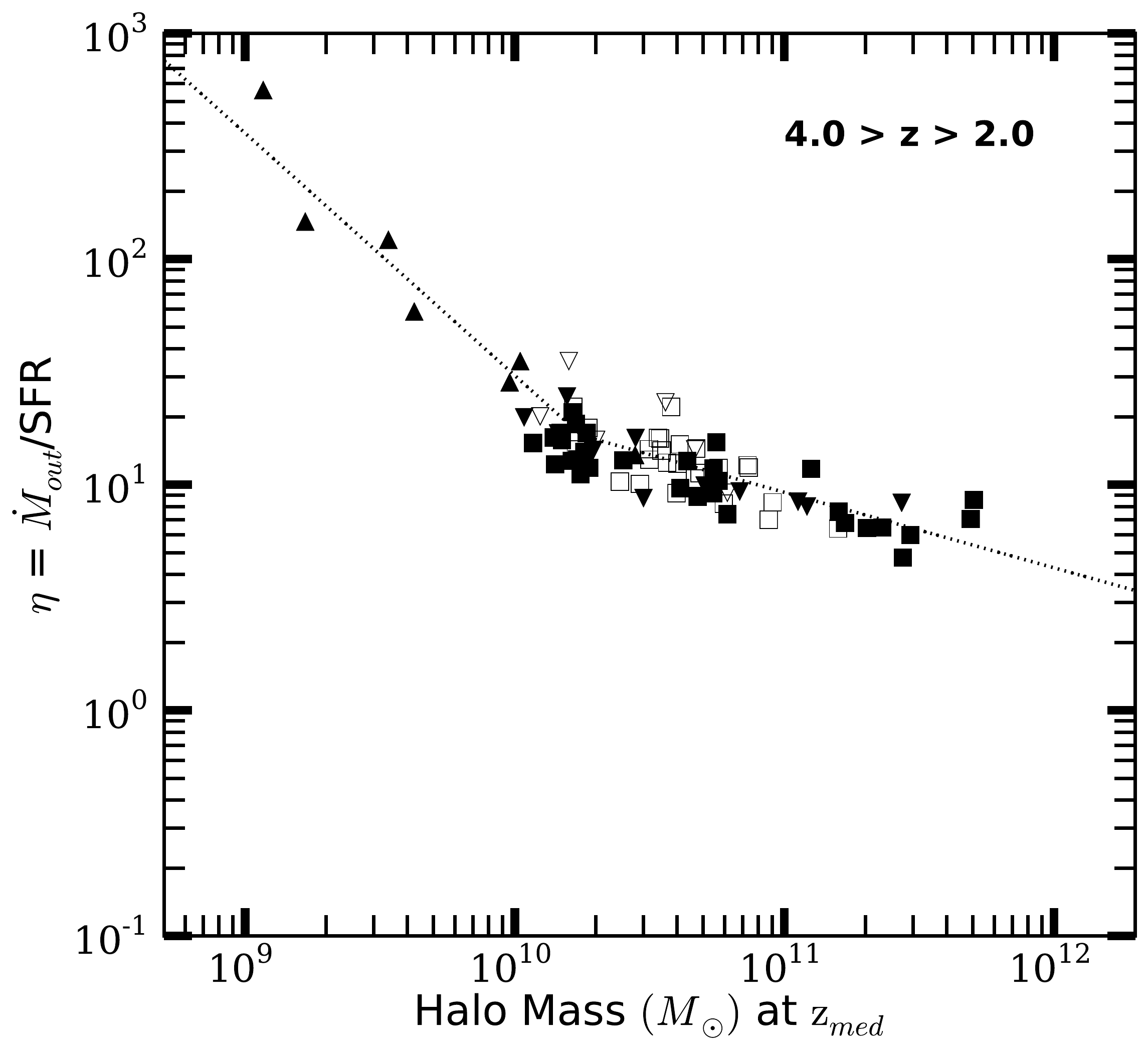}\\ \vspace{-0.10cm}
\includegraphics[width=\textwidth]{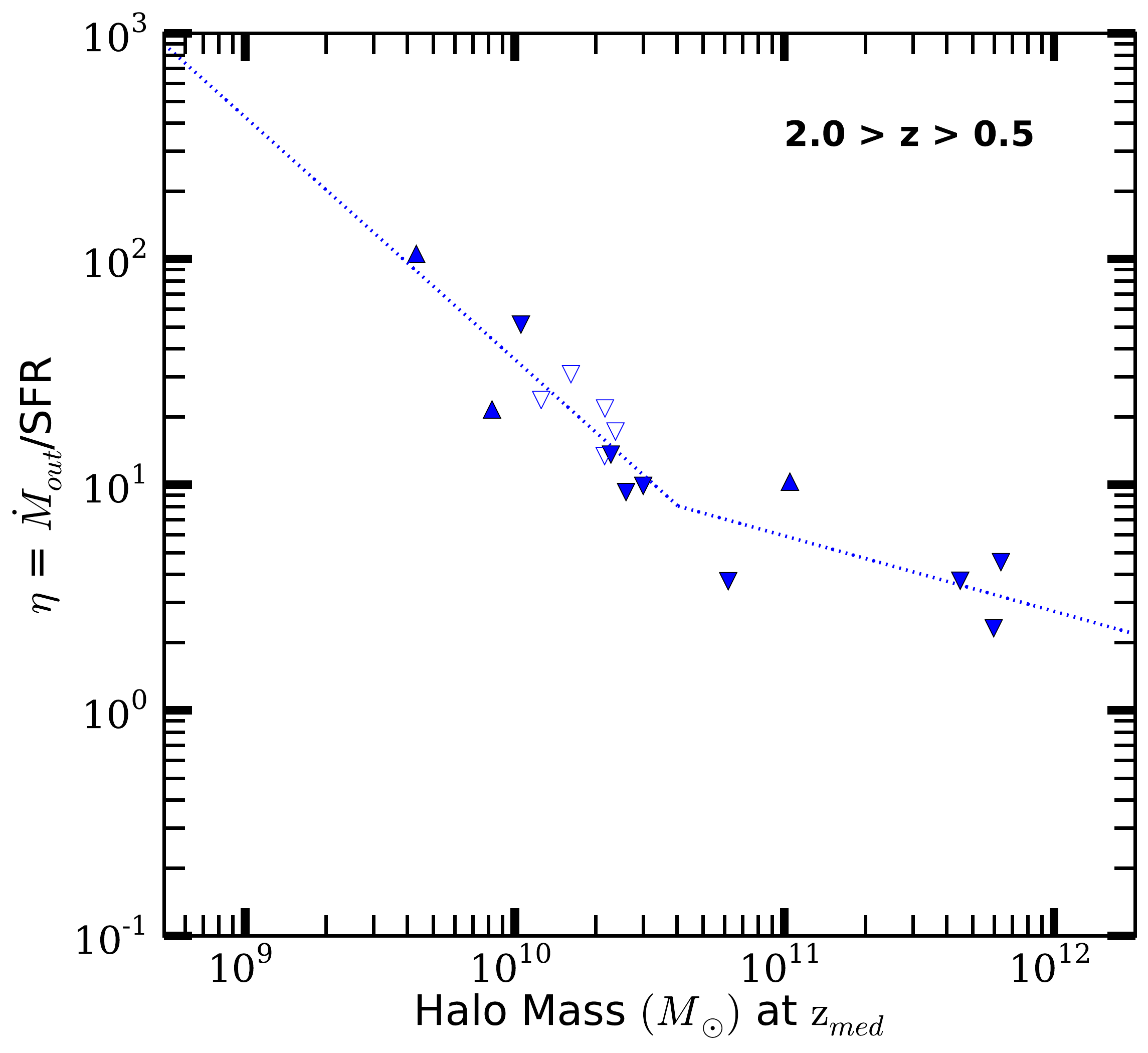}\\ \vspace{0.15cm}
\includegraphics[width=\textwidth]{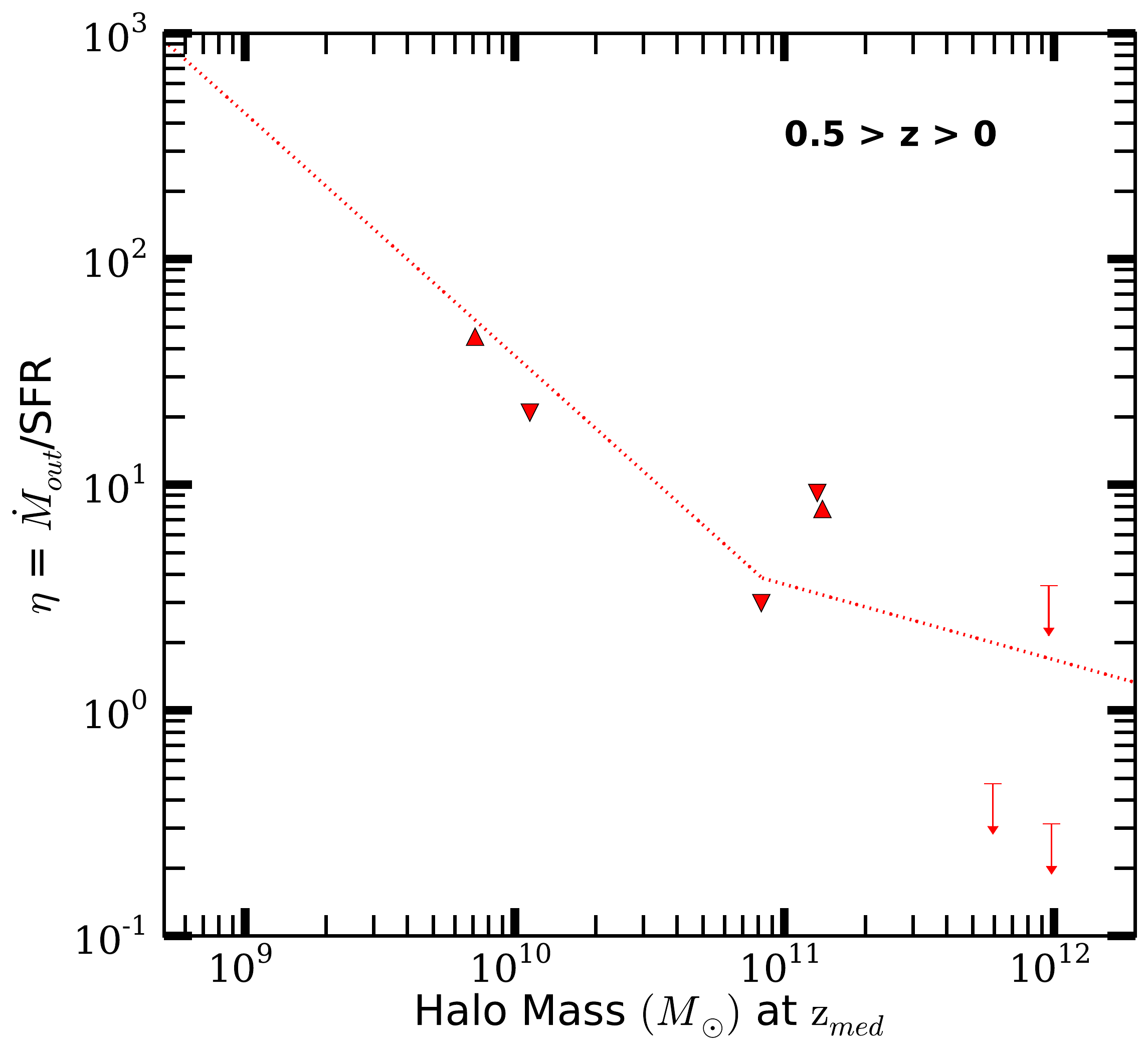}
\end{minipage}
\caption{Average mass-loading factor at  4.0 > z > 2.0 (top, black), 2.0 > z > 0.5 (middle, blue), and 0.5 > z > 0 (bottom, red) vs circular velocity (left) and halo mass (right) as they are at the midpoint of the redshift interval  (zmed = 3, 1.25 and 0.25 for high, medium and low-z). Right side up triangles show the halos in the zoom in regions of \textbf{m09, m10,} and \textbf{m11}. Upside down triangles, show \textbf{m12} halos, except at low-z, where their ``main'' halos are shown as upper limits (see text). Squares show {\bf z2h} halos. Open symbols indicate halos that survived at least as long as the midpoint of the interval, but not until the end of the interval. For $\eta$ as a function of $v_c$, we provide a broken power law fit (dotted line) including a redshift evolution term (Equations 4 and 5), and evaluate it at $z = 3$ (top), $z = 1.25$ (middle), and $z = 0.25$ (bottom). Fits were generated with high and intermediate redshift data and extrapolated to low redshift. The $\eta$ vs $M_h$ fit is derived from the $\eta$ vs $v_c$ fit (Equations \ref{eq:Mh_fit_little} and \ref{eq:Mh_fit_big}).}
\label{fig:MassLoading5}
\end{figure*}

We focus here on average cumulative values of $\eta$ over various intervals for halos in our simulations. We choose three such intervals to divide the $4.0 > z > 0$ evolutionary history, which were already introduced in Section \ref{sec:wiggle} as high-z, med-z, and low-z. Each of these intervals typically provides at least \textasciitilde 5 distinct episodes of outflows from which to make a measurement; however, the halo can gain a very significant amount of mass between the start of the interval and the end. For example, the mass of the ``main'' halo of \textbf{m12i} grows by a factor of 10 between $z=4.0$, the start of the high redshift interval, and $z=2$, the endpoint. This growth affects the physical conditions that affect the launching of the galactic wind. For this reason, we give fits for  $\eta$ as a function of various physical quantities as measured at the midpoint of each interval in redshift ($z_{med}=3$ for high-z, $z_{med}=1.25$ for med-z, and $z_{med}=0.25$ for low-z). We have considered other choices for the representative redshift, such as the epoch when the cumulative time-integrated flux of ejected material in each halo reaches 50\% of its final value, but found that our results were largely unchanged. Within each redshift interval, we elect to use a single epoch for all halos to simplify interpretation.

\begin{figure}
\vspace{-0.2cm}
\includegraphics[width=\columnwidth]{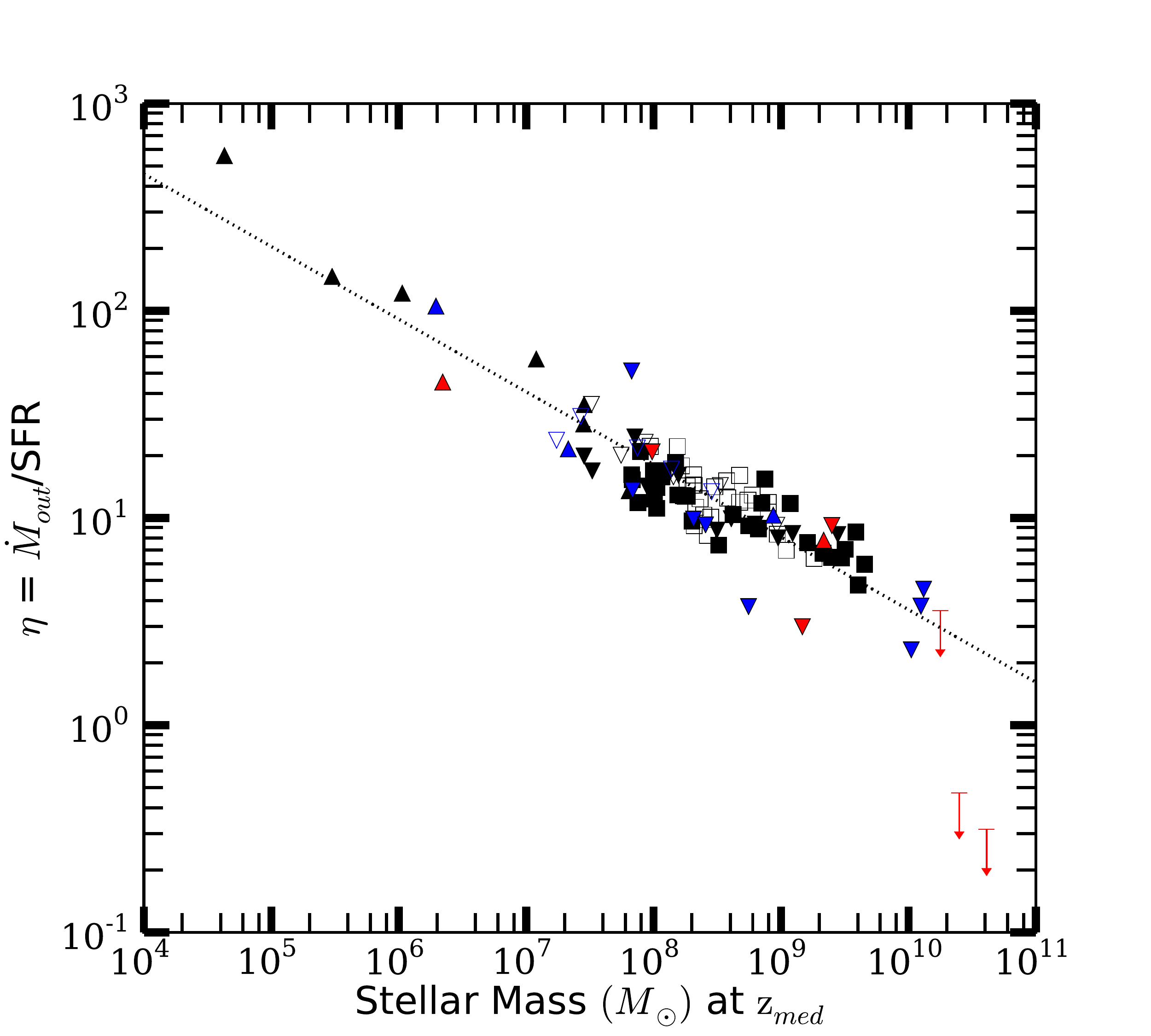}
\caption{ Average mass-loading factor ($\eta$) vs stellar mass ($M_*$), using the same symbol and color conventions as Figure \ref{fig:MassLoading5}. A single power law fit  with no redshift dependence (Equation \ref{eq:Mstarfit}, dotted black line) describes the data well, except for massive halos at low redshifts, where outflows are diminished (red upper limits).}
\vspace{0.3cm}
\label{fig:MassLoading5a}
\end{figure}

In the figures and fits provided in the sections below, we present $\eta$ as measured by the ratio of integrated outflow and star formation rates over the entire considered interval. Outflow rates themselves were measured with the \textit{Instantaneous Mass Flux} method, and a radial velocity cut of $v_{cut} = 0$ is used to define outflows. We also provide Table \ref{tbl:bigtable}, which shows  average values of $\eta$ for the ``main'' halos in each simulation at various epochs using various measurement methods. All outflow rates were measured in the $0.25 \Rvir$ shell. Section \ref{sec:radii} shows how these measurements differ at various halo-centric radii. An alternative approach would be to instead use a shell at a fixed physical radius at all times (i.e. a few tens of kpc). However, using such a threshold would probe rather different spatial regions when applied to our dwarf galaxies (potentially outside $\Rvir$), and to our most massive halos (close to galactic edge). For now, we stick to using shells at a fixed fraction of $\Rvir$, as they can consistently be adapted to all halos at all epochs.

\subsection{Fits of $\eta$ for individual halos}
\label{sec:etafit}

We start by considering the relationship between $\eta$ and the halo circular velocity ($v_c = \sqrt{GM_h/\Rvir}$), which evolves more slowly with redshift than other halo properties (as previously mentioned, the halo mass of \textbf{m12i} increases by a factor of \textasciitilde10 between $z=4$ and $z=2$, while $v_c$ only increases by a factor of \textasciitilde2). We show the average value $\eta$ vs $v_c$ in the left panel of Figure \ref{fig:MassLoading5}. We can immediately see that halos with low $v_c$ and halos with high $v_c$ may be best described by different slopes. Our method for constructing the fit for $\eta$ vs. $v_c$ is as follows: We divide the sample into two distinct populations, $v_c < 60 \mathrm{km/s}$ and $v_c > 60 \mathrm{km/s}$. The choice to use 60 km/s was determined to produce the lowest $\chi^2$ statistics compared to other dividing points. We then "anchor" both fits to the approximate mean value of $\eta$ at $v_c = 60 \mathrm{km/s}$, which is computed by taking the values of eta for all halos between $50\mathrm{km/s} < \eta < 70 \mathrm{km/s}$ and averaging them in log space. We use this anchor to fit a broken power law for low $v_c$ and high $v_c$ halos. We find that the reduced $\chi^2$ statistic for the broken power law fit is significantly lower than the reduced $\chi^2$ statistic for a single power law fit. All fitting is done in log-log space, and the errors quoted come from the diagonalized covariance matrix from the minimized residuals. First, we consider only our high-z sample, which is the most complete (black points in Figure \ref{fig:MassLoading5}). Taking measurements of $v_c$ at $z=3$, halos with $v_c < 60 \mathrm{km/s}$ are fit by the relation $\eta =  17 \left( \frac{v_c}{60 \rm km/s} \right) ^{-3.2}$. For halos with $v_c > 60 \mathrm{km/s}$, we find $\eta = 17 \left( \frac{v_c}{60 \rm km/s} \right) ^{-1.0}$.

Because the \textbf{z2h} sample ends at $z=2$ and halos begin to coalesce and grow to encompass the zoom-in regions, the number of halos that we are able to measure is smaller at $z<2$. Although our data at lower redshift is relatively limited, we also provide a fit for the med-z ($2.0 > z > 0.5$) sample. We again take measurements of $v_c$ at the midpoint of the redshift interval, in this case $z=1.25$, and use 60 km/s as a dividing point. We find $\eta =  8.8 \left( \frac{v_c}{60 {\rm km/s}} \right)^{-3.0}$ for  $v_c < 60 \mathrm{km/s}$, and $\eta =  8.8  \left( \frac{v_c}{60 {\rm km/s}} \right)^{-1.1}$ for  $v_c > 60 \mathrm{km/s}$. Since the slopes measured for the high-z and med-z samples are similar in both $v_c$ regimes, we combine the datasets and provide a single unifying fit including an additional term to account for redshift evolution. For  $v_c < 60 \mathrm{km/s}$, we find: 

\begin{equation}
\eta = 2.9 \left(1 + z \right)^{1.3} \left( \frac{v_c}{60 \rm km/s} \right) ^{-3.2}.
\label{eq:vc_fit_little}
\end{equation}

For halos with  $v_c > 60 \mathrm{km/s}$, we find 

\begin{equation}
\eta = 2.9 \left(1 + z \right)^{1.3} \left( \frac{v_c}{60 \rm km/s} \right) ^{-1.0}.
\label{eq:vc_fit_big}
\end{equation}

The derived errors are $\sigma=0.2$ and $\sigma=0.07$ for the low-$v_c$ and high-$v_c$ sides, respectively. The error on the redshift evolution is $\sigma=0.1$. The root-mean squared error around the fit is $0.1$ dex. If we fit halos with $v_c < 60\mathrm{km/s}$ and $v_c > 60\mathrm{km/s}$ separately and allow the normalization to be another free parameter, the errors for the two slopes are larger by a factor of \textasciitilde1.5, and the normalizations of both fits have errors of $0.2$ dex, suggesting a factor of \textasciitilde2 uncertainty. We stress that these fits apply only to the mass-loading factor as measured at $0.25 \Rvir$, and would likely be somewhat different if we instead measured $\eta$ at a fixed physical radius.\footnote{We have also measured $\eta$ at a fixed physical radius of 25 kpc for all halos at all redshifts, and verified that $\eta$ at fixed $v_c$ has redshift evolution consistent with Equations \ref{eq:vc_fit_little} and \ref{eq:vc_fit_big}.} Fits using $\eta$ computed only from outflows with $v_{rad} > \sigma_{1D}$ show similar slopes for both sides, but are normalized lower by \textasciitilde25\%. 

We choose not to include the limited dataset we have from the low-z ($z < 0.5$) regime in the fitting procedure, but note the most significant outliers are two of the three ``main'' L*-progenitor halos. In fact, the outflows seen in the L*-progenitors at this epoch are probably not caused by star formation, so we therefore mark them as upper limits on the figures. This discrepancy will be explored further in Section \ref{sec:etaloz}.

One notable feature of Figure \ref{fig:MassLoading5} is that the simulations plotted were run with different resolutions, yet low-mass halos from the \textbf{z2h} sample as well as those from the L*-progenitor runs (\textbf{m12v}, \textbf{m12i}, \textbf{m12q}) are found to have similar values of $\eta$ as the highly-resolved dwarf galaxies (\textbf{m09}, \textbf{m10}, \textbf{m11}). This demonstrates a degree of convergence in our simulation sample. 

We convert the broken power law fit derived for $v_c$ into a fit for halo mass, $M_h$ by using the analytic relationship between the two. The result is shown in the right panel of  Figure \ref{fig:MassLoading5}. This fit works just as well as the $v_c$ fit by construction, owing to the one-to-one correspondence between $v_c$ and $M_h$ as measured at a particular epoch. We only need to consider the redshift evolution of $M_h$ at the fixed value of $v_c = 60 \mathrm{km/s}$, which we represent as $M_{h60}$. 

For $M_h < M_{h60}$, we find:
\begin{equation}
\eta = 2.9 \left(1 + z \right)^{1.3}\left( \frac{M_h}{ M_{h60}} \right) ^{-1.1}.
\label{eq:Mh_fit_little}
\end{equation}
For $M_h > M_{h60}$, it becomes:
\begin{equation}
\eta = 2.9 \left(1 + z \right)^{1.3} \left( \frac{M_h}{ M_{h60}} \right) ^{-0.33}.
\label{eq:Mh_fit_big}
\end{equation}

It is straightforward to derive $M_{h60}$ as a function of redshift for any given cosmology and choice of definition of virial overdensity. We find $M_{h60} = 1.8 \times 10^{10} \Msun$ at $z=3$,  $M_{h60} = 4.0 \times 10^{10} \Msun$ at $z=1.25$, and $M_{h60} =  8.3 \times 10^{10} \Msun$ at $z=0.25$. 

Figure \ref{fig:MassLoading5a} shows the relationship between $\eta$ and stellar mass, $M_*$. Unlike the fitting method used for $v_c$ and $M_h$, we use a single power law fit that describes $\eta$ as a function of $M_*$. We have confirmed that the reduced $\chi^2$ statistic for this fit is low, validating our single power law approach. We again combine the data for the high-z and med-z samples, and include an additional term in the fitting function for a redshift dependence. The best-fit relation for the redshift dependence is $\eta \propto (1+z)^{0.02}$, which is consistent with no dependence, given our errors. Hence, we present a redshift-independent fit for $\eta$ as a function of $M_*$:

\begin{equation}
\eta = 3.6 \left( \frac{M_*}{10^{10} \Msun} \right) ^{-0.35}.
\label{eq:Mstarfit}
\end{equation}

The error on the fit for the power-law indices is $\sigma=0.02$, suggesting a better fit than what was found for $v_c$, though the scatter is still significant. The error on the normalization is $0.2$ dex. The L*-progenitors at low redshift are again the most notable outliers from the fit, while the rest of the low-z data is somewhat better described than it was by the $v_c$ fit. 

We note that when a halo has multiple long-lived massive progenitors that coalesce by the end of the interval, we include each progenitor as a separate track in the figures and fits. We employ a few other criteria for inclusion of progenitor halos:\\
1.) They are either the ``main'' progenitor for the entire interval, or are detected as central (non-satellite) halos until the midpoint of the interval in redshift space ($z = 3$ for high-z  $z=1.25$ for med-z, and  $z=0.25$ for low-z. )\\
2.) Consist of at least 98\% high-resolution dark matter (low contamination from low-resolution particles). We make an exception in the case of \textbf{m11} and \textbf{m12q}, as these two runs feature low-resolution DM particles that are about as massive as hi-res DM particles from other runs (\textbf{m12v} and \textbf{m12i}). Each consists of more than 98\% high-resolution DM particles within 0.1 $\Rvir$. \\
3.) Contain at least 50,000 high-resolution dark matter particles at the end of the interval, or at the last epoch at which they are counted as isolated halos. \\
4.) Form at least 50 new stellar particles over the interval. 

Our statistics at the high-z interval are sufficient to obtain a statistically meaningful fit even if we consider only halos that survive as their own ``main'' progenitors to $z=2$, but inclusion of the other progenitors does not appear to alter the fit. We mark halos that did not survive until the end of the given interval as open symbols on the figures.  Although simulations of different resolutions are combined on these figures, it can be seen that resolution has no discernible effect on $\eta$ as long as the stated resolution standards are maintained.

We consider whether the measured correlations for $\eta$ vs circular velocity can inform whether energy or momentum are the conserved quantities during the blowout. Recall that ``energy-driven'' (energy conserving) winds follow the scaling $\eta \propto v_c^{-2}$, and ``momentum-driven'' (momentum conserving) winds follow the scaling $\eta \propto v_c^{-1}$ \citep{murray05}. It is expected that low-mass galaxies are more affected by energy-driven winds, and high-mass galaxies by momentum-driven winds \citep{murray_etal11, hopkins12b}. We find that at high redshift, our low-$v_c$ halos have a dependence steeper than $v_c^{-2}$, while those at high mass have a dependence that matches $v_c^{-1}$. Nonetheless, a clear transition is seen between the two regimes. Stellar feedback on small scales in our simulations includes both energy and momentum input from several processes in a complex environment. In order to explain the dependence of mass-loading factor on halo properties, we therefore plan a more detailed study of the nature of wind-driving in our simulations in future work.

\subsection{$\eta$ at low-z: $z < 0.5$}
\label{sec:etaloz}

We have elected thus far to show data at $z<0.5$, but not to use them in our fitting procedure. As we mentioned previously, our sample of $z<0.5$ non-satellite halos is limited, as it mainly consists of the ``main'' halos from each simulation. Furthermore, the three L*-progenitors - \textbf{m12i}, \textbf{m12v}, and \textbf{m12q} - no longer appear to have outflows in the CGM that correlate with star formation. We have concluded that although, a positive integrated flux of material was measured through the $0.25 \Rvir$ shell, it is misleading to report this as a proper measurement of $\eta$. This is because our methods for measuring outflow rates cannot always discriminate between stellar feedback-driven winds and other sources of gas motion, such as random motion or close-passages of satellites and mergers.

At epochs $z>0.5$, most of the major outflow events coincide with high star formation rates seen 50-100 Myr prior. At $z<0.5$, in the case of the L*-progenitors, star formation rates and outflow rates are continuous rather than episodic, and appear to be uncorrelated. Furthermore, using a $\sigma_{1D}$ radial velocity threshold to discriminate against random motion generally reduces the outflow rate to near-zero values (see Figure \ref{fig:wiggle-loz}), while at earlier times, the reduction was considerably more modest (see Table \ref{tbl:bigtable}). 
In addition, we have visually checked all significant outflow episodes in our three L*-progenitors in this redshift interval and found these to originate in a close passage of gas rich satellites or the material stripped from a passing satellite galaxy. This is in contrast to high-redshift where we can connect all significant outflow episodes to the star formation bursts.

Even if we simply include all material with $v_{rad} > 0$ and make a measurement of $\eta$, \textbf{m12v} and \textbf{m12i} clearly have $\eta < 1$, and \textbf{m12q} has $\eta < 1$ if we exclude the close passage event. A truly discerning measurement to only include outflows generated by stellar feedback would suggest $\eta \ll 1$ for all L*-progenitors, as can be seen from the $v_{cut} = \sigma_{1D}$ outflow measurement in Figure \ref{fig:wiggle-loz}. We find this encouraging, as galaxies like the Milky Way that only have low to moderate star formation at present day do not appear to typically launch winds with high mass-loading factor into the CGM \citep{veilleux_etal05, rubin_etal13}.  This contrasts with the values of $\eta$ that are used for halos of this mass in some large-volume simulations which are able form disk galaxies with correct mass, but assume that the mass-loading factors remain high at low redshift, as we discuss in the next subsection.

\begin{figure}
\vspace{0.2cm}
\includegraphics[width=\columnwidth]{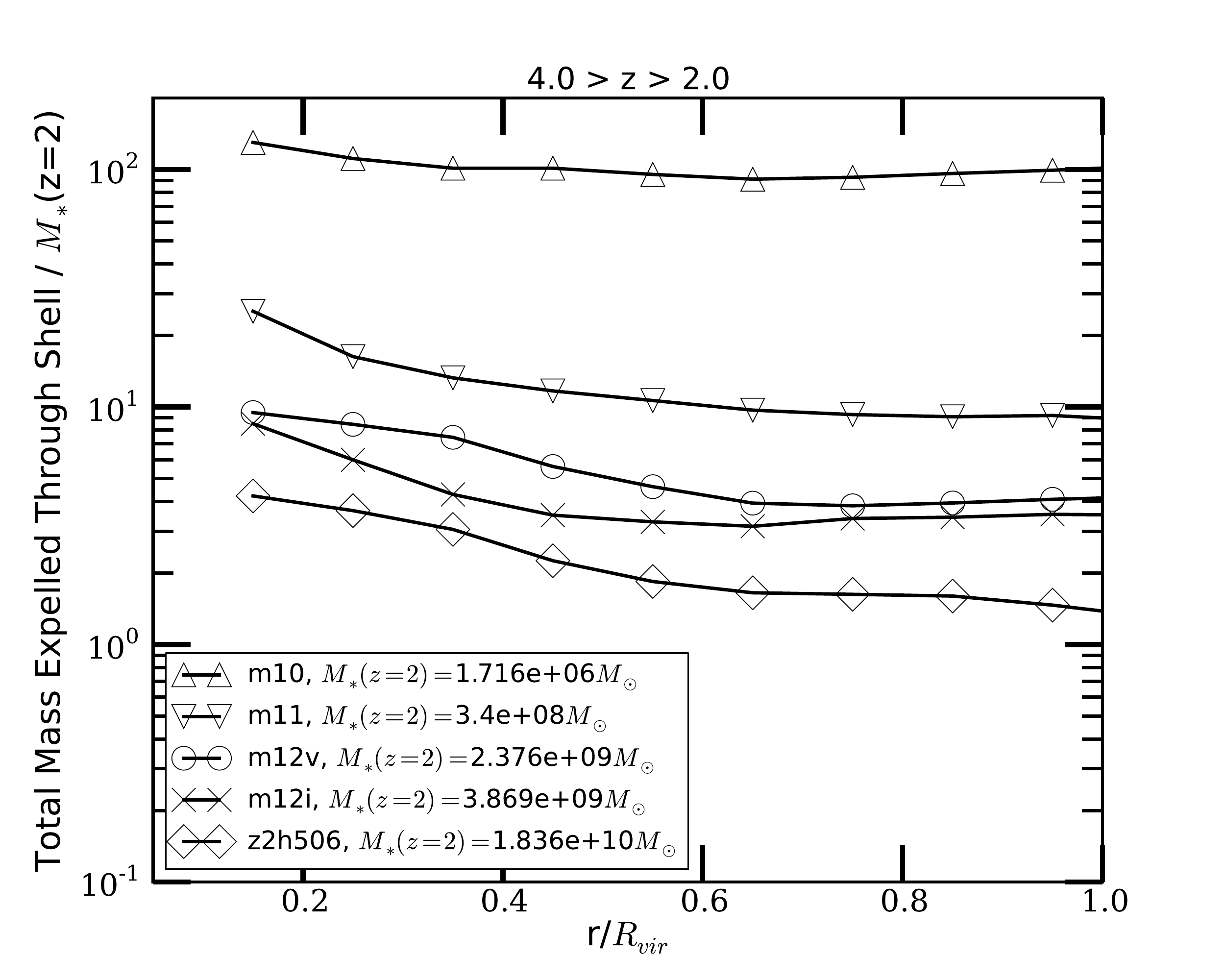}
\vspace{-0.0cm}
\includegraphics[width=\columnwidth]{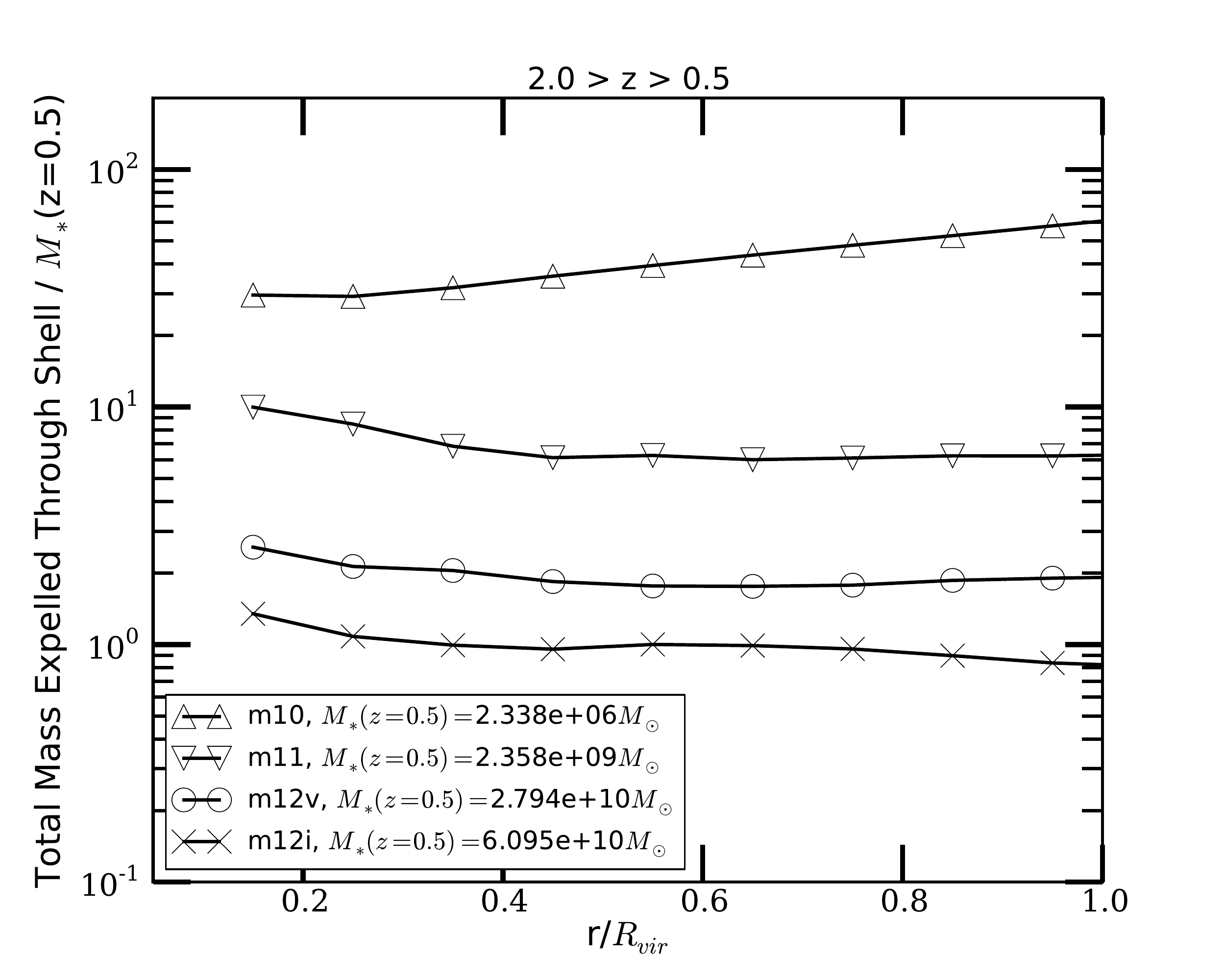}
\vspace{-0.0cm}

\caption{ Top: Total mass of outflows that traversed shells of fixed fraction of virial radius between $z=4.0$ and $z=2.0$ for \textbf{m10, m11, m12v, m12i,} and \textbf{z2h506} respectively. Each value is normalized by the stellar mass of the halo at $z=2$, which is given in the legend. This quantity generally decreases with radius, suggesting that a large fraction of ejected material can either stay in the CGM or recycle back to the galaxy, while entrainment is not significant. Bottom: The same calculation done for \textbf{m10, m11, m12v,} and \textbf{m12i} between $z=2.0$ and $z=0.5$, normalized by stellar masses at $z=0.5$. Less material is recycled and retained in the CGM at these intermediate redshifts. In the case of \textbf{m10}, entrainment may play a significant role. }
\vspace{0.3cm}
\label{fig:ShellOutflow}
\end{figure}

\begin{figure*}
\centering
\begin{minipage}{0.48\textwidth}
\includegraphics[width=\textwidth]{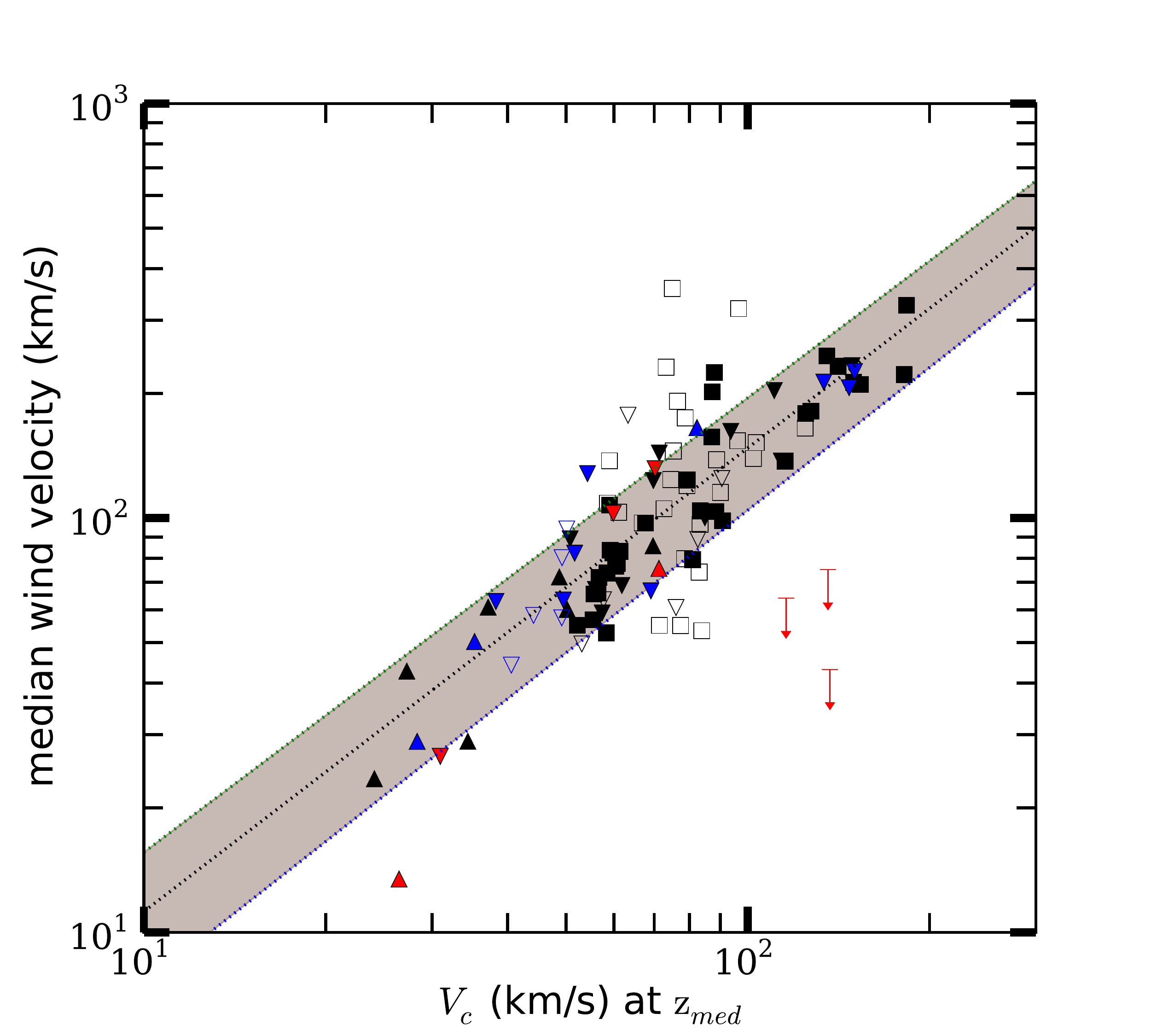}\\
\end{minipage}
\begin{minipage}{0.48\textwidth}
\includegraphics[width=\textwidth]{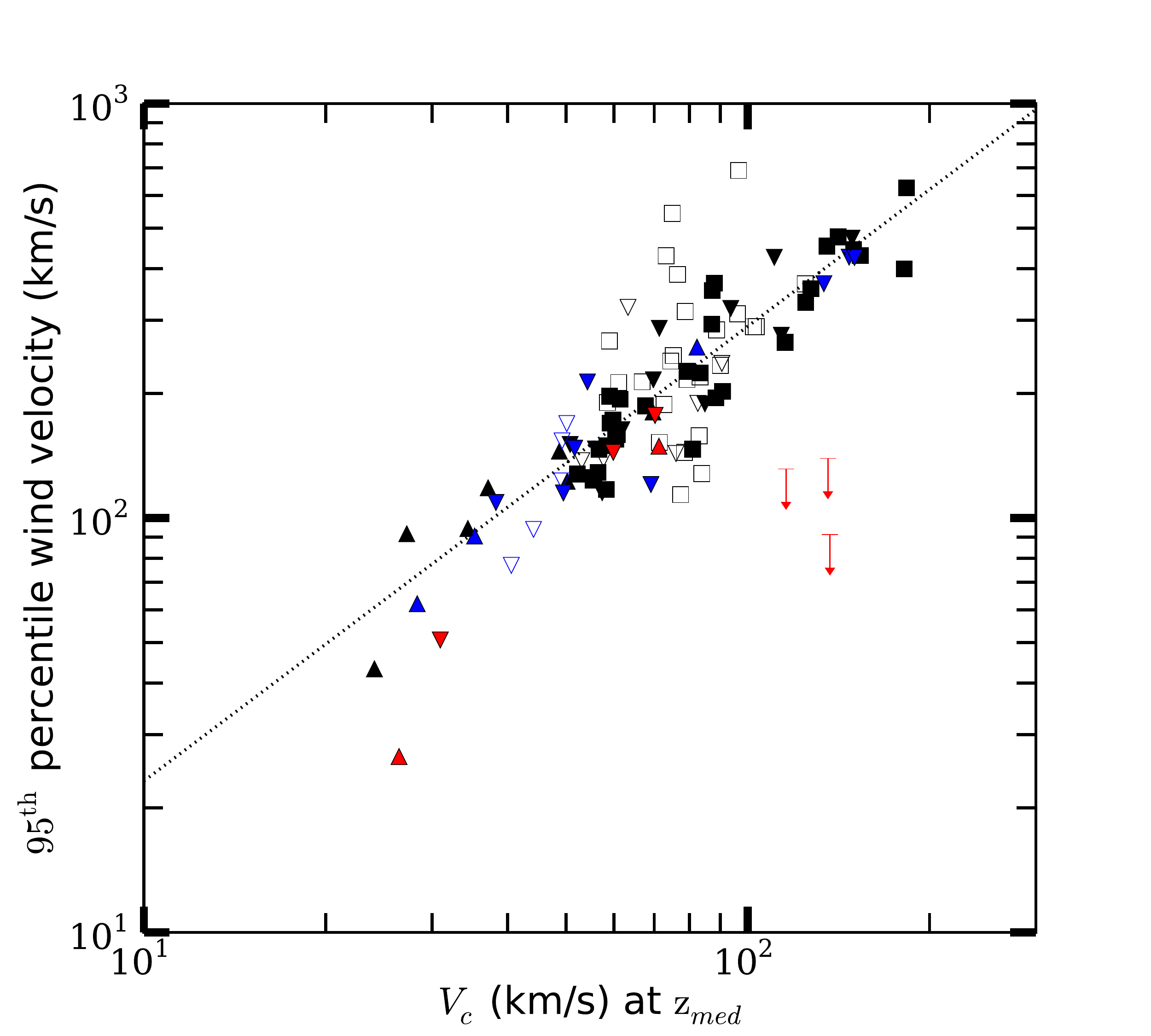}\\
\end{minipage}
\caption{Flux-weighted average 50th (left) and 95th (right) percentile wind velocity vs $v_c$ for all halos in our sample measured at $0.25 \Rvir$. Same coloring convention as Figure \ref{fig:MassLoading5}. Dotted lines show the fits given in Equations \ref{eq:vwind50} and \ref{eq:vwind95}. Both 50th and 95th percentile wind velocities scale slightly superlinearly with $v_c$, and are normalized such that typical wind velocities range from \textasciitilde1-3$v_c$. The shaded region in the left panel shows the range of velocities between best-fit lines for 25th and 75th percentile wind velocities. }
\label{fig:windfall}
\end{figure*}

\subsection{Mass-loading as a function of radius within halo}
\label{sec:radii}

In order to understand how the flux of material traverses different spatial regions of the halo, we provide Figure \ref{fig:ShellOutflow}. The top panel of this figure shows the total integrated mass of gas that has crossed various thresholds in the $4.0 > z > 2.0$ interval, normalized by the total stellar mass of each halo at $z=2$. The values from this plot can be used to estimate mass loading at different radii using the values of $\eta$ at 0.25 $\Rvir$ given in Table \ref{tbl:bigtable}.

It is striking that the total  gas mass expelled through inner region of galactic halos at high redshift is tens or even hundred times larger than the total stellar mass accumulated by $z=2$. Two of the L*-progenitors, \textbf{m12v} and \textbf{m12i}, as well as the massive dwarf \textbf{m11}, and the LBG-like \textbf{z2h506} only have about 33\% of the total material ejected into the CGM (to $0.15 \Rvir$) eventually leave the halo, while \textbf{m10} loses nearly all of the outflows. This leads us to conclude that at high redshifts, the majority of CGM outflows in sufficiently massive galaxies stays within the CGM where a larger fraction of this gas can recycle back into the galaxy or contribute to the gaseous reservoirs of halos.

The bottom panel of Figure \ref{fig:ShellOutflow} reveals that the outflow properties differ at intermediate redshift ($2.0 > z > 0.5$). In this interval, the L*-progenitors and \textbf{m11} now lose 60-70\% of material that is ejected. This suggests that although their gravitational potentials are deepening, these halos are actually more efficient at expelling baryons into the IGM. The implications will be discussed further in Section \ref{sec:discussion_galev}. \textbf{m10} is particularly peculiar, as it loses significantly more mass in the outer regions than the amount ejected from the inner regions. The mass of this halo is below the filtering scale induced by the UV background \citep{thoul96, gnedin00, faucher-giguere11b}, which means that the halo may be gradually heated and unbound. Alternatively, this may imply entrainment of loosely bound material in the outer region of the CGM during the outflow episodes as they propagate outward. We have verified that most of the mass loss follows bursts of star formation, suggesting that entrainment is the dominant mechanism that produces this behavior. These calculations do not account for the fact that the physical virial radius can be much larger at $z=0.5$ than it was at $z=2$, which means that outflow rates we compute here are only valid with respect to the instantaneous position of the virial radius but not all of this material actually leaves the growing virial radius of a halo.

\subsection{Wind velocity}
\label{sec:windvel}
As can be seen from Figure \ref{fig:traceoutflows1}, our simulations produce winds with a broad range of velocities. The detailed kinematic structure of winds will be studied in subsequent work, but here we briefly characterize the typical wind velocities seen in our simulation.

We use the same redshift intervals and criteria for inclusion as for measurements of $\eta$. At each snapshot, we calculate the 50th percentile (median) outflowing ($v > 0 {\rm km/s}$) radial wind velocity from the distribution of flow rates in the 0.25 $\Rvir$ shell (e.g. left panel of Figure \ref{fig:traceoutflows1}). We then average these wind velocities over time, weighting the value calculated at each snapshot by the outflow rate. This ensures that the average wind velocity is a characteristic of the epochs when the most significant outflows are likely to be observed. We use this same procedure to also compute the interval-averaged 95th percentile velocity to give an estimate of some of the fastest winds generated by star formation in our simulations. 


We plot the flux-weighted average 50th and 95th percentiles in Figure \ref{fig:windfall}. We find strong evidence for a slightly super-linear correlation between wind velocity and $v_c$ that is well described by a single power law. Correlations between wind velocity and galaxy mass are in fact found in observational campaigns that sample a sufficiently broad range of galaxy masses \citep{martin_etal12}. We note that while the 50th percentile velocities are sometimes close to the escape velocity of the halo, the 95th percentile velocities are considerably faster, and approach $1000$ km/s for the most massive halos in our sample. Winds with such velocities could be confused for winds generated by black holes. It remains to be seen if in massive compact galaxies that are not present in our sample, stellar feedback-driven outflows can be even faster, as observed in some extreme cases \citep{diamond-stanic12}.  
\begin{figure*}
\centering
\begin{minipage}{0.48\textwidth}
\centering
\includegraphics[width=\textwidth]{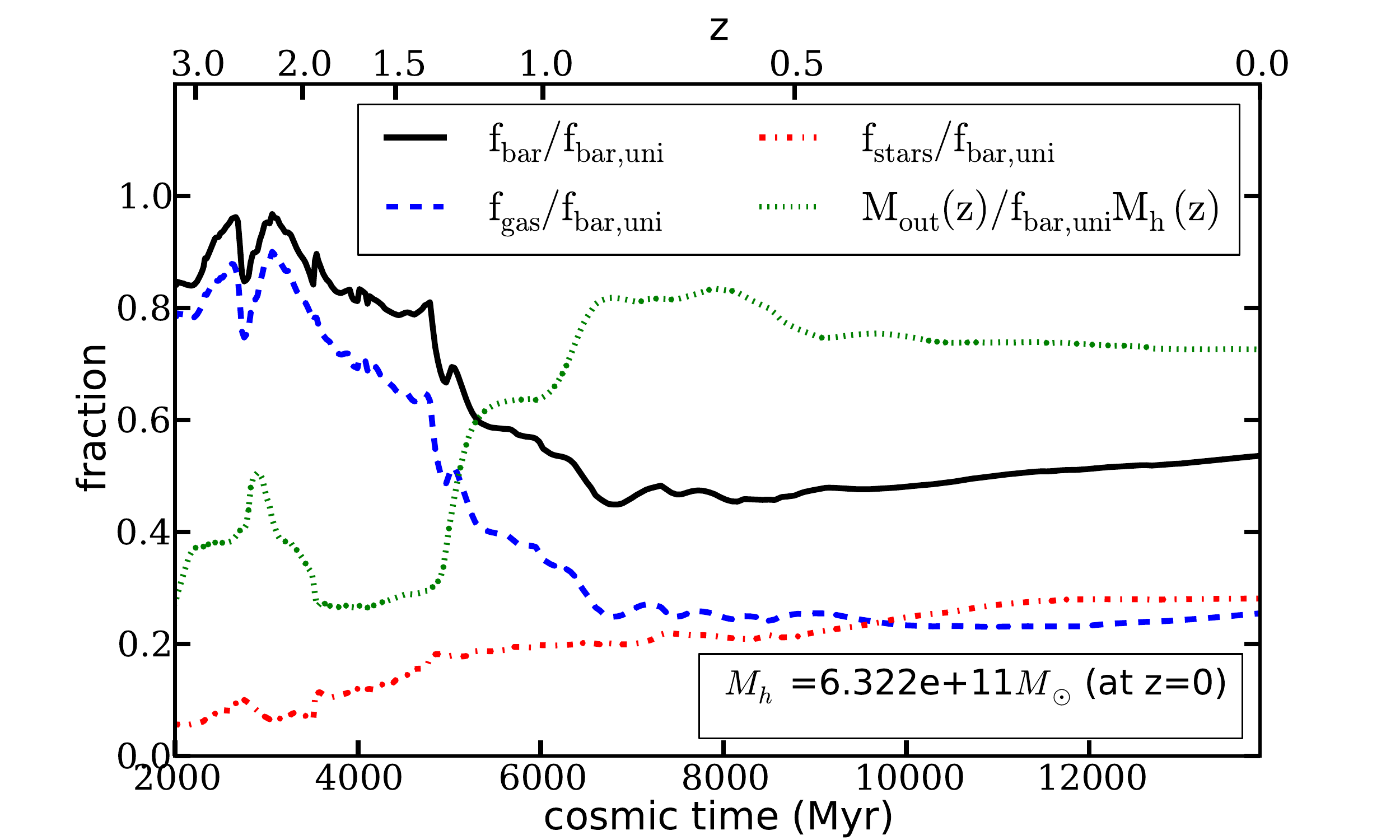}\\
\includegraphics[width=\textwidth]{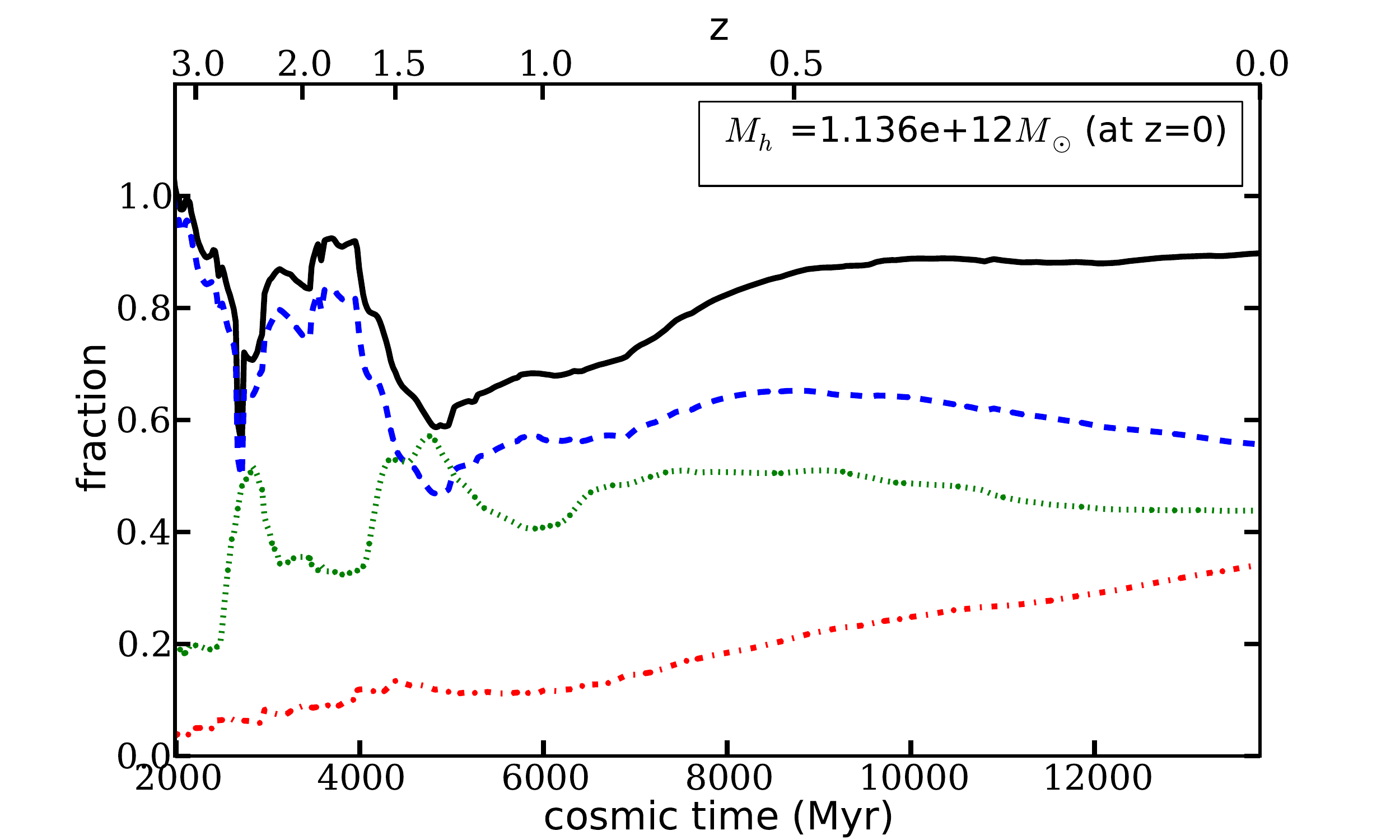}\\

\includegraphics[width=\textwidth]{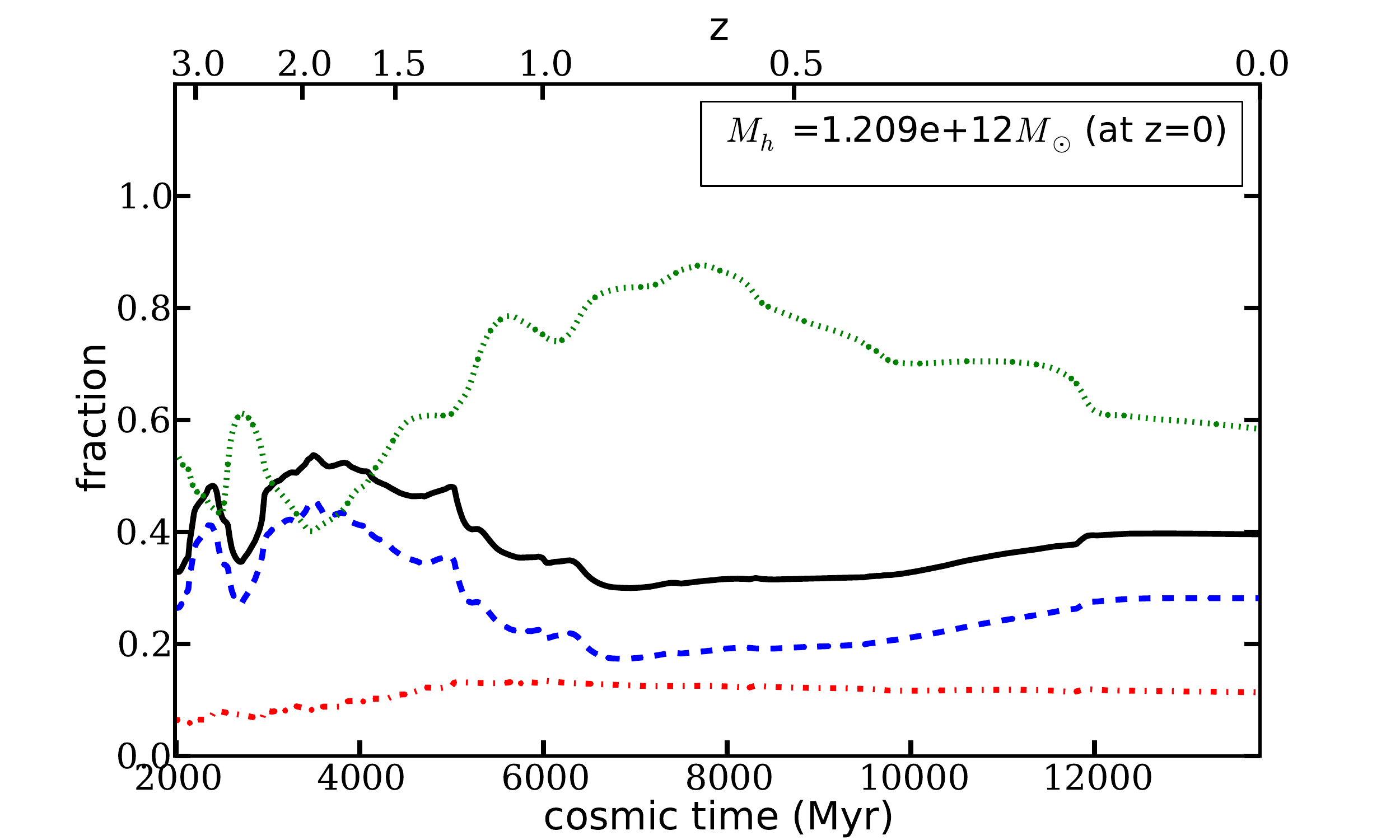}
\end{minipage}
\begin{minipage}{0.48\textwidth}
\includegraphics[width=\textwidth]{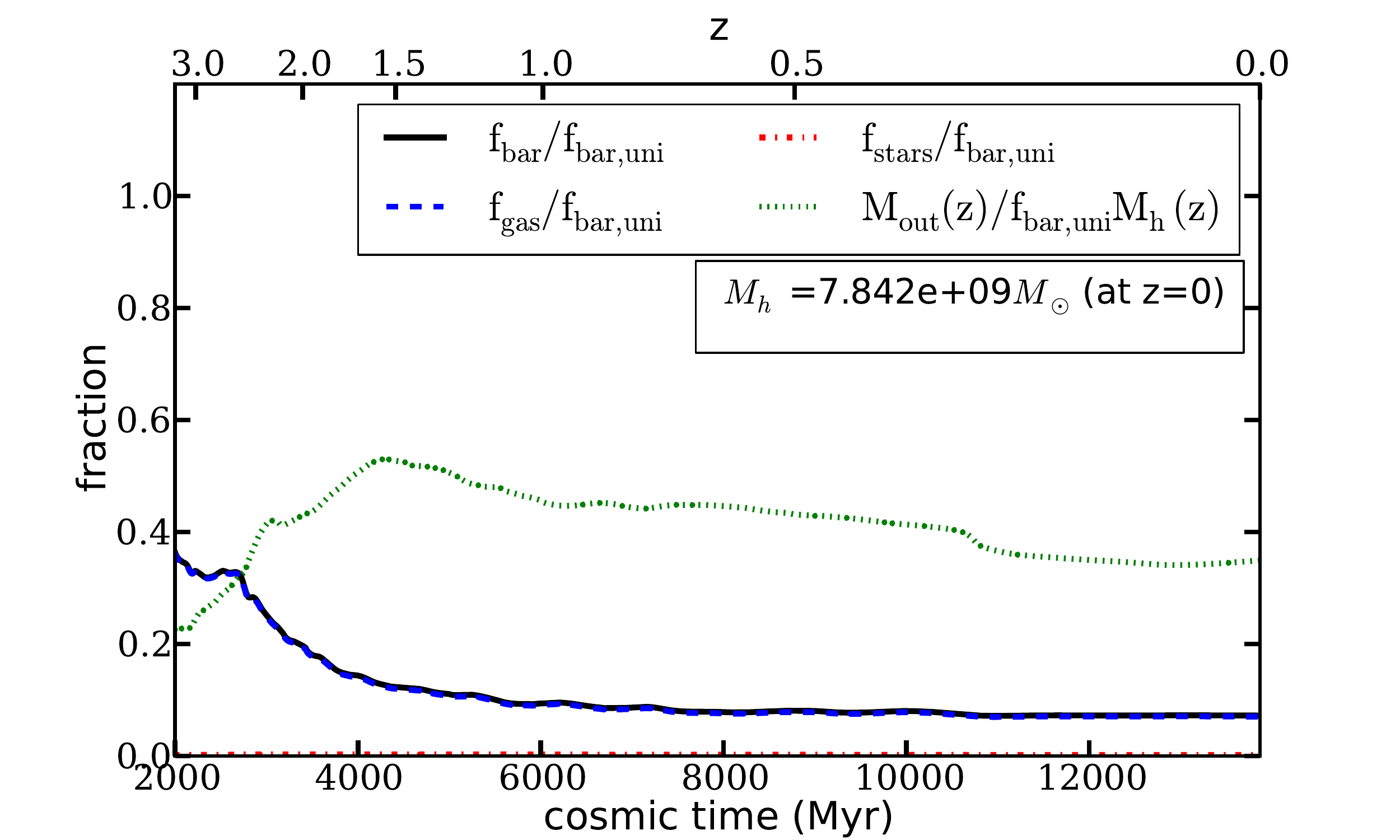}\\
\includegraphics[width=\textwidth]{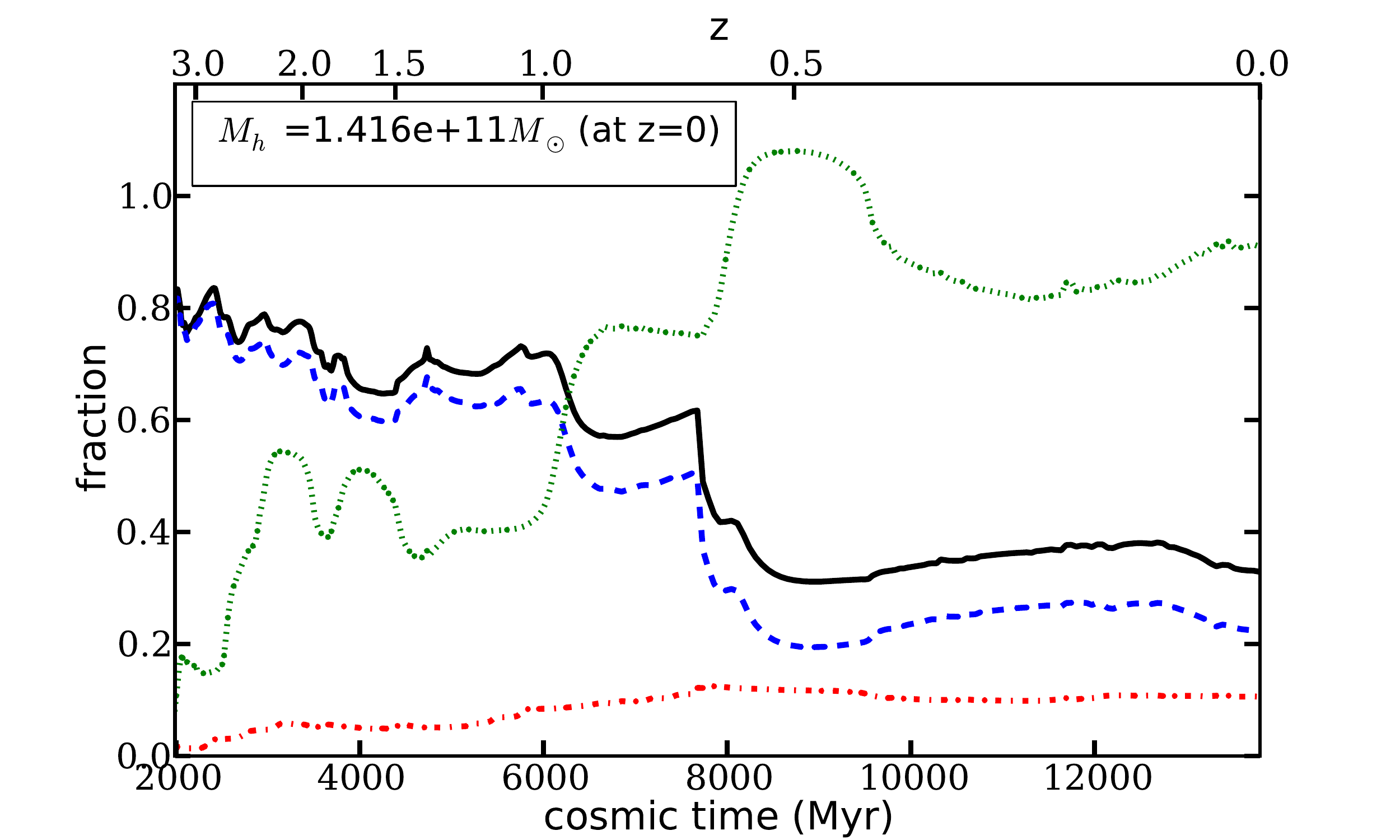}\\
\includegraphics[width=\textwidth]{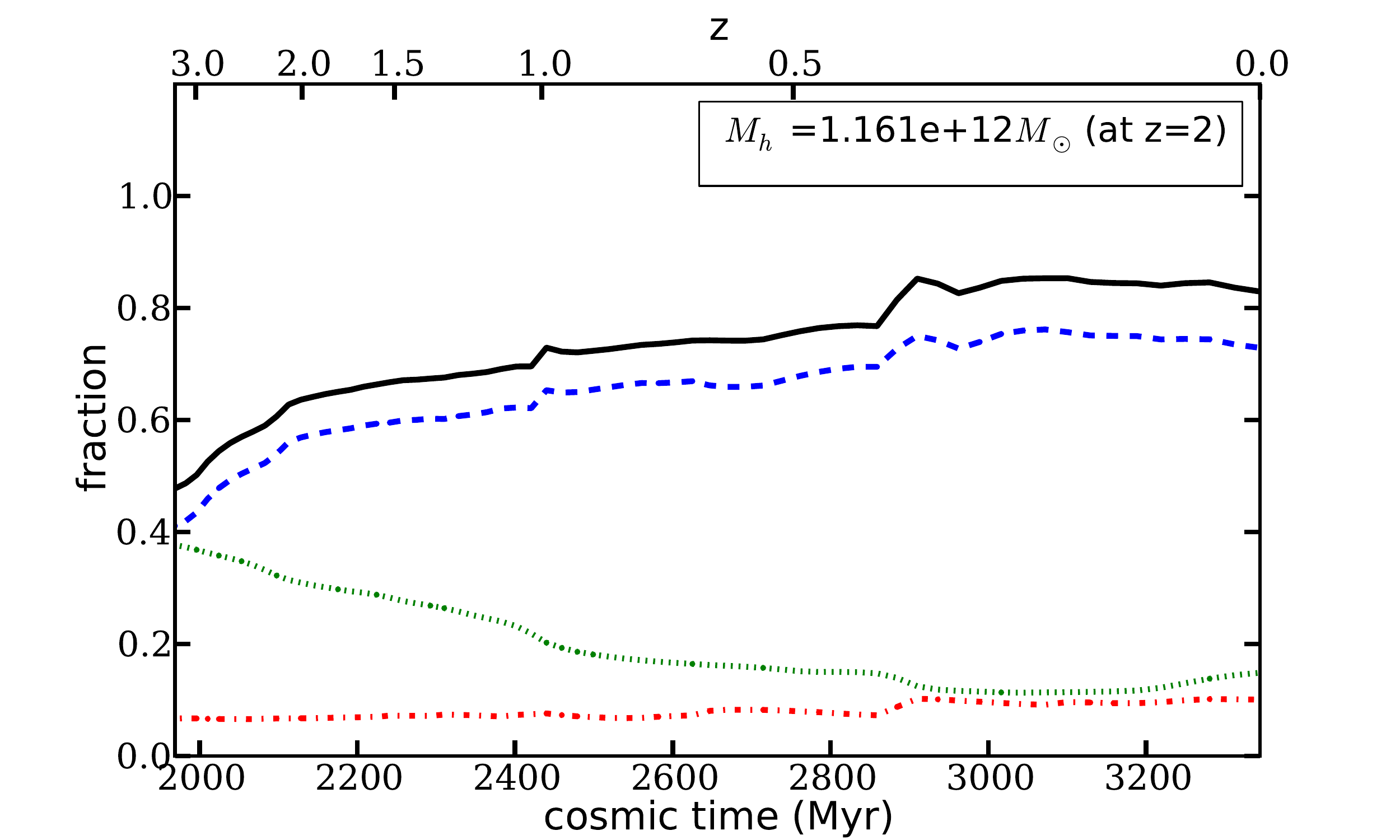}
\end{minipage}
\caption{ Mass fraction of baryons, gas, and stars within $\Rvir$ relative to the cosmic mean for several runs. The left column shows the three L*-progenitor halos  (\textbf{m12v} upper left, \textbf{m12i} middle left, \textbf{m12q} lower left). The right column shows \textbf{m10} (upper right), \textbf{m11} (middle right), and \textbf{z2h506} (lower right). All halos are shown from $3.0>z>0$ except \textbf{z2h506}, which stops at $z=2$. Also shown is the expelled fraction, which is the ratio of $M_{out}(z)$, the cumulative mass of outflowing material which has traversed $1.0 \Rvir$, and $M_h(z)f_{bar,uni}$, the expected baryonic mass of the halo assuming that it contains an exact proportion of the universal baryon budget.  This ratio is meant to indicate when the outflows played a significant role in depleting the halo of gas. A low value implies that outflows played no significant role in the depletion, while a high value implies that outflows were important. Integration for $M_{out}(z)$ starts at $z=9.0$ and is only calculated for the most massive progenitor.}
\label{fig:baryonfraction}
\end{figure*}

Again, the three L*-progenitor halos are exceptions at the low-z interval, and appear to have significantly slower winds than similar halos of their mass did at higher redshift. We interpret this as further evidence that the outflow rates measured for the L*-progenitors at low redshift are not driven by star formation. We find the 50th percentile wind velocity is fit by the relation: 

\begin{equation}
v_{{\rm wind}, 50} = 0.85 v_c^{1.1}. 
\label{eq:vwind50}
\end{equation}

The fit for 95th percentile wind velocity is fairly similar, differing mainly in normalization, which is approximately a factor of \textasciitilde2 higher:

\begin{equation}
v_{{\rm wind}, 95} = 1.9 v_c^{1.1}.
\label{eq:vwind95}
\end{equation}

We have again only used data from the high-z and med-z regimes to construct these fits, and consider no redshift evolution, as it is not apparent from the figure. We caution that the velocities discussed here were measured at $0.25 \Rvir$ for each halo considered. Observations of CGM winds, whether through quasar absorption lines or down-the-barrel spectra, cannot easily determine the precise distance of outflowing material from the wind's point of origin. This means that a typical observation may include fast-moving material within $0.25 \Rvir$ that is not considered in our measurements, though it may also include slow-moving material in the outer reaches of the halo. In future work, we plan to more closely mimic an observational approach and characterize what outflow velocities would be measured at a given SFR and galaxy mass. 

We have verified that the fits are not significantly different when we use a $v_{rad} > \sigma_{1D}$ threshold for outflowing gas. This is because our flux weighting already filters out material that is technically outflowing, but too slow to be considered a wind. 


\section{Discussion}

\label{sec:discussion}
\subsection{Comparison to subgrid prescriptions}
\label{sec:brief_eta_compare}
We present a brief comparison between our results and the subgrid prescriptions that have been employed by several other groups (see also \citealt{zahid_etal14, lu_etal15}). For more details, see Appendix \ref{sec:appendix:Compare_Eta}. We focus on comparing our results to the Illustris project (\citealt{vogelsberger13}, V13), as well as the simulations presented in \citealt{ford_etal14} (F14). Wind velocity measurements from the FIRE simulations discussed here are described in Section \ref{sec:windvel}, and converted to approximate velocities at the time of wind launching, using methodology described in Appendix \ref{sec:appendix:Compare_Eta}.

The results of our comparison are given in Table \ref{tbl:comptable}. There are interesting similarities and differences between the wind velocities and mass-loading factors found in our work and those from F14 and V13. At high redshift, the values of $\eta$ used in V13 are systematically higher than ours. The agreement is better for the most massive halos in our simulations, but worsens gradually down to $v_c = 60 \mathrm{km/s}$. This is likely because they assume a fixed $\eta \propto v_c^{-2}$ scaling at all masses, while we find $\eta \propto v_c^{-1}$ for $v_c > 60 \mathrm{km/s}$. The launch velocities of winds in V13 are consistent with ours. F14 use systematically lower $\eta$ at all masses than we do, but launch their winds at much higher velocities, suggesting that their winds are much less likely to ever recycle.\footnote{The velocities from our simulation quoted in Table \ref{tbl:comptable} use an NFW potential to translate CGM velocities discussed in Section \ref{sec:windvel} into launch velocities.  We have also tried this exercise using a singular isothermal sphere potential, which may be more appropriate for describing massive (L*) halos that are baryon dominated in the center at low redshift \citep{chan_etal15}. This results in launch velocities up to 30\% faster than the NFW results, bringing us to closer agreement with F14 for sufficiently massive halos.}

At low redshift, F14 use values of $\eta$ that are in better agreement with our results, while V13 continue to have systematically higher values of $\eta$. For example, V13 use winds with $\eta \approx 7$ at $z=0$ for a Milky Way-mass galaxy ($v_{max} \approx 200$), while our simulations find that these halos typically have $\eta \ll 1$. Observations generally show that galactic winds are weaker at low redshift (e.g. \citealt{heckman01, heckman_etal15}), which is generally consistent with our results, but in tension with V13. F14 uses $\eta\approx1$ for a $10^{12} \Msun$ halo. As demonstrated in Figure \ref{fig:wiggle-loz}, the outflows seen in \textbf{m12i} at low redshift are most likely not related to winds generated by stellar feedback, and are generally some combination of random gas motion in the halo and close passages of satellites. The values for velocity and $\eta$ discussed here should be treated as upper limits. 

Although the winds in our simulations have lower values of $\eta$ than those of V13, and lower velocities than those of F14, the FIRE simulations - like V13 and F14 - nonetheless roughly reproduce the $M_*-M_h$ relation \citep{hopkins_etal14}. The key to understanding how this is possible may be the burstiness of star formation in the FIRE simulations. Since the consequence of each burst of star formation is the dispersal of the ISM, the resultant wind not only carries out the gas available for star formation, but also has strong dynamical effects on the halo and galaxy. In other words, although the galactic wind bulk properties in the FIRE simulations are different than the prescriptions used in V13 and F14, their detailed dynamics (as well as their phase structure) may result in different dynamical states for the halo and galaxy, thus limiting the efficiency of star formation.

\begin{footnotesize}
\ctable[
caption={{\normalsize Comparison of galactic wind mass-loading factors and wind velocities in our simulations to other subgrid models}\label{tbl:comptable}},center,star
]{lcccccccccccccccccccl}{
\tnote[ ] {
\textbf{(1)} Name: simulation used\\
\textbf{(2)} z: redshift considered\\ 
\textbf{(3)} $M_h$: halo mass in $\Msun$\\
\textbf{(4)} $\sigma_{1/3} $: radial velocity dispersion within $\frac{1}{3} \Rvir$ as estimated from our simulations (km/s)   \\
\textbf{(5)} $v_{max}$: maximum circular velocity in halo as measured in our simulations (km/s)   \\
\textbf{(6)} $v_{wind}$: wind launch velocity from the FIRE simulations, computed by using direct measurement of median and 95th percentile wind velocity at $0.25 \Rvir$ and re-scaled to launch velocity using the NFW potential (km/s) \\
\textbf{(7)} Same as \textbf{(6)} but median and 95th percentile wind velocity at $0.25 \Rvir$ are computed with our fitting formula (Equations \ref{eq:vwind50} and \ref{eq:vwind95}) and then re-scaled to launch velocities (km/s).
\textbf{(8)} $v_{wind, F14}$: approximate wind launch velocity computed using \citet{ford_etal14} prescriptions (km/s).  \\
\textbf{(9)} $v_{wind, V13}$: approximate wind launch velocity computed using \citet{vogelsberger13} prescriptions (km/s). \\
\textbf{(10)} $\eta$: Direct measurement of mass-loading in FIRE halos \\
\textbf{(11)} $\eta_{fit}$: mass-loading factor computed using our fit formula (Equations \ref{eq:vc_fit_little} and \ref{eq:vc_fit_big}). \\
\textbf{(12)} $\eta_{F14}$: approximate mass-loading factor computed using \citet{ford_etal14} prescriptions \\
\textbf{(13)} $\eta_{V13}$: approximate mass-loading factor computed using \citet{vogelsberger13} prescriptions. 
}
}{
\hline\hline
\multicolumn{5}{c}{halo properties} &
\multicolumn{5}{c}{wind velocity} &
\multicolumn{5}{c}{mass-loading } \\
\hline
\multicolumn{1}{l}{Name } &
\multicolumn{1}{l}{z} &
\multicolumn{1}{l}{$M_h$}  &
\multicolumn{1}{l}{$\sigma_{1/3} $}  &
\multicolumn{1}{l}{$v_{max}$}  &
\multicolumn{1}{c}{\ \vline } &
\multicolumn{1}{l}{$v_{wind}$}  &
\multicolumn{1}{l}{$v_{wind, fit}$}  &
\multicolumn{1}{l}{$v_{wind, F14}$}  &
\multicolumn{1}{l}{$v_{wind, V13}$}  &
\multicolumn{1}{c}{\ \vline} &
\multicolumn{1}{l}{$\eta$}  &
\multicolumn{1}{l}{$\eta_{fit}$}  &
\multicolumn{1}{l}{$\eta_{F14}$}  &
\multicolumn{1}{l}{$\eta_{V13}$}  \\
\hline
m10 & 0.25 & 7.1e9  &  20  & 33 &  & 73-76 & 79-99 & 170 & 80 &  & 45 & 54 & 28 & 270 \\
m11 & 0.25 & 1.4e11 & 57 & 83 &  & 200-230 & 210-270 & 410 & 210 &  & 7.8 & 3.2 & 3.5 & 42 \\
m12i & 0.25 & 9.8e11 & 120 & 180 &  & 340-350 & 400-530 & 770 & 460 &   & 0.31 & 1.6 & 1.3 & 8.9 \\
m10 & 1.25 & 4.3e9 & 21 & 33 & & 68-87 &  71-95 & 180 & 84 &  & 110 & 89 & 26 & 270  \\
m11 & 1.25 & 1.0e11 & 60 & 90 &  &230-310 & 200-290 & 460 & 230 &  & 10 & 5.8 & 3.1 & 36 \\
m12i & 1.25 & 6.0e11 & 100 & 150 &  &350-510 & 370-530 & 750 & 380 &  &2.3 & 3.3 & 1.5 & 13 \\
m11 & 3.0 & 2.8e10  & 49 & 70  &  & 140-210 & 150-230 & 410 & 180 &  & 14 & 14 & 4.7 & 59 \\
m12i & 3.0 & 1.2e11 & 63 & 110 &  &230-330 & 250-390 & 500 & 280 &  &8.0 & 8.7 & 2.8 & 24  \\
z2h506 & 3.0 & 2.3e11 & 86 & 130 &  &330-530 & 320-490 & 670 & 330 &  & 6.5 & 7.0 & 1.7 & 17 \\
z2h506 & 2.0 & 1.2e12 & 140 & 220 &  & 430-600 & 500-760 & 1100 & 560 &  & 6.5 & 3.3 & 1.0 & 6.0 \\
\hline\hline\\
}
\end{footnotesize}

\subsection{Implications for galaxy evolution}
\label{sec:discussion_galev}
Our analysis has primarily focused on measuring the mass-loading factor via outflow rates in the inner regions of the CGM (at $0.25 \Rvir$). In Section \ref{sec:etafit} we briefly devoted attention to demonstrating that in the L*-progenitors at high redshift, \textbf{m12v and m12i} specifically, about a third of material that is ejected into the CGM and crosses $0.15 \Rvir$ eventually flows out through the virial radius of the halo.

The fact that these numbers are not 100\% implies that there is a significant amount of material that is initially ejected into the CGM, but later able to recycle back into the inner halo and the galaxy. Even gas that flows out of the virial radius is not necessarily permanently unbound from the halo. This is true by construction, as our fiducial choice of $v_{cut}=0 \mathrm{km/s}$ is below the local escape velocity. In addition, the gravitational potential and virial radius of all central halos continue to grow with time, and may re-absorb the ejected material. Our results imply that the CGM of massive halos hosts a vast reservoir of gas that has been enriched by local high-redshift and intermediate-redshift outflows which helps fuel later stages of galaxy formation. 

It is interesting to estimate the amount of gas pushed out of halos, and to study the evolution of the halo baryon fraction, which is a good tracer for the aforementioned reservoir. In Figure \ref{fig:baryonfraction} we show the baryon fraction within $\Rvir$ over an interval $3.0 > z > 0$ for \textbf{m12v, m12q, m12i, m11,} and \textbf{m10}, as well as $3.0 > z > 2.0$ for \textbf{z2h506}. We plot the baryon fraction  alongside a quantity which describes the cumulative mass of outflowing gas which has traversed the virial radius since $z=9$, relative to the baryon budget of the halo if it were to contain the cosmic mean fraction of baryons. We call this quantity the expelled fraction. This figure can be used to trace the role that galactic winds play in the overall evolution of the halo. We note that for many of the halos considered, the sum of the baryonic mass fraction within the halo at $z=0$ and the expelled fraction add up to a number that is near unity. This is not necessarily true by construction, as we do not consider whether outflows that cross $\Rvir$ are actually unbound form the halo, do not account for the growth of $\Rvir$ through cosmic time, and only follow a single massive progenitor. A complete understanding of what sets the baryon fraction of a given halo requires us to consider all of these processes, as well as any other process that modifies the ratio of accretion rates for gas and dark matter. Nonetheless, the expelled fraction is a useful diagnostic.

A low-mass dwarf like \textbf{m10} already has a depleted baryon fraction by $z=3$ and it becomes further depleted by a series of powerful outflows between $z=3$ and $z=1.5$. At lower redshift, outflow rates are weaker but infall is only able to replenish a small fraction of what has been lost. Infall rates may be diminished due to heating of the IGM by the UV background \citep{thoul96}. The sum of the baryon fraction and the expelled fraction are far below unity at all times, suggesting that the UV background plays a role in preventing accretion continuously, starting at very high redshifts \citep{gnedin00, faucher-giguere11b}. The somewhat more massive dwarf (\textbf{m11}) can more efficiently accrete the UV heated gas, allowing it to have a high baryon fraction while still maintaining powerful outflows until late times ($z\approx0.6$), when a merger-driven burst expels a huge amount of gas, apparently sealing its fate. 

In the L*-progenitor \textbf{m12v}, star formation is more efficient at early times, building up the stellar fraction to \textasciitilde15\% of the baryonic budget by $z\approx1.5$ while maintaining an expelled fraction of under 30\%. However, this changes during the interval $1.5 > z > 1.0$, when the halo undergoes its most prominent burst of star formation, reaching $\dot{M}_{*} \approx 40 \Msunyr$. After this powerful burst, as in the late-time burst of \textbf{m11}, we see a rapid rise in the expelled fraction. Similar events occur sometime during the med-z interval in both \textbf{m12q} and \textbf{m12i}. In the case of \textbf{m12v}, what follows is 7 Gyr of evolution where the expelled fraction and baryon fraction stay relatively constant. Although accretion does not cease at late times, it is slow, and only able to partially repair the damage done by the outflows of the intermediate redshift era. This implies that some of the ejected material never returns back into the halo that ejected it, suggesting that such gas can transport metal polluted gas into regions of the IGM that are far from galaxies.

Following the intermediate-redshift major outflow events, the L*-progenitors are left with enough gas to continue star formation, but with a somewhat depleted reservoir in the CGM. The three halos exist on a spectrum of their capacity to form stars at late times. \textbf{m12q} has the most efficient star formation at early times, and violent outflows that did permanent damage to the baryon fraction prior to $z=3$, followed by more violent activity at intermediate redshift. At late times, star formation rates are the lowest of the three L*-progenitors, and it is structurally most similar to an elliptical or lenticular galaxy (Van de Voort et al., in preparation). \textbf{m12v} establishes a stable gaseous disk at late times, but earlier outflows were able to deplete the CGM significantly at intermediate redshifts. Finally, \textbf{m12i} is able to keep the largest reservoir of gas in the CGM, allowing it to have sustained and active star formation at late times. We also find that the central galaxy of \textbf{m12i} has a significant thin gaseous and stellar disk at late times.


The feedback model employed in our simulations starts at local scales where the energy and momentum injection rates are taken from stellar evolution models and simple approximations to feedback processes including supernovae and radiation pressure are used to couple the energy and momentum to the ambient gas. Coupling with the local gas properties and galactic conditions then influences the prevalence, strength, and nature of the emergent phenomenon of global galaxy-scale winds. We have identified that variations in halo mass and cosmological redshift of the galaxy at the time of wind-launching systematically influence the velocities and mass-loading factors of the winds. The underlying mechanisms that translate local feedback from star-forming regions to galactic winds will be better explored in a future, more focused study. We refer readers to \citet{hopkins_etal14} for evidence that neither radiation pressure nor supernovae alone are sufficient to reproduce the results  of the fiducial FIRE feedback model - namely, star formation histories and stellar-mass halo-mass relations. It is evident that hydrodynamic processes, gravitational deceleration, as well as the multitude of phenomena included in our feedback model affect galaxies of different masses and epochs at different efficiencies.  



\section{Conclusions}
\label{sec:conclusion}

We have explored the relationship between star formation and outflows in the CGM in the FIRE hydrodynamic cosmological zoom-in simulations.  Our key results include: \\


\noindent (i) We have shown that star formation at high and intermediate redshifts ($z>1$) occurs almost exclusively in episodic bursts. The feedback from each burst of star formation drives a "gusty" galactic superwind through the CGM. These superwinds are sometimes capable of ejecting the entirety of the dense ISM from the galaxy and can temporarily suppress gas infall in the inner halo (e.g. Figure \ref{fig:outandinflows_m12v}). However, at lower redshifts we have found that in massive halos with $M_h \approx 10^{12} \Msun$, star formation becomes less bursty and significant outflows largely cease  (e.g. Figure \ref{fig:wiggle-loz}).\\

\noindent (ii) We have quantified $\eta$ - the mass-loading factor - which is defined as the ratio of gas mass carried out in galactic winds to the mass of stars formed in the galaxy for many of the halos in our simulations over specified redshift intervals. At $z>2$, we find values of $\eta$ as measured in the inner CGM ($0.25 \Rvir$) are above unity for massive LBG-progenitors, while $\eta \approx 10$ is typical for L*-progenitors, and $\eta \approx 100$ are found in low-mass dwarfs. We found that $\eta$ varies systematically with halo mass ($M_h$), stellar mass ($M_*$), circular velocity ($v_c$), and redshift. The relationship between $v_c$ and $\eta$ (Figure \ref{fig:MassLoading5}) at low mass is steeper than expected for the simplest energy conserving wind model, and at high masses it is consistent with the simplest momentum conserving wind models (both assume that wind velocities are proportional to $v_c$).\\

\noindent (iii) We provide convenient fit functions for $\eta$ and outflow velocity, which can be used as a starting point of galactic wind implementation in large-volume simulations and semi-analytic models (see Equations \ref{eq:vc_fit_little} - \ref{eq:vwind95} and Figures \ref{fig:MassLoading5} - \ref{fig:windfall}). \\

\noindent (iv) Gas throughout the halo is dynamically disturbed following particularly powerful outflow events, and star formation can be suppressed for significant lengths of time.  This regulatory mechanism helps to keep halos on the $M_*-M_h$ relation throughout cosmic history (see \citealt{hopkins_etal14}). Large-volume simulations which use subgrid models with continuous star formation to match the $M_*-M_h$ relation must assume higher values of $\eta$ and/or wind velocity than what was found in our work.  \\ 

\noindent (v) During the intermediate-redshift epoch ($2 > z > 1$) that coincides with the peak of cosmic star formation, the dwarfs and L*-progenitors in our sample drive winds that carry material beyond the CGM, into the IGM. As cosmic accretion rates begin to decrease at these times, many halos can no longer easily replenish their supply of baryons. \\

\noindent (vi) The baryon fraction at low redshift ($z<1$) remains low in dwarf galaxies, but reaches 40-90\% of the cosmic mean for the L*-progenitors by $z=0$ (Figure \ref{fig:baryonfraction}).\\ 

\noindent (vii) At low redshift ($z<1$), dwarfs with a sufficient supply of gas can continue to episodically form stars and drive winds with $\eta > 5$. At the same time, L*-progenitors no longer drive outflows into the CGM through stellar feedback, and stable gaseous disks develop. Stars still form at low to moderate rates within these disks, but the dense ISM is not dispersed by outflows. This behavior demonstrates that our feedback model, which treats early radiative feedback from young stars as well as supernovae, is simultaneously capable of producing powerful starburst-driven outflows at early times and the formation of gaseous disks at late times.  \\

\noindent In companion papers, we will address the phase structure of the CGM at all times, provide a more detailed history of its composition and origin, determine its kinematic structure and study observational signatures of galactic winds.\\

\noindent We thank Tsang-Keung Chan, Xiangcheng Ma, Daniel Angl{\'e}s-Alc{\'a}zar, Freeke van de Voort, Robert Feldmann, Chris Hayward, and Michael Anderson for various constructive suggestions. We thank Romeel Dav{\'e} for providing details on wind implementation for our simulation comparison section. We thank the anonymous referee for providing suggestions that improved this work. DK was supported by an Hellman Fellowship and NSF grant AST-1412153, and by funds from the University of California San Diego. Support for PFH was provided by the Gordon and Betty Moore Foundation through Grant 776 to the Caltech Moore Center for Theoretical Cosmology and Physics, by the Alfred P. Sloan Foundation through Sloan Research Fellowship BR2014-022, and by NSF through grant AST-1411920. CAFG was supported by NSF through grant AST-1412836, by NASA through grant NNX15AB22G, and by Northwestern University funds. EQ was supported by NASA ATP grant 12-APT12-0183, a Simons Investigator award from the Simons Foundation, the David and Lucile Packard Foundation, and the Thomas Alison Schneider Chair in Physics at UC Berkeley. The simulations analyzed in this paper were run on XSEDE computational resources (allocations TG-AST120025, TG-AST130039, and TG-AST140023). We would like to thank the Simons Foundation and the participants of the {\it Galactic Super Winds} symposium for stimulating discussions. We would also like to thank the Kavli Institute for Theoretical Physics and the participants of the {\it Physics of Star Formation Feedback} program for interactions that improved this work.\\

\bibliography{galaxies}

\begin{appendix}

\section{Methods for calculating outflow rates}\label{sec:appendix:outflowrate}

\subsection{Outflow rate calculation}

Here, we discuss the nuances of the method we used to measure the outflow rate, which was introduced in Section~\ref{sec:measure}. In this section, we refer to our primary method as the \textit{Instantaneous Mass Flux} method. We describe it in detail, and compare it to another method, \textit{Interface Crossing}, which relies on particle tracking. This comparison serves as a consistency check for the methods.  

\subsubsection{Method 1: Instantaneous Mass Flux}
\label{sec:appendix:flux}
If outflow rates were time-steady, and followed a constant velocity distribution, this method would robustly capture the outflow rate at a given radius. In reality, outflows are time-variable and the velocity distribution of outflowing particles is unique to each episode. This implies outflow rates may not be sufficiently sampled when $\frac{dL}{\Delta t} < v_{SPH}$. Here, $\Delta t$ is the time interval between snapshots, and $v_{SPH}$ is the velocity of an SPH particle in the outflow, and $dL$ is the shell length. In this regime, when an outflowing particle that originated interior to the shell traverses the entirety of the shell between snapshots, it is not included as in the rate measurement.  On the flip side, it is possible to overemphasize the prevalence of fast-moving particles if the shell happens to encapsulate a narrow, coherent burst at the time of sampling - this transient, high outflow rate will be extrapolated for the entirety of the time interval. Over a large enough sample of bursts, these two effects should average out, and average measured outflow rates would reflect true values.

Despite the limitations of this method, we show that it generally provides accurate estimates of the outflow rate as long as  $\frac{dL}{\Delta t}$ is not too far from the typical velocity of SPH particles in outflows. In our high-redshift regime, the sampling of our simulation snapshot interval (\textasciitilde 30 Myr) and choice of $dL$ ($0.1 \Rvir$, which is 3-5 physical kpc for the L* progenitors) would be ideal for sampling particles with $v_{rad}$ between $100-200 \mathrm{km/s}$. We find that this is a typical velocity for much of the outflowing material in these halos, though it may miss the fastest material.

Measuring outflow rates with this method gives \textit{instantaneous} results. This is most relevant when contrasted to particle tracking methods that apply no velocity cut to particles deemed outflows, instead relying only on position differences (e.g. \citealt{ford_etal13a}). Outflows found through particle tracking methods are not necessarily confined in a narrow range of radii.

\subsubsection{Method 2: Interface Crossing}
\label{sec:appendix:cross}
In this method, particles that are marked as outflows via Equation  \ref{eq:lala} at a given snapshot are traced back exactly one snapshot. We track the change in each particle's radial distance from the halo center, to determine if it has crossed one of several pre-determined radial ``interfaces''. We determine the location of the interfaces by using the midpoints of the $dL = 0.1\Rvir$ shells employed in the \textit{Instantaneous Mass Flux} method. The outflow rate at each interface is simply computed by dividing the total mass of all outflowing particles that passed this interface since the previous snapshot, $\sum m_{SPH}$, by the time elapsed since the previous snapshot, $\Delta t$. Each particle may cross more than one interface per snapshot, and will be counted as part of the outflow rate at every interface it traverses.  When this method is employed, fast-moving particles are always accounted for, regardless of their distance from the halo center.

\begin{figure}
\vspace{-0.2cm}
\includegraphics[width=\columnwidth]{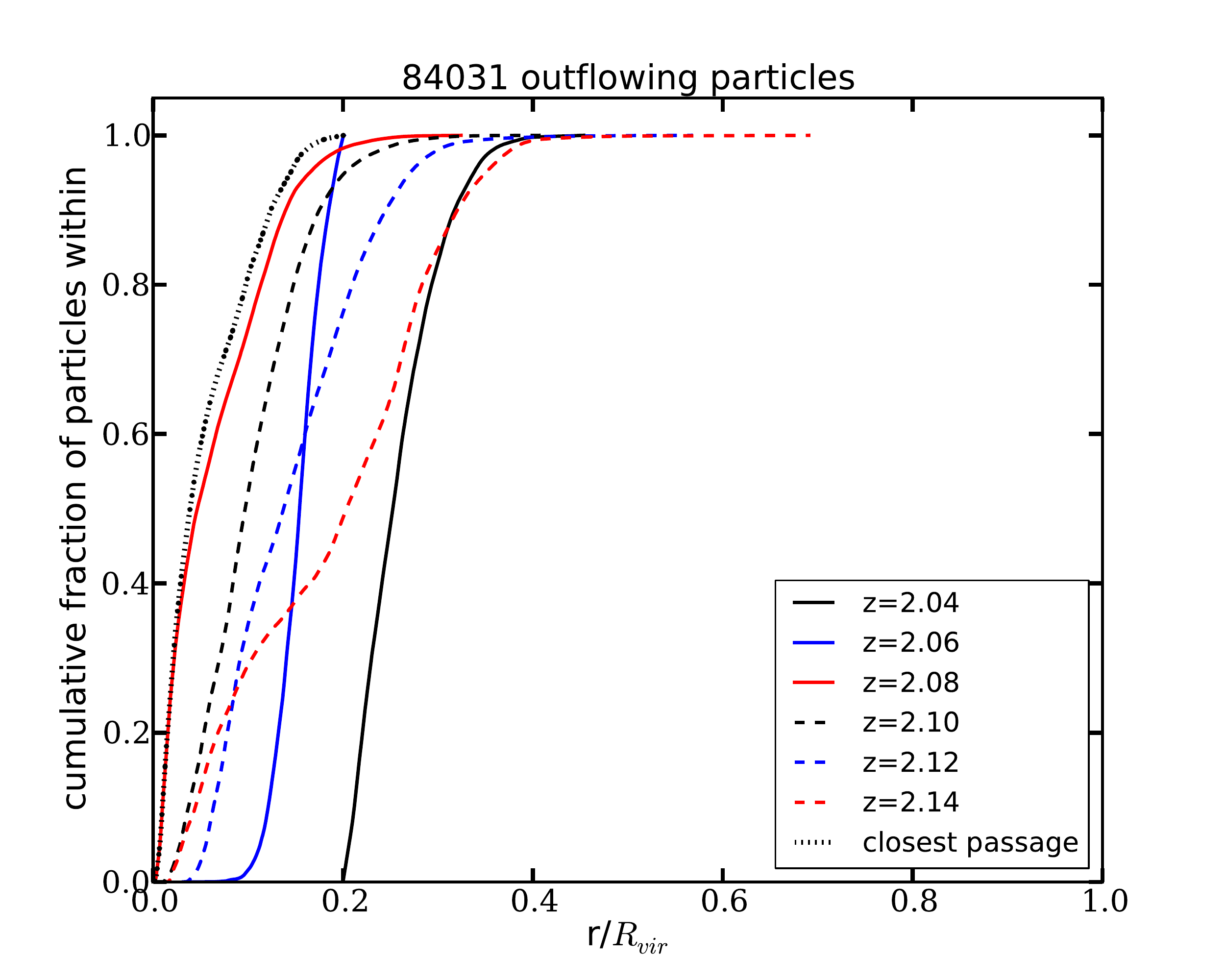}
\vspace{-0.0cm}
\caption{ Cumulative distribution of outflowing particles during a major outflow event which crossed the $0.2 \Rvir$ threshold since the last timestep (black). These particles are counted towards the outflow rate according to the \textit{Interface Crossing} method. We follow these particles backwards in time to the previous 5 snapshots from $z=2.04$ to $z=2.14$, and confirm that the majority were within $0.1 \Rvir$ two snapshots (\textasciitilde 60 Myr) ago. The closest halo-centric passage, for each particle within the last 5 snapshots (\textasciitilde 150 Myr) is also shown (dotted line). At the earliest time considered, we see that many of the particles now in the outflow were actually accreted form outside of $0.1 \Rvir$ only to be ejected by stellar feedback. }
\vspace{0.3cm}
\label{fig:outflow_origins}
\end{figure}

The capacity of this method to pick out outflowing material is shown in Figure \ref{fig:outflow_origins}. Here, we see the outflow following a major burst of star formation in the central halo of \textbf{m12v} at $z\approx2$ using $v_{cut} = \sigma_{1D}$. The outflowing particles are picked out by the interface-crossing method using a threshold at $0.2 \Rvir$. At this epoch, the particles are mostly contained between $0.2 \Rvir$ and $0.4 \Rvir$, but they all resided in the inner regions of the halo and the ISM before beginning their outward trajectory two snapshots (60 Myr) ago. This implies that outflows usually start in the galaxy and ISM, and usually do not contain much additional swept-up material from the CGM.

This method has a few other caveats. As slow-moving and fast-moving particles are weighted equally, the velocity cut from Equation \ref{eq:lala} could influence the measured flux more drastically than in the case of the \textit{Instantaneous Mass Flux} method. In addition, because of the specific crossing thresholds, centering errors in the halo catalog from snapshot to snapshot can generate spurious signal, particularly when $v_{cut} = 0$ is used. 

\begin{figure}
\vspace{-0.2cm}
\includegraphics[width=\columnwidth]{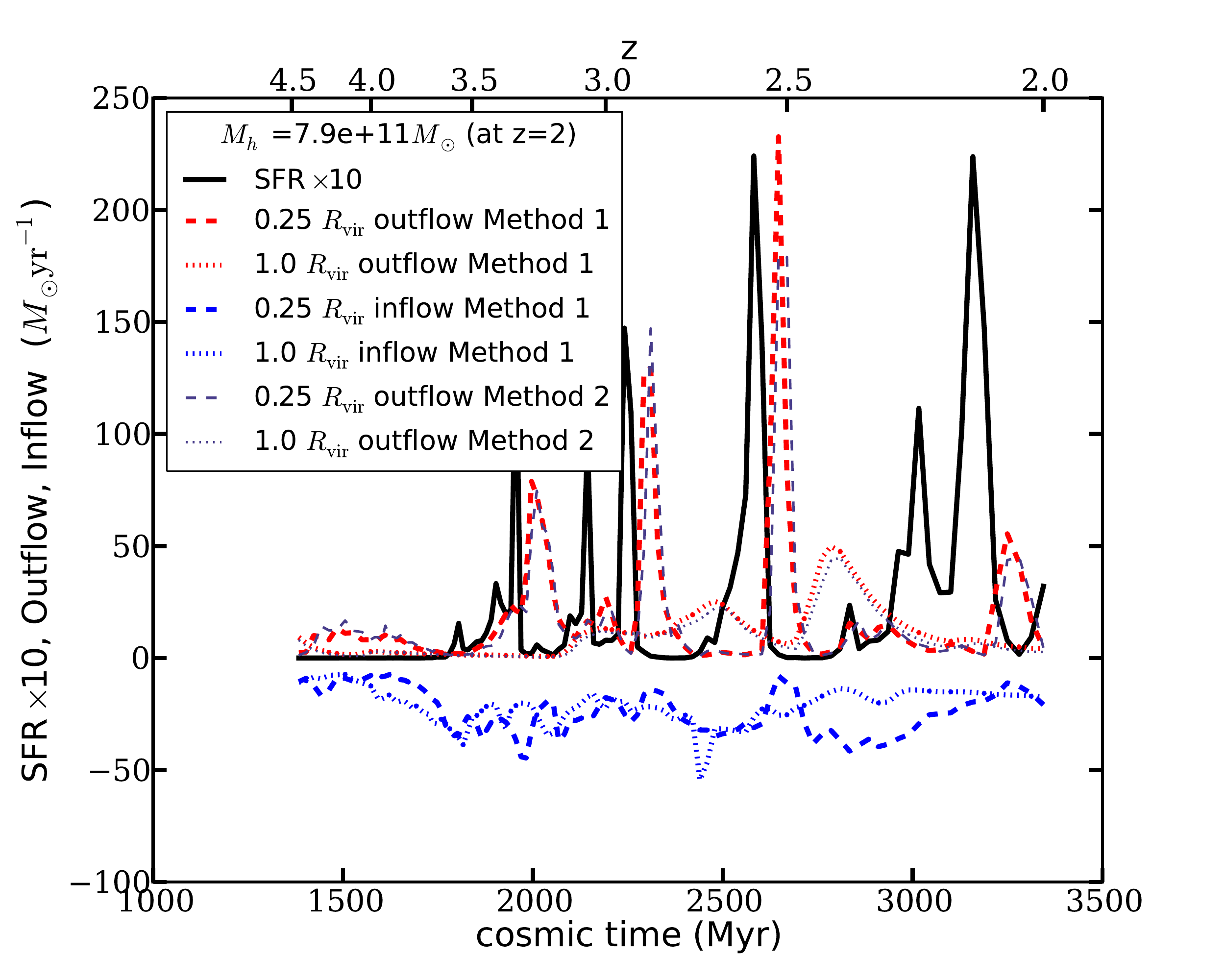}
\vspace{-0.0cm}
\caption{ Time evolution diagram for \textbf{z2h400} from $4.5 > z > 2.0$ using the same format as Figure \ref{fig:wiggle-hiz}. Red outflow rates and light blue inflow rates were computed using the standard \textit{Instantaneous Mass Flux} method (Method 1). We also show the outflow rate at $0.25 \Rvir$ and $1.0 \Rvir$ as computed by the \textit{Interface Crossing} Method (Method 2) in dark blue. We see that for the most part, the methods produce remarkably similar values for outflow rate at both radii. The star formation rate is plotted in black.}
\vspace{0.3cm}
\label{fig:wiggle_grand}
\end{figure}

A comparison between the outflow rate derived from using \textit{Instantaneous Mass Flux} and \textit{Interface Crossing} on \textbf{z2h400} is shown in Figure \ref{fig:wiggle_grand}. The similarity between the outflow rates as measured by both methods confirms their consistency. The measured values of $\eta$ for this halo differ by only 5\% between the two methods.

\subsection{Alternative velocity cuts}
\label{sec:appendix:sigma}

As mentioned in Section \ref{sec:measure}, the majority of measurements quoted in the text define outflows as all particles with $v_{rad} > v_{cut}$ with $v_{cut} = 0$. While using this cut ensures that we account for all gas mass that is crossing our specified radii, it may not be the best definition for what should count as a galactic wind. Observationally, a wind should be distinct from the randomized dispersive motions of gas in the halo. For this reason, we provide additional measurements of $\eta$ using $v_{cut} = \sigma_{1D}$, the 1-dimensional velocity dispersion of the halo.  We assume isotropic velocity dispersion such that $\sigma_{1D} = \frac{\sigma_{3D}}{\sqrt{3}} $.  We take $\sigma_{3D}$ from the halo finder as calculated explicitly using all species (gas, stars, and dark matter). This cut provides a simple scaling with halo mass that  filters out random motion not associated with galactic winds. We note that there is nothing that prohibits gas accretion with radial velocity $|v_{rad}| < \sigma_{1D}$, meaning that the total inflow rate may exceed the "infall rate". On the other hand, it is far less likely that the halo may launch powerful coherent winds with $v_{rad} < \sigma_{1D}$, as the local escape velocity in the galactic region will prohibit this material from reaching outer regions of the halo. 

In Table \ref{tbl:bigtable}, we demonstrate that typically, use of this threshold reduces measured outflow rates by \textasciitilde25\% in L*-progenitors compared to measurements with $v_{cut}=0$. However, the discrepancy may is more significant at low redshift: we have argued that $\eta$ is difficult to measure in the L*-progenitors at $z<0.5$ because the outflows appear uncorrelated to star formation, and this is supported by the low outflow rates of material with $v_{cut}=\sigma_{1D}$ in Figure \ref{fig:wiggle-loz}. 

\section{Instantaneous Methods for measuring $\eta$}\label{sec:appendix:eta}

\subsection{Instant rates}
\label{sec:appendix:instant_app}

Here, we discuss how we correlate outflow rates and star formation rates on the finest possible sampling timescale that is available in post-processing of our simulation suite, which is the interval between snapshots. 

As the star formation rates and outflow rates are highly variable, and are occasionally both measured to be zero, we impose a few restrictions on which snapshots we use as data points. First, we employ a requirement that the outflow rate must be above zero and the  specific star formation rate must be above $10^{-10} \mathrm{yr}^{-1}$. Such cuts are necessary to produce a statistically meaningful fit to data, as outlying points could differ by several orders of magnitude, and are likely the result of mismatched events rather than physical effects. We impose one other cut that is somewhat more arbitrary: snapshots are only counted if the outflow rate is more than 1\% of the maximum measured outflow rate for the given redshift interval in the halo. This restriction excludes additional data points produced by relatively insignificant outflows set by the halo's own activity pattern. We choose not to impose a symmetric requirement for star formation rate, as it excludes too much data when star formation is more continuous over an interval. While this could imply that $\eta$ is artificially boosted, we find measurements to be consistent with other methods. 

\begin{figure}
\vspace{0.2cm}
\includegraphics[width=\columnwidth]{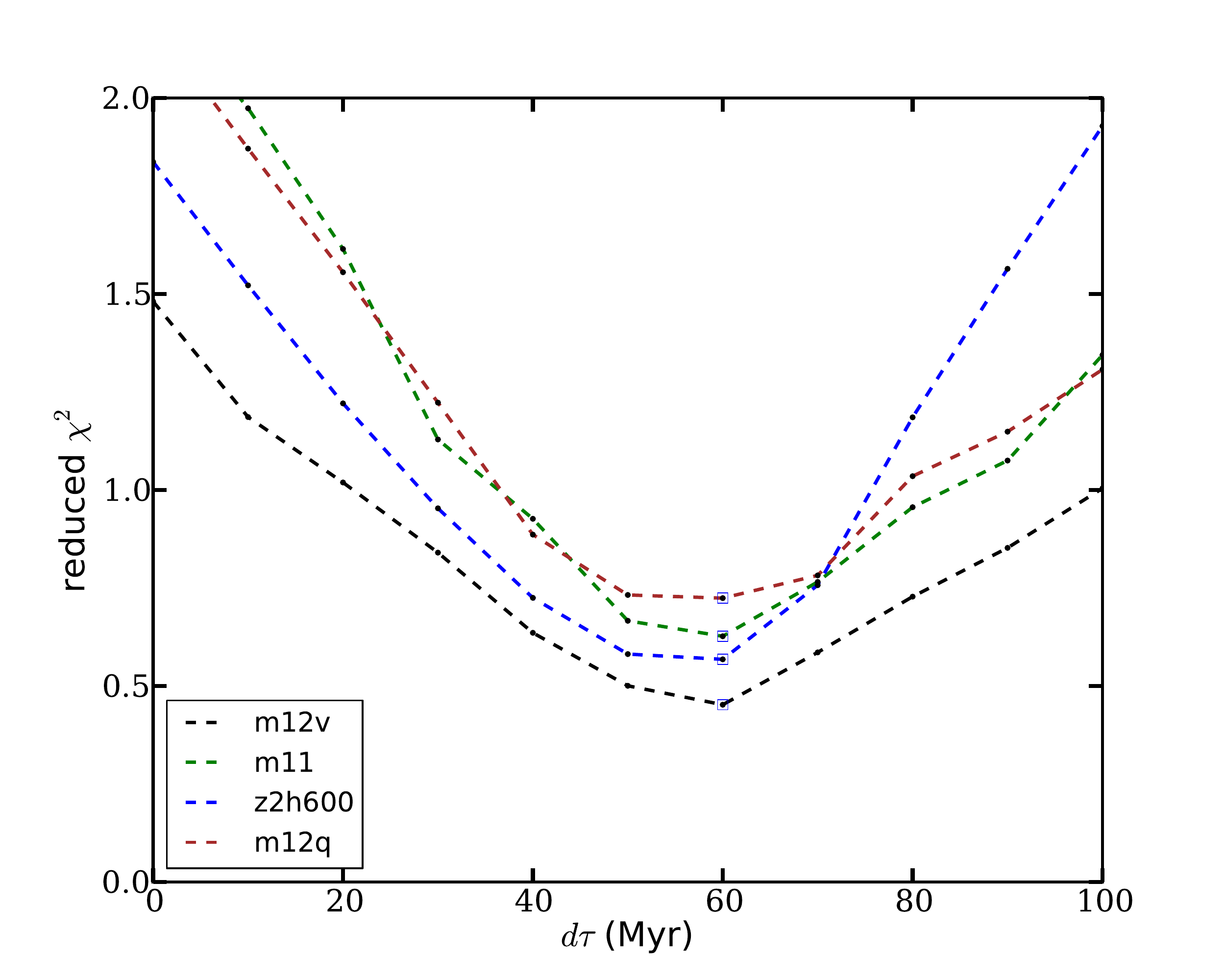}
\vspace{0.0cm}
\includegraphics[width=\columnwidth]{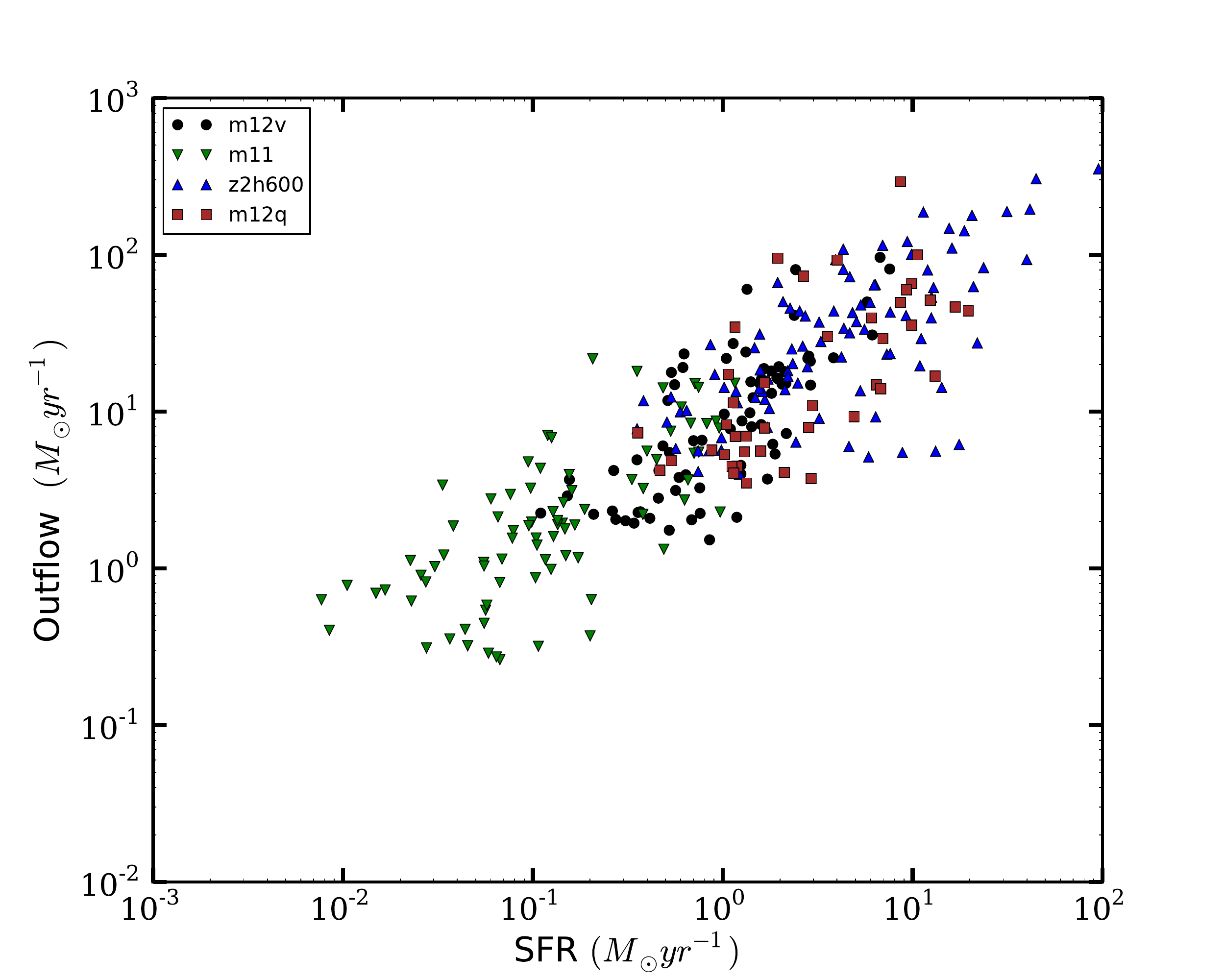}
\caption{ Top: Reduced $\chi^2$ statistic for fits to data points on the $\dot{M}_{out}(t+d\tau)$ vs $SFR(t)$ relation as a function of $d\tau$. The slope of the relation is fixed such that  $\dot{M}_{out}(t+d\tau) \propto SFR(t)$. i.e. linear relation. Each line represents one of the ``main'' halos in various simulations (\textbf{m12v, m11, m12q,} and \textbf{z2h600}) during the $4.0 > z > 2.0$ (high-z) interval. The minimized value of $\chi^2$ is marked for each halo with a box around the point. In all cases shown here, $d\tau=60$Myr is optimal. Bottom: instantaneous outflow rate vs. star Formation rate using $d\tau=60$Myr, using the same four halos over the high-z interval. While correlation for each halo is apparent, the scatter is still high.  }
\vspace{0.4cm}
\label{fig:instant_chisq}
\end{figure}

Each snapshot's star formation rate is matched with an outflow rate at a time $d\tau$ later. The timescale $d\tau$ provides a measurement of the delay between the formation of a population of stars, presumably near the halo center, and the time that the bulk of outflowing material arrives to a given location in the halo, in this case $0.25 \Rvir$. To understand which value of $d\tau$ provides the cleanest measurement of $\eta$, we perform an error analysis on fits for outflow rate vs. star formation rate for a range of values of $d\tau$ ($\dot{M}_{out}(t+d\tau)$ vs $SFR(t)$). We correlate points for each halo separately, over the three redshift intervals (high-z $4.0 > z > 2.0$, med-z $2.0 > z > 0.5$, low-z $0.5 > z > 0$). 

The results of our correlations for several of the ``main'' halos in the high-z interval are shown in Figure \ref{fig:instant_chisq}. A range of halo masses are represented in this simulation, but we see that  $d\tau \approx 60$ Myr provides the best correlation for all halos shown. Analysis of a larger, complete sample including all of the available simulations reaffirms that $d\tau \approx 60$ Myr is the mean, median, and mode of "most correlated" delay value for halos of all mass in the high-z interval. We have verified that this is remarkably close to the expected length of time it takes a gas parcel to travel between the halo center and $0.25 \Rvir$, if one assumes that the wind travels with constant velocity predicted from Equation \ref{eq:vwind50} (62 Myr, 57 Myr, and 54 Myr for \textbf{m11}, \textbf{m12i} and \textbf{z2h600} at $z=3$, respectively).  Performing the same analysis for the med-z and low-z intervals, we find best that the average preferred value of $d\tau$ shifts to 90 and 120 Myr, respectively. 

The emergence of this trend with redshift is expected because of the growth of physical $\Rvir$ relative to comoving $\Rvir$ with redshift (we consistently use shells of fixed comoving fraction of $\Rvir$ when computing outflow rates) for all halos, while the velocity distribution of outflowing particles during significant bursts does not change significantly. The lack of a significant trend for $d\tau$ as a function of halo mass suggests that the timescale derived from the ratio of $\Rvir$ to some characteristic wind speed is constant for all halos, and is well represented by $d\tau$.  We adopt 60 Myr, 90 Myr, and 120 Myr as uniform values of $d\tau$ at $0.25 \Rvir$ for the high-z, med-z, and low-z regimes, respectively. 

As can be seen from the bottom panel of Figure \ref{fig:instant_chisq}, the scatter about the $\dot{M}_{out}(t+d\tau)$ vs $SFR(t)$ relation is high even for the optimized value of $d\tau$. This is indicative of the complex physics that goes into wind generation, and shows that each burst of star formation and outflow episode must be studied individually. On top of this, we emphasize that our measurements of outflow rate was meant to capture material that is flowing into the CGM (which we define as $0.25 \Rvir$). In reality, it is possible for outflow rates to be high elsewhere in the halo even when it is low at this interface. We therefore caution against the interpretation that our optimized values of $d\tau$ represent uniquely important timescales on which winds should be observed following star formation.

It is even more difficult to correlate winds in the CGM of the outer halo ($1.0 \Rvir$) to star-forming events in the galaxy than at the inner halo ($0.25 \Rvir$). This is because, as we have shown in Section \ref{sec:wiggle}, outflows that reach the virial radius are typically moving more slowly than those at $0.25 \Rvir$, and leave the halo at a wider range of times. Using a similar $\chi^2$ optimization technique, we find an average value of $d\tau=200{\rm Myr}$ (during the high-z interval) to be optimal for correlating the SFR to outflow rate at $1.0 \Rvir$, with no obvious trend for halos to have higher or lower values of $d\tau$ depending on mass or $v_c$. 

We used this method to derive average values of $\eta$ for halos in our sample (i.e. measure the y-intercept of the best-fit correlation for each halo as shown in the bottom panel of Figure \ref{fig:instant_chisq}), and provide measurements in Table \ref{tbl:bigtable}.

\subsection{Episodic Rates}
\label{sec:appendix:episodes_app}

In this approach, the outflow and star formation rates are integrated over an interval sufficient to capture the entire stellar mass formed during a single burst of star formation, and the entire mass of the gas blown out in the outflow episode. If done properly, this method provides nearly the same statistical flexibility as correlating instantaneous rates, while also mitigating the stochasticity with using a single fixed value of $d\tau$. 

However, in order to best identify the episodes over which to integrate, we must employ a parametric algorithm that allows for flexibility in the duration of the bursts. This episode-finding algorithm must follow the time evolution of the halo (e.g. Figure \ref{fig:wiggle-hiz}), look for the tell-tale signs that a burst is beginning, integrate the star formation for only as long as the burst lasts, then integrate the outflow rate for only as long as the outflow is associated with said burst of star formation. We constructed such an algorithm, and verified its validity by ensuring that it was efficient in segregating the entire outflow and star formation history into distinct episodes. Parameters were adjusted to ensure that winds were not misattributed to star formation that was unlikely to be physically related to the winds and vice versa. Interestingly, we found that the best fits were achieved when we mandated that the starting times for integrating star formation and outflow rates were separated by timescales comparable to the optimal $d\tau$ measurements with the \textit{instant} method. 

Using this detection algorithm we find $\eta$ for each episode by dividing the total mass of gas expelled per episode by the total mass of stars formed. We then again combine the data for all halos in the same redshift interval and derive a correlation. We used this method to derive average values of $\eta$ over three redshift intervals for halos in our sample, and provide the measurements in Table \ref{tbl:bigtable}.

The true strength of the instantaneous methods, both \textit{instant} and \textit{episodic}, could be harnessed by aggregating simulation data and statistically analyzing $\eta$ and its dependence on halo properties at the epoch of a burst using data points from as many simulations as possible. However, we found that our simulation sample was not sufficiently large or complete to derive useful results with this approach. Future work with statistically representative halo populations, aided by analysis which investigates the origin of the systematic scatter, can make optimal use of these measurement techniques to thoroughly quantify galactic wind generation.

\section{comparison with other works}
\label{sec:appendix:Compare_Eta}
Here, we attempt to compare our findings for best-fit values of $\eta$ and wind velocity to the subgrid models employed in work by other groups. We focus on representative stat-of-the-art large-volume cosmological simulation projects that did not resolve star-forming regions within galaxies, and instead use parametric models to prescribe $\eta$ and wind velocity. Namely, \citealt{ford_etal14} (F14), and \citealt{vogelsberger13} (V13)\footnote{Illustris project; see also \citealt{vogelsberger_etal14, genel_etal14}.}. Both of these works assign wind velocities based on halo properties, and then decouple the wind particles from hydrodynamic interactions to ensure escape from the local environment. The following expands on the comparison summarized in Section \ref{sec:brief_eta_compare} and Table \ref{tbl:comptable}.

We emphasize that we cannot precisely reproduce the wind prescription employed in these simulations. While each describes prescriptions for $\eta$ and wind velocity at the time of launch, the results shown in this work have focused on measurements in the CGM. Furthermore, the subgrid prescriptions used by those authors involve properties such as the local velocity dispersion and stellar metallicity, which we did not consider in our fits. Since each group used different codes and initial conditions for their simulations, the resulting halos are also intrinsically different. As such, we can only compare approximations of their models to our results.

First, we convert our measured wind velocities in the CGM to approximate launch velocities near star forming regions. We take each outflowing wind SPH particle's kinetic energy at the time it is detected as an outflow at $0.25 \Rvir$ and assume that it has conserved total energy since being launched at $r_{launch} = min(0.02 \Rvir, 1 {\rm kpc})$. Therefore, we consider the difference in gravitational potential energy between the two locations. As an approximation for the potential, we assume that each halo follows an NFW profile with a concentration parameter given by \citet{dutton_maccio14}. We find that launch velocities at high redshift are typically different from CGM velocities at $0.25 \Rvir$  by a factor of \textasciitilde2 at the most, but they can differ by a factor of \textasciitilde5 or more at low redshift. In addition to NFW, we have also considered a singular isothermal sphere (SIS) potential, which better describes the inner regions of massive halos in the FIRE simulations (e.g. \textbf{m12i}, \citealt{chan_etal15}). We normalize the profile such that the circular velocity is equal for both NFW and SIS at $r = 0.25 \Rvir$. We have also tried normalizations based on the enclosed mass within $0.25 \Rvir$ in our simulations.  Launch velocities are found to increase by between 10-30\%, with the greatest increase seen in massive halos that are baryon-dominated at low redshift. 

On the other hand, our low-redshift estimate for \textbf{m12i} is also potentially misleading, since the outflow velocity measured at $0.25 \Rvir$ is likely to be random motion rather than a coherent star formation-driven galactic wind. As we saw in Section \ref{sec:radii}, $\eta$ may also differ between 0.25 $\Rvir$ and the inner regions of the halo. However, since F14 and V13 both consider only winds that can escape the innermost regions of halos, and re-couple winds when they are outside galaxies, we assume that our measurements of $\eta$ in the CGM constitute an appropriate analog. 

V13 determine bulk properties of galactic winds using the local 1-D dark matter velocity dispersion of the regions where star formation takes place. We have no way of reproducing this quantity precisely. However, they suggest that this quantity correlates to the halo's $v_{max}$ according to the work of  \citet{okamoto_etal10}\footnote{\citet{okamoto_etal10} actually suggest that $v_{max}$ is related to the velocity dispersion of gas.}. We can therefore measure the $v_{max}$ of halos in our simulation and approximate $\sigma_{1D,DM}$. $\eta$ is prescribed based on $v_w$, the wind velocity:  $\eta \propto v_w^{-2}$. For our comparison, we employ the same parameters used in the fiducial model of V13, which is clearly outlined in that work.

F14 use the velocity dispersion of the tightly bound "galactic" component of the halo. We approximate the relevant region to be $\frac{1}{3}  \Rvir$, based on the linking length used in their galaxy finding algorithm as described in \citet{oppenheimer08}. Wind velocity is set by this velocity dispersion multiplied by several boost factors (see \citealt{oppenheimer08, ford_etal13a, ford_etal14}). One of the boost factors depends on metallicity, which we compute using a fit relation for gas-phase metallicity as a function of galactic stellar mass for the FIRE simulations \citep{ma_etal15}. Another is a uniformly sampled random number between 1.05 and 2, which we simply take to be 1.5. They calculate $\eta$ for winds also using the inner velocity dispersion, with a scaling of $\eta \propto v_c^{-2}$ for $v_c < 75 {\rm km/s}$ and $\eta \propto v_c^{-1}$ for $v_c > 75 {\rm km/s}$\footnote{It is worth noting that the break in the power law employed in F14 occurs at $75 km/s$ which is close to the $60 km/s$ break we favor in our fit.}. The exact feedback formula employed in F14 has been provided to us by R. Dav{\'e} (private communication). We refer to the radial velocity dispersion within  $\frac{1}{3}  \Rvir$ as $\sigma_{1/3}$, which we measure directly from our simulations, and plug into their formulae. Although velocity dispersion fluctuates throughout a given redshift interval, we find our typical $\sigma_{1/3}$ are consistent within 10\% of values given by fits from \citet{hoeft_etal04}.  Values of $\eta$ are quoted for $10^{12} \Msun$ and $10^{11} \Msun$ halos in \citet{ford_etal14}, and we roughly reproduce these results.

We perform the comparison by using three halos in our sample at $z=3$, $z=1.25$ and $z=0.25$. We consider \textbf{m10}, \textbf{m11}, \textbf{m12i}, and \textbf{z2h506}. We evaluate \textbf{z2h506} at both $z=3$ and $z=2$ for completeness, as it greatly increases its stellar mass at this interval. We measure $v_{max}$ directly from our halo finder to enable the comparison.  


\end{appendix}

\end{document}